\documentclass[submission, Phys]{SciPost}

\usepackage{url}
\usepackage{siunitx}
\usepackage{amsfonts}
\usepackage{amssymb}
\usepackage{tikz}
\usepackage{bm}
\usepackage{braket}
\newcommand{\I}{\mathrm{i}}
\newcommand{\E}{\mathrm{e}}
\newcommand{\D}{\mathrm{d}}
\DeclareMathOperator{\Tr}{Tr}
\newcommand{\PL}{{\!+}}

\definecolor{pred}{HTML}{800000}

\begin{document}

\begin{center}{\Large \textbf{
    Quantum walk on a graph of spins: magnetism and entanglement
}}\end{center}

\begin{center}
  Kevissen Sellapillay,\textsuperscript{1} and Alberto D.\ Verga\textsuperscript{1*} 
\end{center}

\begin{center}
  {\bf 1} Aix-Marseille Université, CPT, Campus de Luminy, case 907, 13288 Marseille, France
  \\
  * alberto.verga@univ-amu.fr
\end{center}

\begin{center}
  \today
\end{center}


\section*{Abstract}
{\bf
We introduce a model of a quantum walk on a graph in which a particle jumps between neighboring nodes and interacts with independent spins sitting on the edges. Entanglement propagates with the walker. We apply this model to the case of a one dimensional lattice, to investigate its magnetic and entanglement properties. In the continuum limit, we recover a Landau-Lifshitz equation that describes the precession of spins. A rich dynamics is observed, with regimes of particle propagation and localization, together with spin oscillations and relaxation. Entanglement of the asymptotic states follows a volume law for most parameters (the coin rotation angle and the particle-spin coupling).
}

\vspace{10pt}
\noindent\rule{\textwidth}{1pt}
\tableofcontents\thispagestyle{fancy}
\noindent\rule{\textwidth}{1pt}
\vspace{10pt}

\section{Introduction}

The concept of information was introduced as a physical quantity measured in bits by Leo Szilard in 1929 \cite{Szilard-1929kb}, and further explored by Rolf Landauer (1961) \cite{Landauer-1961uq} who establishes that erasing a bit is an irreversible process costing \(\ln 2\) of entropy (in units of the Boltzmann constant). However, as the laws of nature are quantum, information must ultimately be related to the quantum state. The quantum extension of a bit was proposed by Benjamin Schumacher (1995) \cite{Schumacher-1995uq} in order to demonstrate a quantum coding theorem analogous to the Shannon one in classical information theory \cite{Shannon-1948fj}. The quantum coding theorem shows that the von Neumann entropy  \(S = -\Tr \rho \log \rho\) is a measure of the information content of a general quantum system in a state described by the density matrix \(\rho\). In spite of the apparent similarity between Shannon and von Neumann entropies, quantum information is essentially different to its classical counterpart: general quantum states are entangled. 

Quantum information \cite{Bennett-1998jk,Preskill-1998cl,Watrous-2018} is a fundamental physical concept, now part of the curriculum \cite{Vedral-2006,Schumacher-2010zl}, which has a profound impact in vast domains, well beyond its obvious application to computing, as for instance in condensed matter through the phenomenon of many-body entanglement \cite{Amico-2008zj,Laflorencie-2016fk,Abanin-2019}. An interesting field of investigation opened at the frontier of matter and information \cite{Zeng-2019}. It is in this context that we propose here a model based on a quantum walk interacting with a network of spins, in order to manipulate the quantum state by local unitaries to create entanglement.

Discrete quantum walks \cite{Kempe-2003fk,Kitagawa-2012fk,Portugal-2013}, although initially proposed as an extension of the classical random walk \cite{Aharonov-1993fk} with possible applications in optics, were first investigated in relation with the theory of quantum simulation by Meyer in 1996 \cite{Meyer-1996sf}. In his seminal paper Meyer demonstrated that a quantum system evolving by discrete steps on a lattice (cellular automaton), cannot be homogeneous and local: in order to keep a unitary evolution one needs to introduce an internal degree of freedom associated with the walking particles and coupled to their motion. Applications of quantum walks range from the implementation of the quantum circuit universal model of computation \cite{Feynman-1986kx,Deutsch-1989,Lovett-2010rm,Kendon-2020} to the study of topological phases of matter \cite{Kitagawa-2010jk}. The experimental realization \cite{Broome-2010} demonstrated the practical feasibility of quantum walks, allowing the investigation of topological edge states \cite{Kitagawa-2012xy} and the measurement of topological invariants \cite{Flurin-2017,Cardano-2017,Xie-2020}. Generalizations from the one particle case to two entangled particles, including possible interactions, were also studied theoretically \cite{Omar-2006vn,Ahlbrecht-2012ec,Bisio-2018,Verga-2018} and experimentally \cite{Sansoni-2012kx,Schreiber-2012fk}; besides, Anderson localization was observed in an array of interferometers \cite{Crespi-2013xe}, and the effect of boson and fermion statistics simulated using the symmetries of the photon's state. Another important extension of quantum walks from regular lattices to general graphs \cite{Aharonov-2001ty,Tregenna-2003rp,Szegedy-2004ul}, led to interesting algorithms like the generalization of the Grover search \cite{Grover-1997}, showing that the square speedup over the classical algorithm applies to structured data \cite{Shenvi-2003fk,Ambainis-2004qq,Stefanak-2016}, or to algorithms aiming at determining graph isomorphism \cite{Berry-2011qq,Wang-2015}, although a solution to this problem is still open \cite{Mills-2019}. Finally, quantum walks are a starting point to explore many-body physics, once generalized to a quantum automaton \cite{Meyer-1996sf,Farrelly-2017,Bisio-2018nr,Arrighi-2019a}.

Loosely speaking, in quantum mechanics the state of a system can be modified by a unitary transformation or by a projection over a smaller Hilbert space. These two possibilities lead to two main models of quantum computing \cite{Nielsen-2006fv}: the circuit model theorized by David Deutsch \cite{Deutsch-1989}, and the measurement driven computation \cite{Raussendorf-2001uq,Raussendorf-2003rm,Nielsen-2003,Nielsen-2006fv}. In the first model a universal set of one and two qubits gates \cite{Barenco-1995,Lloyd-1996,DiVincenzo-2000cr} are used to create a suitable entangled state, while in the second model, a highly entangled state, a cluster state \cite{Briegel-2001fk,Hein-2006eu}, is modified by one qubit measurements to imprint the logical operations. In this fuzzy classification, quantum walks pertain to the `circuit' category, in which suitably local coin and motion operators entangle a state \cite{Ambainis-2003if} in order to coherently enhance the probability of certain configuration (as for example in the search problem). Information processing requires in addition to ressources, communication channels. A physical realization of a transmission line is a spin chain \cite{Bose-2003,Christandl-2004}. Therefore, very schematically a possible computing system is a network of spin chains connecting logical gates, implemented for example by interacting spins.

We present here a model in which the resource, the entangled state, is created by a walker on a network of spins, as already mentioned. In its simplest form, the spins are non interacting, and it is the motion of the particle that induces spin correlations and a global entangled state. This is reminiscent to the protocol used to entangle  carbon nuclear spins by the electron of a nitrogen vacancy in diamond; entanglement is produced by successive interactions of the electron with the neighboring nuclear spins (two qubit gates) \cite{Bradley-2019}. This protocol was successfully applied to experimentally demonstrate teleportation \cite{Pfaff-2014} and error correction \cite{Waldherr-2014}.

One may consider a physical motivation to generalize the usual setup of a quantum walk. Actually, in a simple discrete quantum walk the particle motion is determined by an internal degree of freedom, however, the space geometry in which the particle moves is imposed in the form of a lattice or more generally a graph. At each time step, the position amplitudes in one node of the graph is then distributed over the adjacent nodes. We can make an analogy with a tight binding model, in which an electron jumps between sites of a crystal. The main physical difference is that the crystal is itself a lattice of ions whose intricate interaction with the electron gas (for instance in a semiconductor) gives rise to both, the crystal structure and the characteristic energies of the electron hopping. This observation motivates our choice of assigning a material support to the graph in which the walker moves. A natural way is to associate a spin degree of freedom to the graph nodes \cite{Verga-2019}, or, as we do here, to the links between nodes. At variance to the model of Ref.~\cite{Verga-2019,Verga-2019b} the spins do not interact, are then located on the edges of the graph, and, as a consequence, the particle-spin interaction is independent of the local degree of the graph (which was the case in the node spin model). The interaction of the walker with the geometry gives rise to an extension of the usual model of a quantum walk to an interacting one. The main consequence of this generalization is that we leave the simple world of one particle to the complex many—body interacting quantum system.

Consequently the constituents of our system are a particle and a set of spins. As usual in a discrete quantum walk, the particle's motion between neighboring nodes of the graph is controlled by a coin (an operator acting on the particle internal degree of freedom); the coupling between the walker and the spin network is defined by an exchange interaction, as in a Heisenberg magnet. This model keeps some analogy, albeit with one free particle, with a magnetic system where free electrons interact with local magnetic moments creating an effective interaction between spins, like in the Ruderman-Kittel-Kasuya-Yosida exchange interaction and the Kondo lattice \cite{Ruderman-1954,Kasuya-1956,Nolting-2009}. The idea is that the walker, which may spread ballistically over the graph, a defining feature of the quantum walk already noted in the original paper by Aharonov et al.\ \cite{Aharonov-1993fk}, can create spin entanglement at a linear rate in a many-body system \cite{Lauchli-2008,Zhang-2015,Nahum-2017qf,Hackl-2018}. 

Quantum walks can be used to simulate physical systems; see for instance the recent experimental realization of a periodically driven Chern insulator \cite{DErrico-2020}. Meyer \cite{Meyer-1996sf} showed that the one particle case of his quantum automaton, in fact a quantum walk, reduces to the Dirac equation in the continuum limit. Since then, the continuum limit of quantum walks was extensively explored to simulate neutrino oscillations \cite{Di-Molfetta-2016kn}, Weyl fermions \cite{Bisio-2017}, or the Dirac equation in two dimensions \cite{Arrighi-2018}, to give a few examples.

In the present model the dynamics of spins adds to the particle motion; we investigate the continuum limit and show that, for a choice of parameters, the dynamics can be modeled by a Landau-Lifshitz equation \cite{Landau-1935fk}. The relation with the magnetization dynamics, described by the Landau-Lifshitz equation, can be understood by observing that the exchange interaction of a moving spin with the canted fixed spins should induce a torque in much the same way as a spin polarized current induces a so-called spin-transfer torque \cite{Slonczewski-1996lq,Tatara-2019} (note however that the current is here replaced by the current probability of one quantum particle).

In order to process information the physical system must be able to transfer the information through the network and also to stock it at some location. This means that it is desirable to find regimes of the quantum walk in which the quantum state can spread over the available space or instead, can remain localized. We focus here on the simplest case, in which the quantum walk is defined on a one dimensional lattice with spin interaction absent, to show that even in this geometry, and with only the local particle-spin coupling present, a rich variety of dynamical regimes exists. We observe, both propagation and localization of the particle distribution, together with spin oscillations, irregular dynamics and relaxation to some uniform state. As in condensed matter topological phases allow the control of the conduction to insulator transition, quantum walks exhibit analogous topological properties \cite{Kitagawa-2012fk} in which the interaction may play an important role \cite{Verga-2018}. We investigate the topological properties of the particle-spin quantum walk to identify the effect of the interaction on an otherwise single step Dirac walk, which is known to support edge states at the interface between two non equivalent phases. (We call ``Dirac'' the walk with a \(SU(2)\) coin that converges to the Dirac equation in the continuum limit \cite{Strauch-2007,Di-Molfetta-2012fv}.)

What is special about the quantum state is that we need an exponentially large number of parameters to specify it; this is a straightforward consequence of the superposition principle: while a classical state is defined by a point in phase space, the quantum one is a superposition of all the basis states \cite{Jozsa-1999a}. The amount of information encoded in the quantum state is potentially exponentially larger than in a classical state, as shown by the existence of superdense coding \cite{Bennett-1992}, polynomial algorithm to factorize large numbers \cite{Shor-1994qr}, and fast methods to solve linear equations \cite{Harrow-2009}. These algorithms exploit the informational resources of an entangled state, the question that naturally arises is how to build useful quantum states. We propose, in the sequel of \cite{Verga-2019}, to use a walker on a graph to construct an entangled state of spins. Instead of using single qubit measurements on a previously prepared generic entangled state as in one-way computing, we let the particle to explore the graph creating through its interaction with the spins, the entangled quantum state. The resulting state depends, in addition to the initial condition and parameters, on the graph connectivity and topology. It is worth noting that the analogy with the graph state is nevertheless limited because of the different structure of the Hilbert space; the walk build state possesses besides the spin degrees of freedom, which are localized on the network edges, the walker position and internal degrees of freedom, which spread generally over the network: this structure changes the entanglement characteristics, impeding, for instance, to spatially partition the graph as can be done in pure spin systems (the walker wave function can be delocalized on the whole graph). This difference is exploited here to investigate the case where entanglement of spins is indirectly induced by a moving particle, in particular we can analyze the role of local and nonlocal effect in the multipartite entanglement of spins.

Our aim in this work is to demonstrate that a simple quantum system governed by local unitary rules leads to complex dynamics suitable to investigate the links between entanglement, topological, and magnetic properties. The paper starts with the presentation of the model, we show in particular how to handle the spins lying on the graph edges and their coupling with the itinerant particle. We specializes the model to the simple Dirac walk on the line and explore its phenomenology using exact numerical computations. We discuss the different dynamical regimes and their relation with the entanglement of the quantum state. We observe that the many-body interaction is essential in the setting of well defined magnetic properties and in the evolution towards a stationary regime. These magnetic properties can be described by a semiclassical theory and the corresponding Landau-Lifshitz equation derived. We end by a discussion of the results and a conclusion.

\section{Model}

We implement a quantum walk on a simple, undirected, simply connected graph \(G(V,E)\), where \(V\) is the set of nodes and \(E\) the set of edges. The number of nodes is the cardinal \(|V|\) (we write \(|\cdot|\) for the number of elements of a set). Nodes are denoted \(x,y,\ldots \in V\) and edges are pairs of linked nodes \((x,y) \in E\). The set
\begin{equation}
  \label{e:Vx}
  V_x = \{y \mid  (x,y) \in E\}
\end{equation}
is the set of nodes \(y \in V\) neighbors of \(x\), and \(d_x = |V_x|\) the degree of node \(x\). A particle jumps between neighboring nodes (\(y \in V_x, \; \forall x \in V\))  and interacts with spins located on the edges \((x,y) \in E\). The motion between adjacent vertices depends on an internal degree of freedom, the ``color'' \(c\), taking \(d_x\) values. The color decorates the graph nodes with a label `\(c\)', the ``subnode'', for each incident edge \cite{Kendon-2005fk}. In addition, the color degree of freedom is coupled with the particle motion, and with the edge spin through a spin-particle interaction. This model simulates a kind of itinerant magnetism in which the motion of a particle determines the correlations between localized spins.

\begin{figure}[hbt]
  \centering
  \begin{tikzpicture}
    \coordinate (A) at (-0.5,1.5); 
    \coordinate (B) at (-0.5,-1.5);
    \coordinate (C) at (3.5,1.5);
    \node[circle,fill=gray!20,draw] (x) at (0,0) {$x$};
    \node[circle,fill=gray!20,draw] (y) at (3,0) {$y$};
    \path[-] (x) edge node [pos=0.1, above] {$c_y$}
    node[pos=0.5,below] {$s_e$}
    node [pos=0.9,above] {$c_x$} (y);
    \path[dashed] (x) edge  node [pos=0.1, anchor=east] {1} (A)
    edge node [pos=0.1, anchor=east] {2} (B)
    (y) edge node [pos=0.1, anchor=west] {0} (C);
  \end{tikzpicture}
  \caption{Graph showing the labeling of subnodes (color, $0,1,\ldots$) between nodes (position $x,y,\ldots$) and edges (spin $s_e$).
  \label{f:G}}
\end{figure}
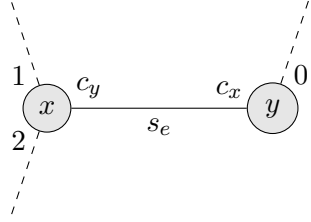

The basis of the graph Hilbert space \(\mathcal{H}_G\) is the set of kets of the form
\begin{equation}
  \label{e:xcs}
  \ket{xcs} = \ket{x} \otimes \ket{c} \otimes \ket{ s_0 \ldots s_{|E|-1} } \in \mathcal{H}_G\,,
\end{equation}
(the canonical basis) where the labels `\(x,c,s\)' stand for:
\begin{itemize}
  \item  the position \(x\), spanning the nodes \(x = 0, \ldots, |V|-1 \in V\);
  \item the color \(c = c(x)\), spanning the incoming edges of each node \(\forall x \in V, \; c = 0, \ldots, d_x - 1\);
  \item and the set of spins \(s=s_0s_1\ldots s_{|E|-1}\), represented by a string of binary numbers \(s_e = 0, 1, \; e = 0,\ldots,|E|-1\) (\(e\) labels the edges) for the up and down states at each edge, and spanning the configurations \(s \in \{0, \ldots, 2^{|E|} - 1\}\).
\end{itemize}
The Hilbert space dimension is
\begin{equation}
  \label{e:dim}
  \mathrm{dim} \mathcal{H}_G = \left(\sum_{x\in V} d_x \right) \times 2^{|E|} \le |V| \times d_\text{max} \times 2^{|E|}
\end{equation}
(\(d_\text{max} = \max_x |V_x|\) is the maximum degree). A node state is a superposition of \(\ket{xcs}\) vectors with fixed \(x\), and the edge state is given by a superposition of pairs of node vectors of the form,
\begin{equation}
  \left\{\ket{xc_ys_e}, \ket{yc_xs_e}\right\}, \quad (x,y) \in E \,,
\end{equation}
where \(c_e = (c_y,c_x)\), formed by the subnodes of the corresponding neighbors, is defined as the edge color (Fig.~\ref{f:G}).

The interacting quantum walk is defined by a unitary operator,
\begin{equation}
  \label{e:U}
  U = V(J)W\,,
\end{equation}
which splits into two parts, the walk \(W\) part and the interaction \(V(J)\) part that depends on a particle-spin coupling constant $J$:
\begin{itemize}
  \item The walk part \(W = MC\), includes a motion operator (\(M\)) that permutes the amplitudes at a node \(x\) with the amplitudes of its neighbors \(y \in V_x\), according to the edge color state, and a coin operator (\(C\)) that actualizes the color state at each node; the motion step \(M\) is, 
    \begin{equation}
      \label{e:M}
      M\ket{x c_y s} = \ket{y c_x s},\; \forall y \in V_x\,,
    \end{equation}
    where the set of values of \(c= c_y\) with \(y\in V_x\), corresponding to the subnodes of \(x\), determines the direction of the particle motion; and the coin step \(C\),
    \begin{equation}
      \label{e:C}
      \ket{xcs} \rightarrow C \ket{xcs}, \quad C = \mathbb{I}^{|V|} \otimes C_c \otimes \mathbb{I}^{2^{|E|}}
    \end{equation}
    (\(\mathbb{I}^D\) is the $D$-dimensional identity matrix) where \(C_c\) can be any operator of dimension \(d_x\) (conveniently completed with zeros up to the total dimension \(d_\text{max}\)). The composite operator \(W\) defines a simple quantum walk in a graph \cite{Aharonov-2001ty,Tregenna-2003rp,Kendon-2011,Berry-2011qq}. The choice of the coin \(C_c\) is somewhat arbitrary but the Grover and Fourier operators possess complementary properties \cite{Kempe-2003fk}. The Grover operator,
    \begin{equation}
      \label{e:grover}
      G(d) = \frac{2}{d} \mathbb{J}^d - \mathbb{I}^d\,,
    \end{equation}
    where \(\mathbb{J}^d\) is the \(d\)-dimensional matrix filled with 1, is not balanced but distributes the walker amplitudes at each node preserving the graph symmetries; the Fourier operator is the quantum version of a balanced coin,
    \begin{equation}
      \label{e:fourier}
      F(d) = \frac{1}{\sqrt{d}} \E^{2\I \pi \bm c \bm c^\textsc{T}/d}\,,
    \end{equation}
    where \(\bm c\) is the vector whose coordinates are the colors at each node.

  \item  And the color-spin interaction part, which superposes the color state of the edge \(c_e = (c_y, c_x)\) with the edge spin \(s_e\)
    \begin{equation}
      \label{e:se}
      \forall e \in E,\; \ket{c_e} \otimes \ket{s_e} \rightarrow V(J) \ket{c_e} \otimes \ket{s_e}\,.
    \end{equation}
    This operator acts then on the set of edges \(E \in G\),
    \begin{equation}
      \label{e:Vop}
      V_{xy}(J) \begin{pmatrix} xc_y 0_e \\
        xc_y 1_e \\
        yc_x 0_e \\
        yc_x 1_e
      \end{pmatrix}\,, \; \forall (x,y) \in E\,,
    \end{equation}
    in an obvious notation, for instance \(0_e\) denotes a configuration \(s=s_0 \ldots s_e \ldots\) such that \(s_e=0\). The choice of \(V(J)\) determines the physics of the system. In the framework of an itinerant electron we can implement an analogous to the so-called \(sd\) exchange interaction \(H_J\) \cite{Berger-1984uq,Tatara-2019}:
    \begin{equation}
      \label{e:VJ}
      V_{xy}(J) = \exp(-\I H_J),\; H_J = -\frac{J}{4} \bm{\tau} \cdot \bm{\sigma}\,,
    \end{equation}
    with \(\bm{\tau}, \bm{\sigma} = (X,Y,Z)\) vectors of Pauli matrices acting on the color and spin spaces, respectively; \(J\) is the color-spin coupling constant. (Remark that, for the sake of simplicity, we define the matrix \(V(J)\) by its action on the edges \((x,y)\), which are not in the basis of the Hilbert space \(\mathcal{H}_G\).) This interaction
    \begin{equation}
      \label{e:VxyJ}
      V_{xy}(J) = \E^{-\I J/4}  \begin{pmatrix} \E^{\I J/2} & 0 & 0 & 0 \\
        0 & \cos(J/2) & \I \sin(J/2) & 0 \\
        0 & \I \sin(J/2) & \cos(J/2) & 0 \\
      0 & 0 & 0 & \E^{\I J/2}  \end{pmatrix} \,,
    \end{equation}
    is related to the swap gate for the particular choice of the coupling \(J=\pi\).  The interaction \(V(J)\) is able to entangle the color and spin degrees of freedom, while \(W\) may entangle the color and particle position; as a result, distant spins generically get entangled through the particle motion (for standard choices of the coin operator and parameters).
\end{itemize}

Therefore, the unitary evolution of the quantum state \(\ket{\psi}\) is governed by the operator \(U\) such that,
\begin{equation}
  \label{e:Upsi}
  \ket{\psi(t+1)} = U \ket{\psi(t)}, \quad \ket{\psi(t)} \in \mathcal{H}_G\,,
\end{equation}
which advances the quantum state by one time step (we chose units such that \(\hbar=1\) and the time step \(\Delta t =1\)). In summary, this equation describes the quantum walk on the nodes of an arbitrary graph of a colored particle interacting with the spins on the corresponding edges. In Appendix~\ref{S:U} we study in more detail the tensor structure of \(U\), to explicitly show its local action on the canonical basis of \(\mathcal{H}_G\) in the specific case of a line graph.

\subsection{One dimensional lattice}

In the following we will investigate the one dimensional lattice. The graph reduces to the set of nodes \(x \in V \subset \mathbb{Z}\) with edges \(e \in E\) (\(e=0,1,\ldots,|E|-1\)) simply given by \((x, x+1)\). The set of neighbors \eqref{e:Vx} of node \(x\) is \(V_x = \{x-1, x+1\}\). The Hilbert space dimension \eqref{e:dim} is \(|V| \times 2^{|E|+1}\).  We consider periodic (\(|E| = |V|\)), and finite lattices (\(|E| = |V| - 1\)), for different initial states. 

We choose a rotation matrix of angle \(\theta\) for the coin operator \(C_c\),
\begin{equation}
  \label{e:R}
  R(\theta) = \begin{pmatrix} \cos \theta & - \sin \theta \\
  \sin \theta & \cos \theta \end{pmatrix}\,,
\end{equation}
which, for the special value \(\theta = \pi/2\), exchanges amplitudes like a two dimensional Grover matrix \(G(2)\), and for \(\theta = \pi/4\), distributes the amplitudes like the balanced Hadamard coin \(F(2)\), thus conveniently interpolating between Grover and Fourier coins. The motion operator reduces to a translation such that the swapping of neighbor amplitudes can be written as,
\begin{equation}
  \label{e:M1d}
  M = \sum_x \left(\ket{x+1}\bra{x}\otimes \ket{1} \bra{0} +
  \ket{x-1}\bra{x}\otimes \ket{0} \bra{1} \right)\,,
\end{equation}
so that the \(0\) color amplitude moves to the right, while its state flips to \(1\). The one step evolution operator is then given by,
\begin{equation}
  \label{e:U1}
  U(J,\theta) = V(J)MR(\theta)\,,
\end{equation}
which depends on two parameters taking values in the relevant ranges \(\theta \in (0, \pi)\) and \(J \in (0,\pi)\). Appendix~\ref{S:U} shows the action of \(U\) on the basis vectors arranged in edges related amplitudes, and Appendix~\ref{S:A} illustrates the case \(|V| = 2\).

Initially the particle can be located on one node, or uniformly distributed on the line (`i'), in different color superpositions, and spin states: all spins \(z\)-polarized (`z'), \(x\)-polarized (`x'), or a single spin up in a background of `x' spins. We label the initial state as follows,
\begin{itemize}
  \item particle at \(x = x_0\) and spins up \(\ket{0}\),
    \[
      \text{`z': } \ket{\text{z}} = \ket{x_000} = \ket{x_00}\otimes \ket{0}^{\otimes |E|}\,, 
    \]
  \item particle at \(x = x_0 \) and spin `right' \(\ket{+} = (1/\sqrt{2})(\ket{0} + \ket{1})\), 
    \[
      \text{`x': } \ket{\text{x}} = \frac{1}{2^{|E|/2}}\sum_{s = 0}^{2^{|E|}-1} \ket{x_00s} = \ket{x_00}\otimes \ket{+}^{\otimes |E|}\,, 
    \]
  \item particle at \(x = x_0\) spin up at \(e = (x_0, x_0 +1)\), and the other spins in \(\ket{+}\), 
    \[
      \text{`zx': } \ket{\text{zx}} = \frac{1}{\sqrt{2^{|E|-1}}}\sum_{\{\forall s \mid s_{e} = 0\}} \ket{x_00s} \,,
    \]
  \item (entangled spins) the particle at \(x = x_0\) is in a superposition of up and down spins,
    \[
      \text{`e': } \ket{\text{e}} = \frac{1}{\sqrt{2}} \left( \ket{x_000} + \ket{x_002^{|E|-1}} \right) \,. 
    \]
\end{itemize}
If initially the particle is uniformly distributed over the position states we add a prefix `i'  to the initial state label (`iz', `ix', etc.).

We also studied different boundary conditions: (`p') periodic (`b') reflective and (`t') with an interface. With reflection of the walker at the line boundary the translation symmetry preserved by the periodicity is broken. The interface is introduced to investigate topological properties of the quantum walk, using different coins on the two sides.

In summary, the system is determined by the boundary and initial conditions and the couple of parameters \((\theta, J)\): an example is \((\pi/3, 1)\), `p', `z'.

\subsection{Observables}

After the \(t\) step, the quantum walk (pure) state is given by the density matrix,
\begin{equation}
  \label{e:rho}
  \rho(t) = \rho(xcs,t) = \ket{\psi(t)} \bra{\psi(t)},\; \ket{\psi(t)} = \sum_{xcs} \psi_{xcs}(t) \ket{xcs}\,, 
\end{equation}
where we explicitly wrote the functional dependency of \(\rho\) on the basis labels \(xcs\). This state is numerically computed iteratively multiplying by \(U\), and then it is exactly known. The knowledge of the quantum state allows us to monitor the physical properties of the system; we focus on the probabilities distribution of position and spin, as well as different measures of the entanglement entropy. We denote \(l \in \{x,c,s\}\) one of the basis state labels and \(\bar{l}\) the complement set of the label \(l\) (i.e.\ if \(l=x\), \(\bar{l} = \{c,s\}\)), and
\begin{equation}
  \label{e:tracel}
  \braket{O}(l,t) = \Tr O \rho(l,t),\; \rho(l,t) = \Tr_{\bar{l}} O \rho(xcs,t)
\end{equation}
the expected value of the magnitude \(O\) as a function of the variables (\(l,t\)), obtained from the partial density matrix \(\rho(l,t)\), which is the partial trace \(\Tr_{\bar{l}}\) over the complement \(\bar{l}\) of the total matrix \(\rho(xcs,t)\).

The particle distribution probability, function of the nodes \(x\) and step \(t\), is given by
\begin{equation}
  \label{e:pxt}
  p(x,t) = \Tr_{\bar{x}} \rho(xcs,t) = \Tr \rho(x,t)\,,
\end{equation}
where we used the notation \eqref{e:tracel}. The expected value of the spin \(\bm s(e,t)\) located at edge \(e\) in the state \(\rho(t)\), is 
\begin{equation}
  \label{e:set}
  \bm s(e,t) = \Tr \bm \sigma \rho(s_e,t)\,,
\end{equation}
where, as noted in the preceding section, \(s_e\) labels the spin at edge \(e\) in the configuration string \(s=s_0\ldots s_e \dots s_{|E|-1}\). We can label the edges simply using the node label and write \(\bm s(x,t)\) for the edge \(e = (x, x+1)\). The spatial mean is then given by,
\begin{equation}
  \label{e:st}
  \bm s(t) = \frac{1}{|E|} \sum_e \bm s(e,t)\,.
\end{equation}
The expected value \(\bm s\) corresponds to the graph magnetization distribution (here the magnetization on the line).

In order to measure the von Neumann entropy we partition the Hilbert space into two parties, selecting the relevant degrees of freedom. The entropies related to the basis degrees of freedom are defined by,
\begin{equation}
  \label{e:Sl}
  S_l(t) = - \Tr \rho(l,t) \log \rho(l,t)
\end{equation}
(we use throughout base 2 logarithms denoted \(\log\)). It is also interesting to measure the entanglement of a subset \(A\) of spins. We take,
\begin{equation}
  \label{e:AB}
  A = \{s = 0 \ldots s_e \ldots \mid e = e_0, e_1, \dots \in E \}, \; \bar{A} = \{xcs \mid s \notin A\}\,,
\end{equation}
where here \(e\) is a list of selected edges. With this definition, the entanglement entropy of \(A\) spins is,
\begin{equation}
  \label{e:SA}
  S_A(t) = - \Tr \rho(A,t) \log \rho(A,t) ,\quad \rho(A,t) = \Tr_{\bar{A}} \rho(xcs,t)\,.
\end{equation}
For a binary partition of the system the von Neumann entropy measures the entanglement of the corresponding reduced state \cite{Schumacher-1995uq,Horodecki-1994,Bennett-1996fr}. It is important to note that \eqref{e:Sl} is a global quantity defined with respect to the Hilbert's space degrees of freedom, while $S_A$ is much a local quantity in the sense that the set \(A\) is spatially defined by the set of edges; as a consequence we may have in general that 
\begin{equation}
\label{e:SASl}
S_A > S_s\,,
\end{equation}
for a large enough \(|A|\) (the number of spins in \(A\)). In particular \(S_A\) can be used to test entanglement for spatially separated spins, using a disconnected set \(A\).

\begin{figure}[tb]
  \centering
  \includegraphics[width=0.5\textwidth]{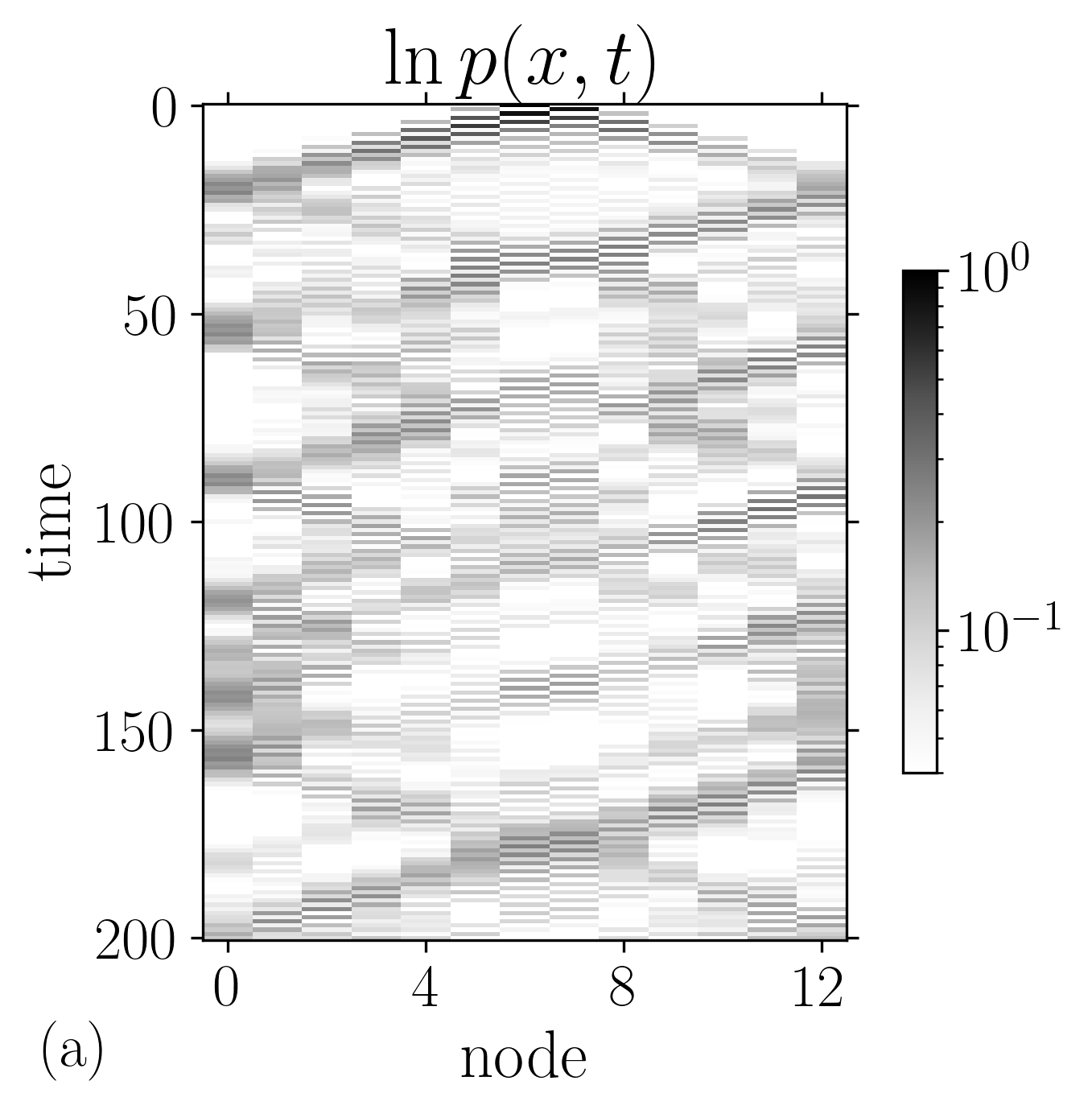}%
  \includegraphics[width=0.5\textwidth]{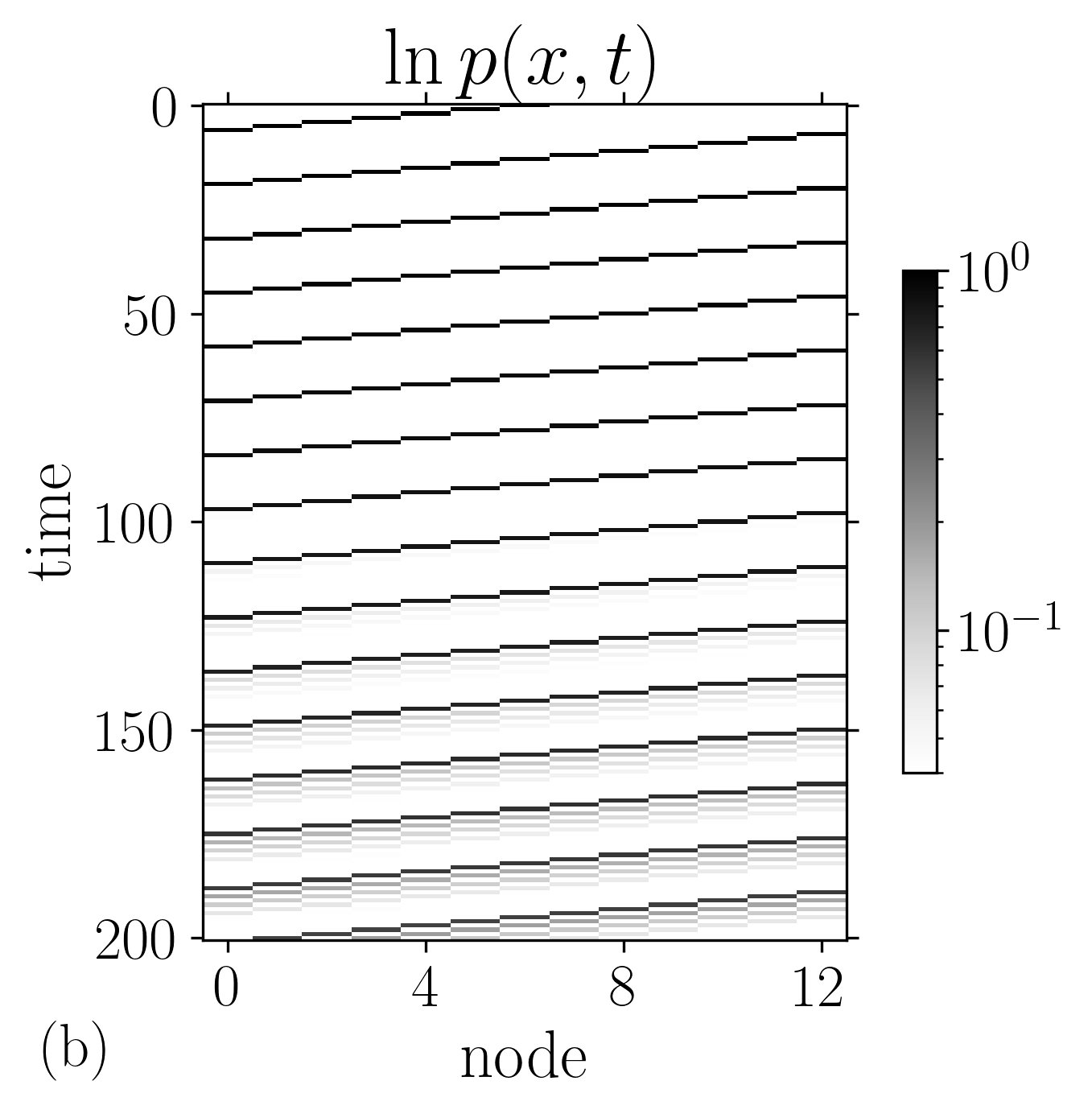}
  \caption{Spatiotemporal plot of the particle density \(p(x,t)\), colors in logarithmic scale. Parameters (a) $(\pi/8, 0.2)$ `x', (b) $(\pi/2, 0.2)$ `x'. (a) Dispersive ballistic propagation; (b) ballistic propagation with weak dispersion.
  \label{f:o54p}}
\end{figure}

\begin{figure}[htb]
  \centering
  \includegraphics[width=0.5\textwidth]{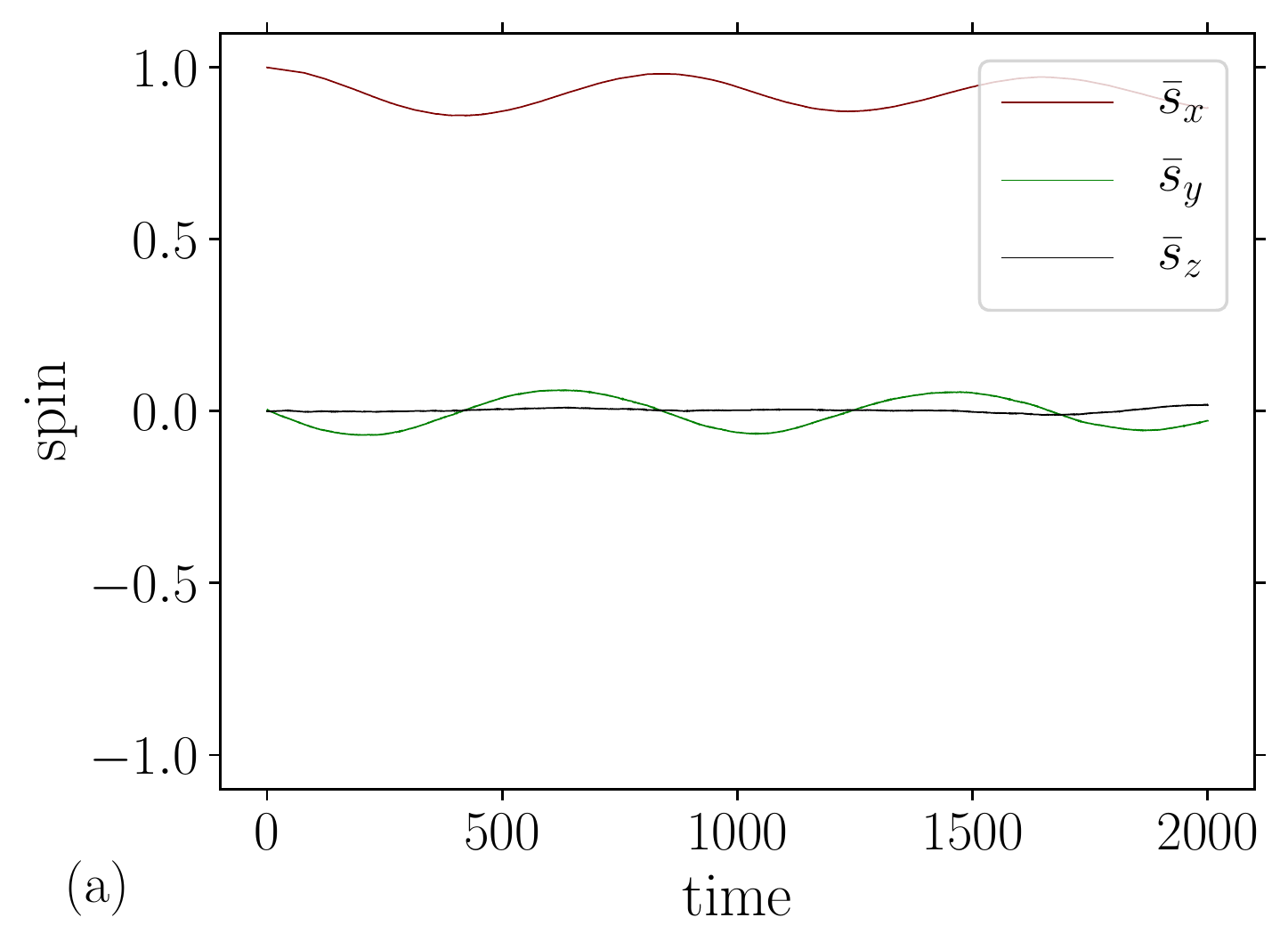}%
  \includegraphics[width=0.5\textwidth]{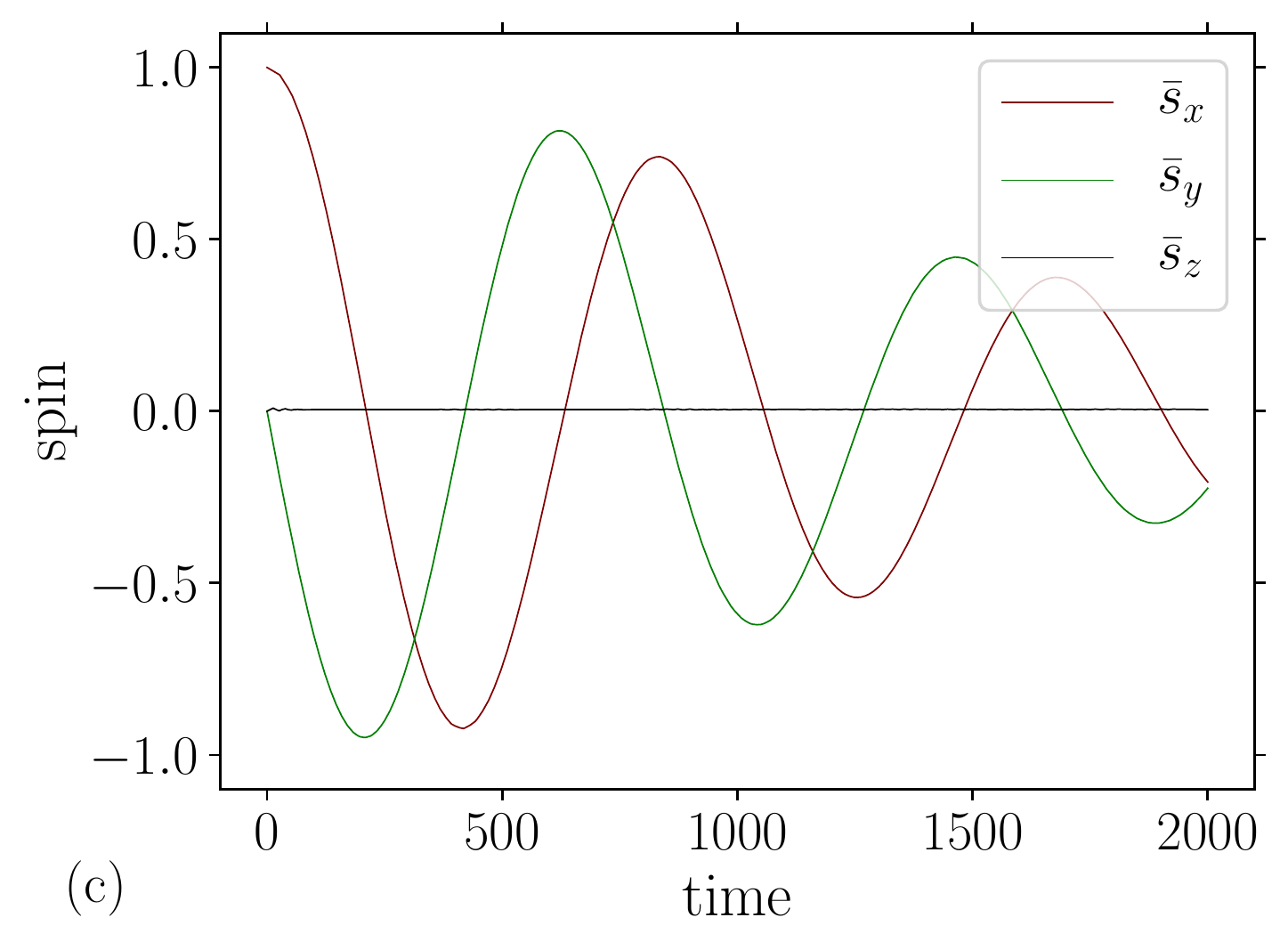}\\
  \includegraphics[width=0.5\textwidth]{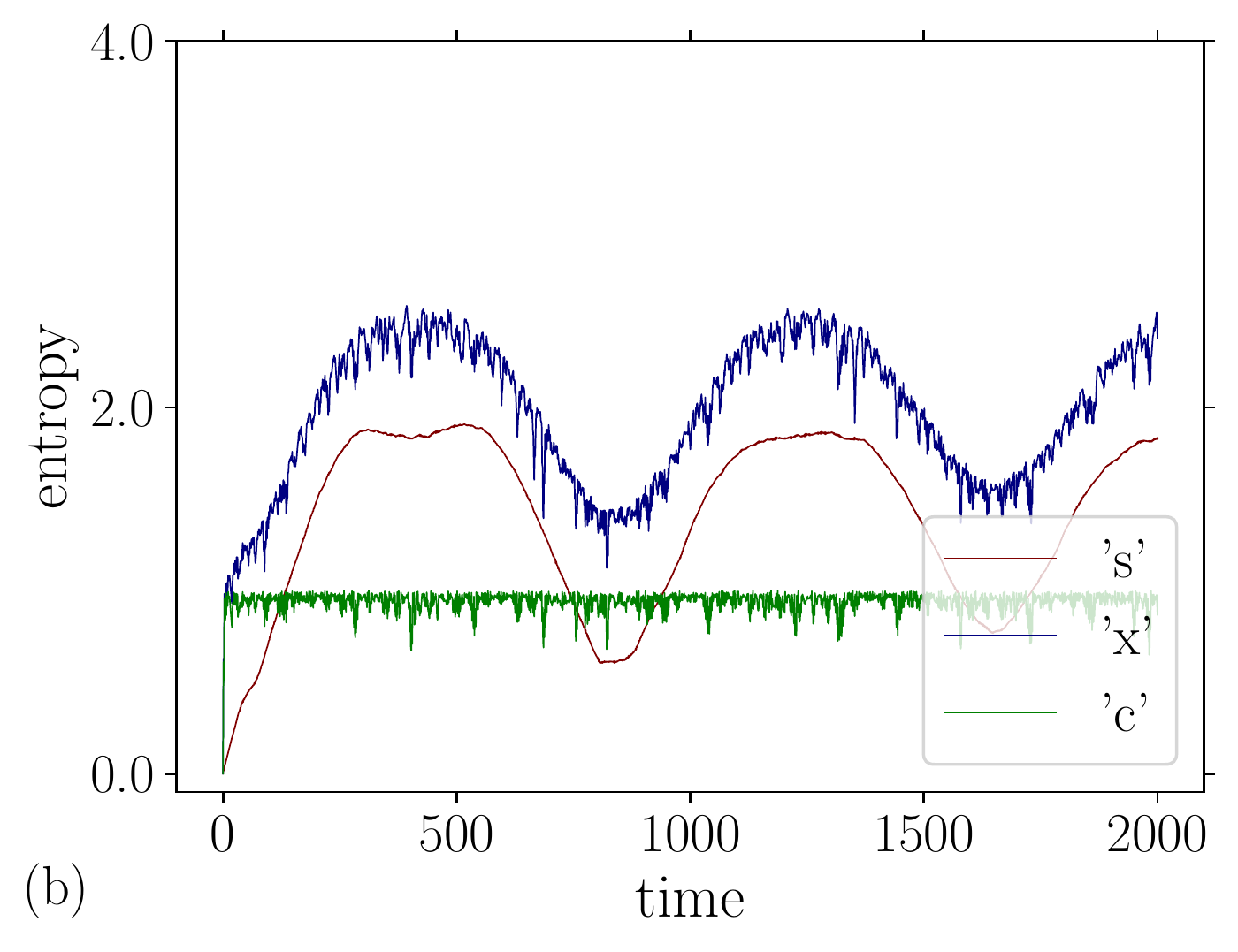}%
  \includegraphics[width=0.5\textwidth]{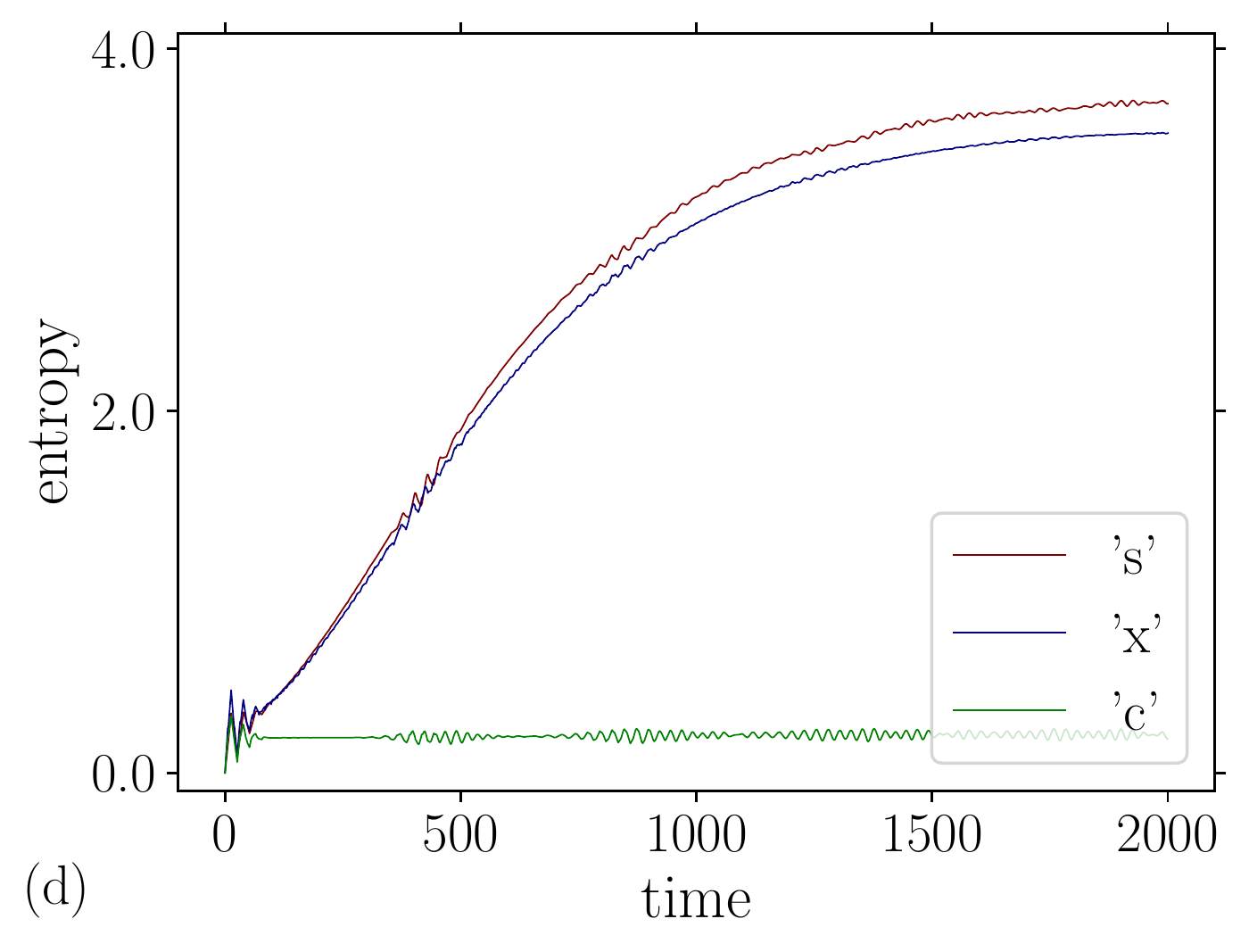}
  \caption{Spin and entanglement. Parameters as in the previous figure, (a,b) $(\pi/8, 0.2)$ `x', (c,d) $(\pi/2, 0.2)$ `x'. (a,c) Average spin \(\bar{\bm s}(t)\); (b,d) entanglement entropy \(S_l(t)\), \(l=x,c,s\). (a) oscillations, (c) oscillations with a weak damping; (b) recursive behavior of the particle and spin entropies, (b) growth of the entanglement.
  \label{f:o54}}
\end{figure}

To establish the entanglement between two spins at edges \((e_1, e_2)\) we use the standard notion of concurrence \cite{Wootters-1998,Horodecki-2009yg}, defined by,
\begin{equation}
  \label{e:Cee}
  C(e_1,e_2) = C[\rho_2] = \max\{0,\sqrt{\lambda_{1}}-\sqrt{\lambda_{2}}-\sqrt{\lambda_{3}}-\sqrt{\lambda_{4}}\},
\end{equation}
where \(\rho_2\) is the reduced matrix of the couple of spins, and \(\lambda_{i}\) are (in decreasing order, \(i=1,\ldots,4\)) the  eigenvalues of the matrix,
\begin{equation}
  \rho_2(Y \otimes Y) \rho_2^\star (Y \otimes Y)\,,
\end{equation}
where \(\rho^\star\) is the conjugate matrix in the canonical basis (the same as the \(\rho_2\) basis). The concurrence measures the entanglement of formation of a two qubits state, it is zero for a separable state and one for a maximally entangled one; in particular, if \(\rho_2\) is diagonal it vanishes, and hence no two-spins entanglement would be present. See Appendix~\ref{S:A} for an application to the case \(|V|=2\) of some of these formulas.

\section{Results}

\begin{figure}[htb]
  \centering
  \includegraphics[width=0.5\textwidth]{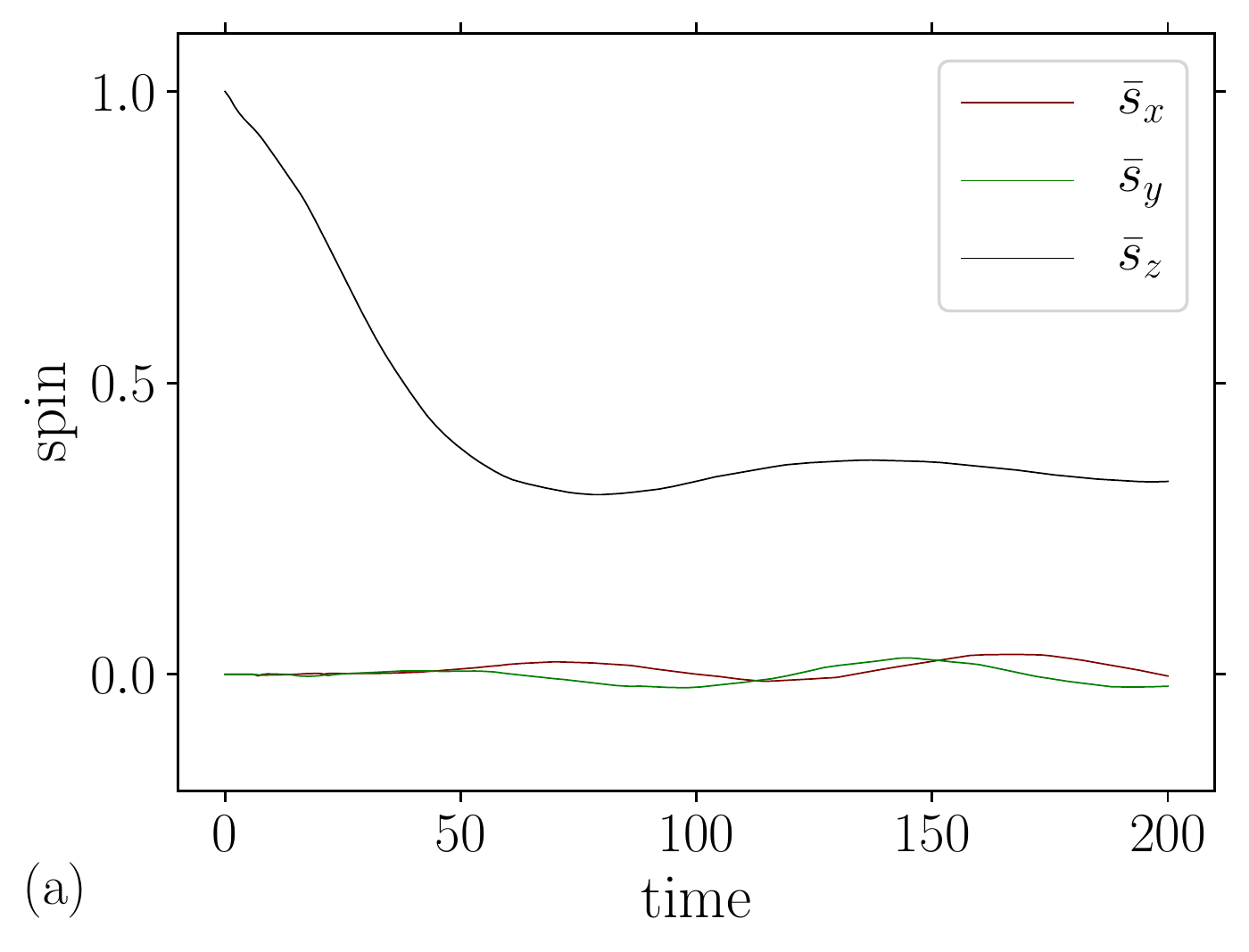}%
  \includegraphics[width=0.5\textwidth]{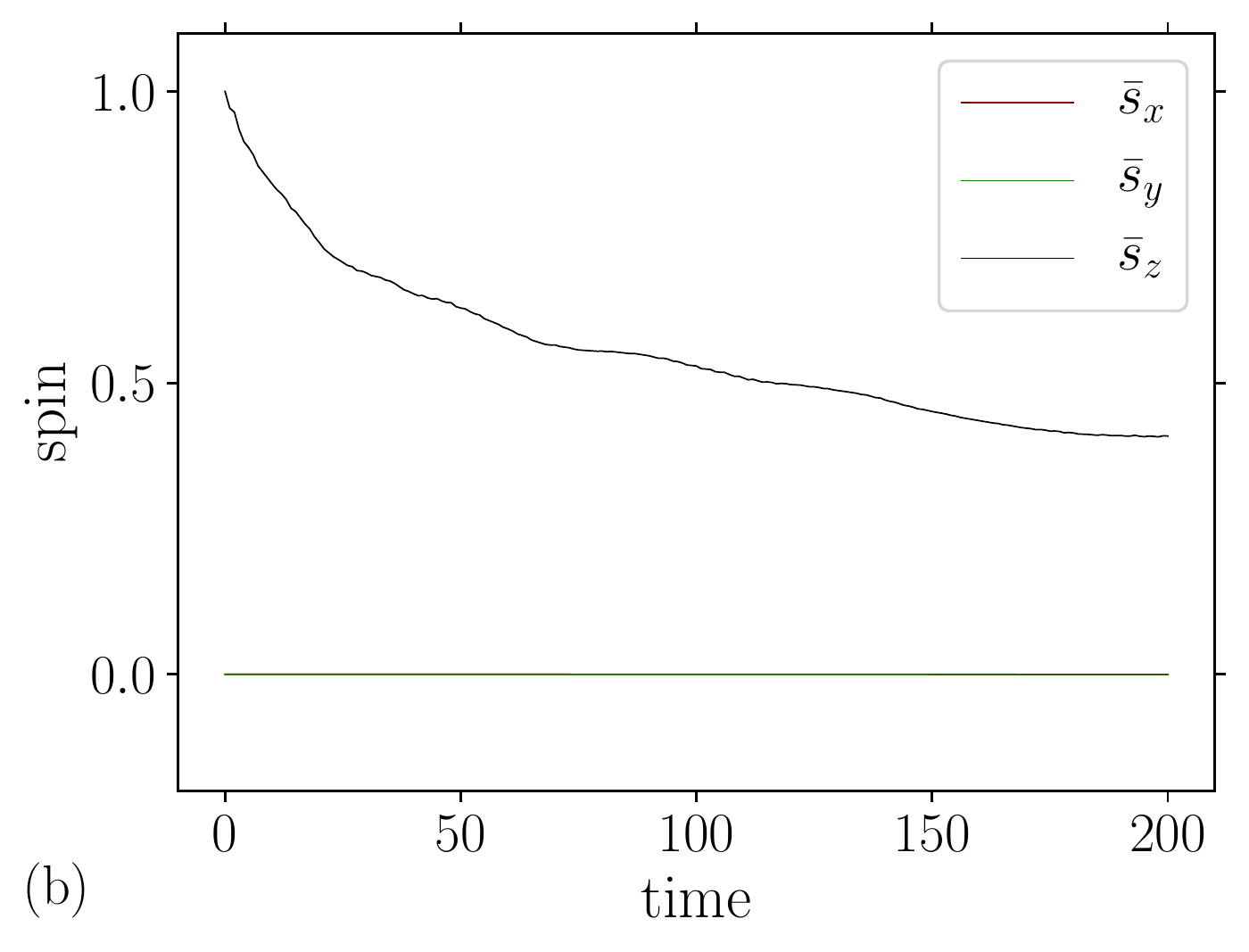}%
  \caption{Spin relaxation for strong coupling. Parameters (a) $(3\pi/8, 1.2)$, $|V|=13$, `z', (b) $(\pi/8, 1.2)$, $|V|=19$, `z'.
  \label{f:zst}}
\end{figure}

\begin{figure}[tb]
  \centering
  \includegraphics[width=0.5\textwidth]{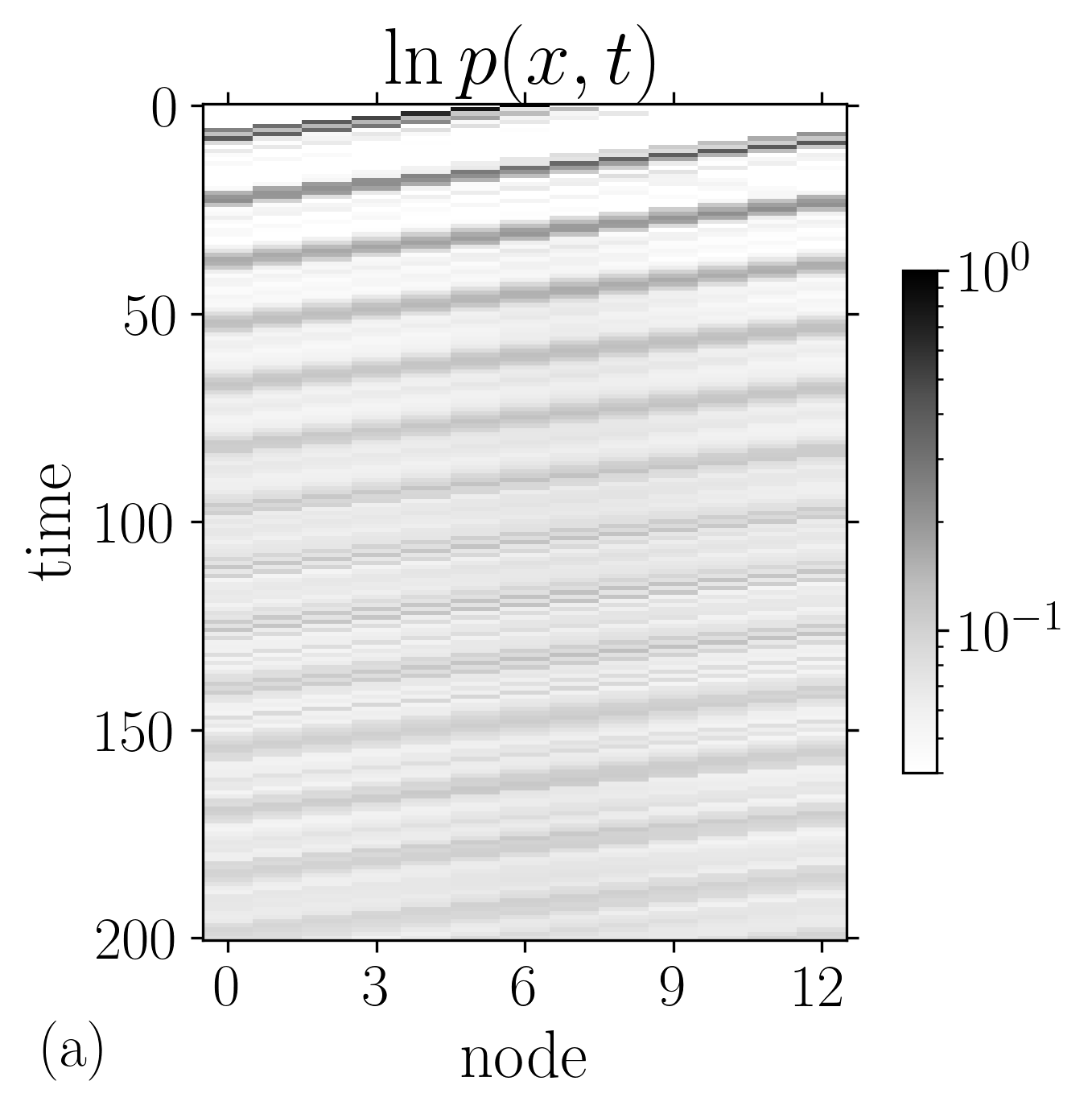}%
  \includegraphics[width=0.5\textwidth]{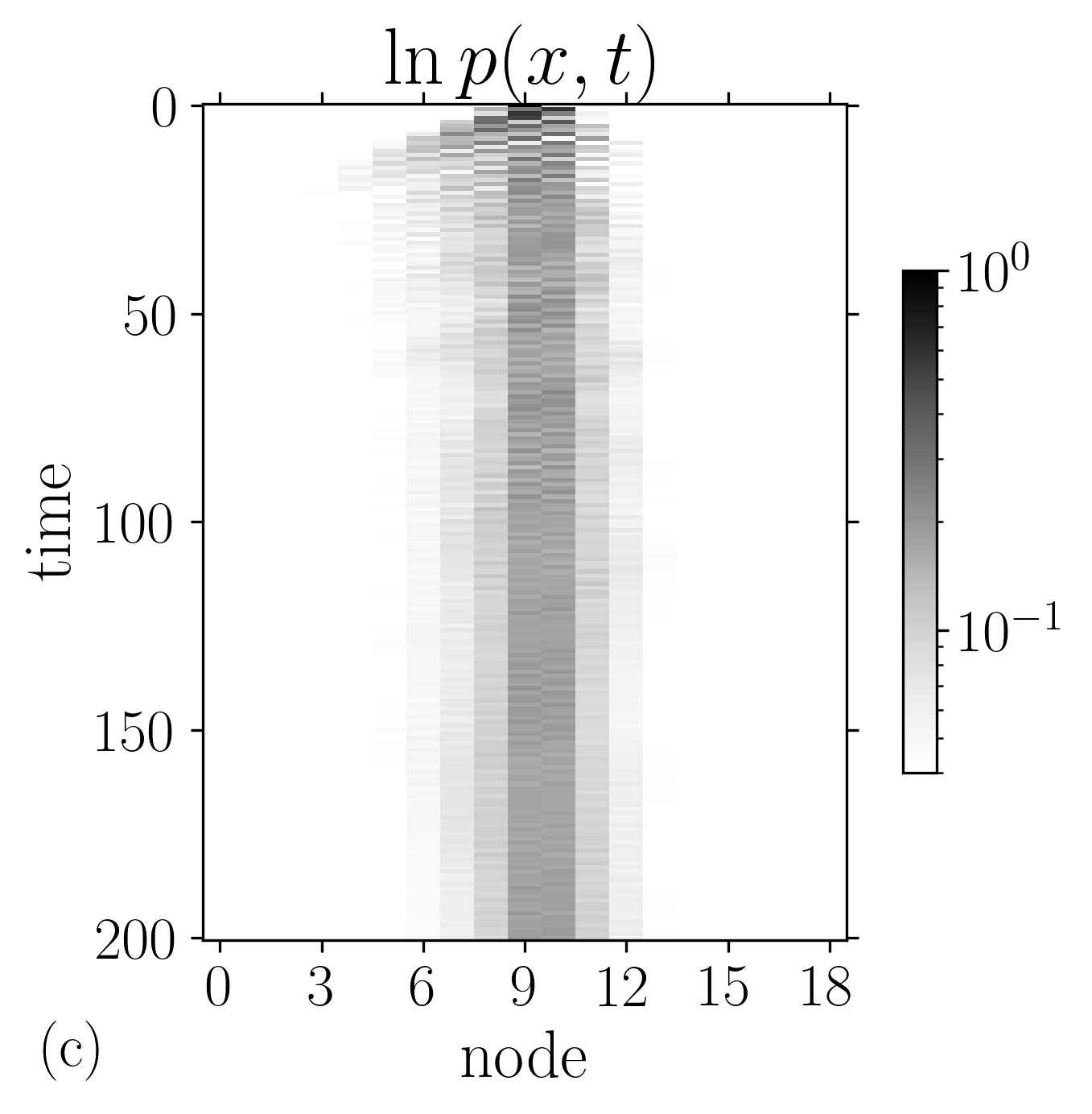}\\
  \includegraphics[width=0.5\textwidth]{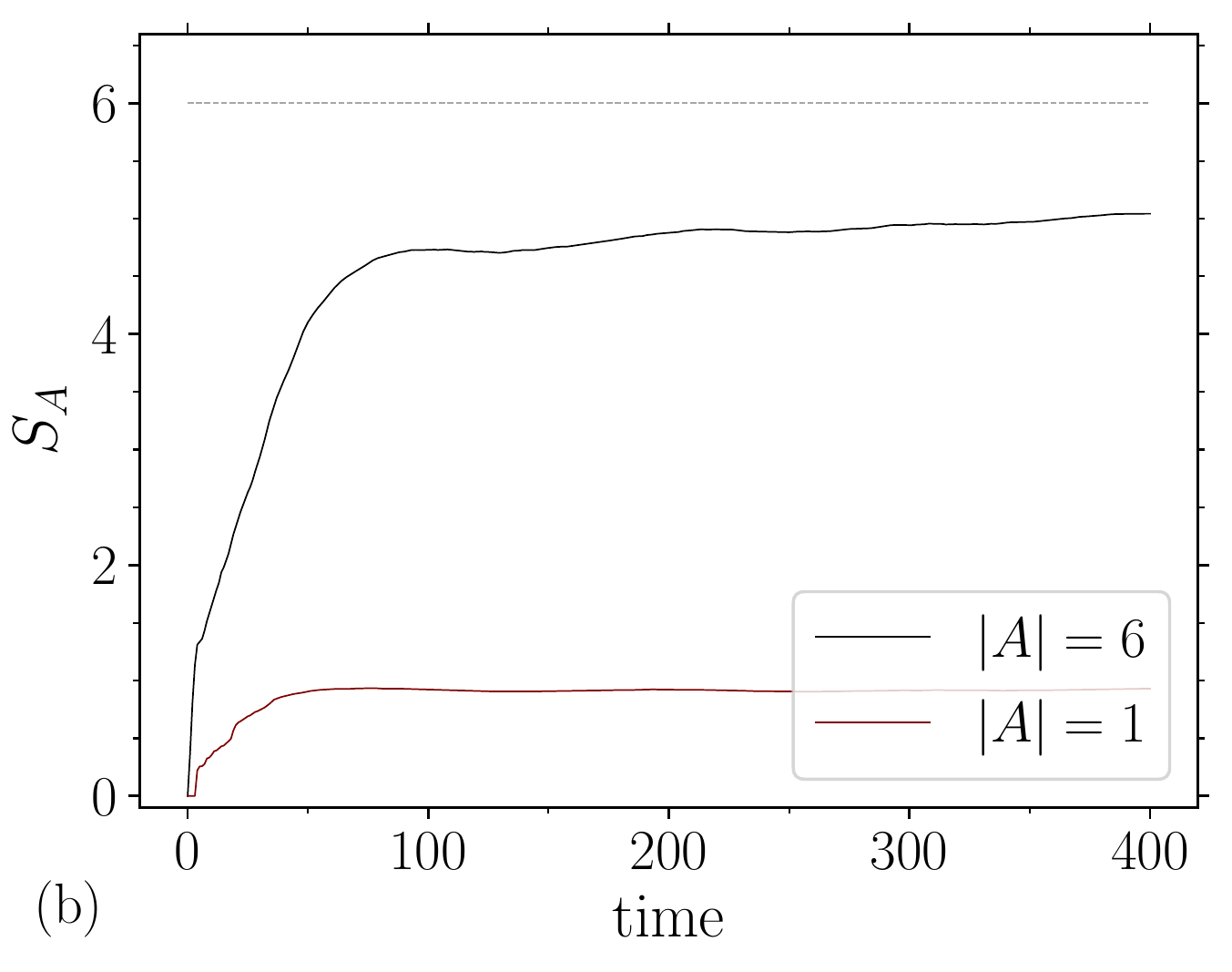}%
  \includegraphics[width=0.5\textwidth]{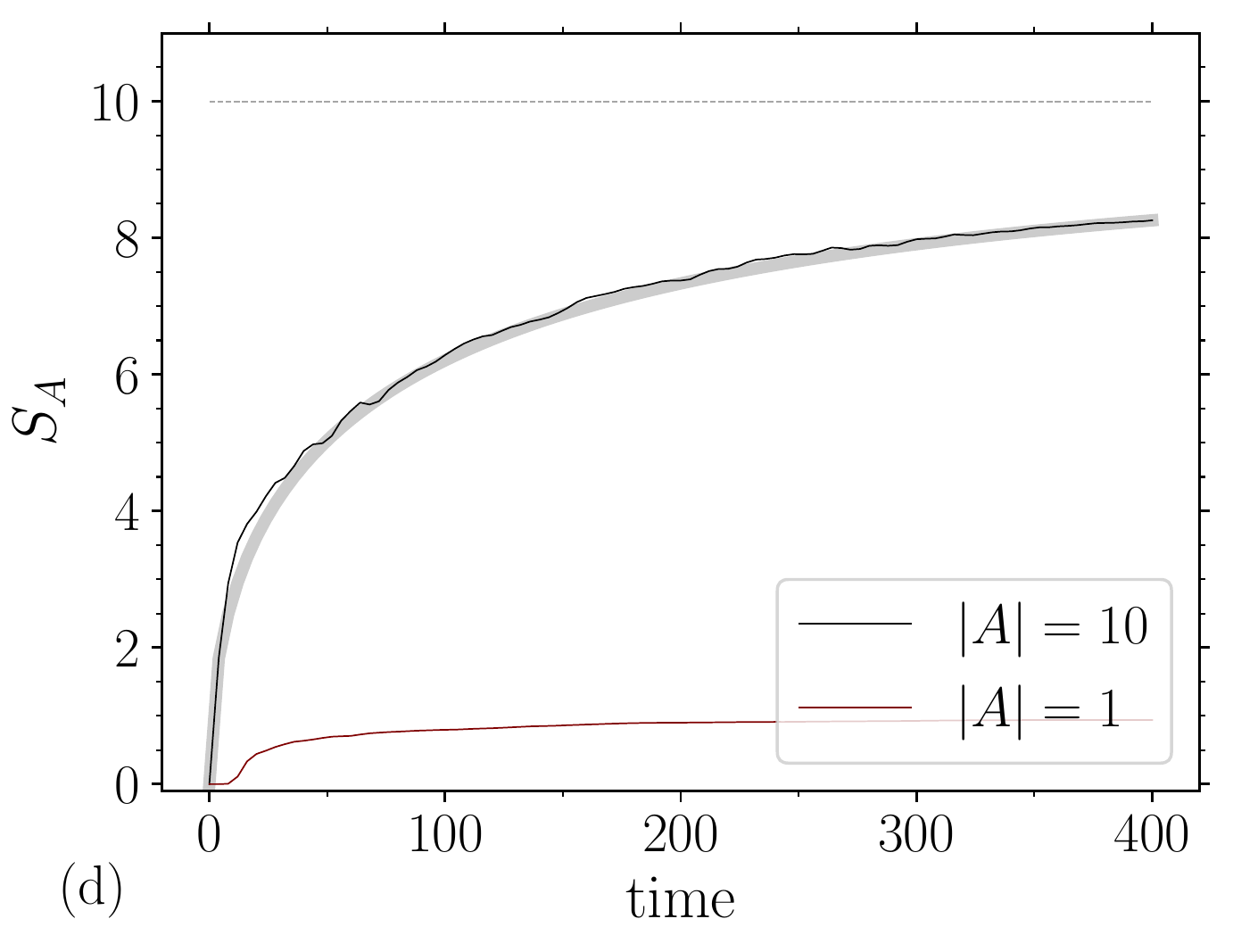}
  \caption{Position density and entanglement. Parameters as in Fig.~\protect\ref{f:zst}, (a,b) $(3\pi/8, 1.2)$, $|V|=13$, `z', (c,d) $(\pi/8, 1.2)$, $|V|=19$, `z'. Propagation (a) and localization (c) regimes. (b,d) Entanglement of a set of spins. (b) The one spin entropy saturates to its maximum value ($S_1=1$); the $|A|=6$ curve shows a linear range. (d) The one spin entropy saturates; $S_A(t)$ for $|A|=10$ slowly increases following a curve that can be fitted by a stretched exponential with exponent $1/2$ (see text).
  \label{f:zp}}
\end{figure}

We investigate now the dynamical evolution of the system in the parameter phase space \((\theta, J)\). We consider systems of different sizes, the typical one being \(|V|=13\), whose Hilbert space dimension is about \(2\,10^5\); some simulations were performed for \(|V| = 19\) (dimension about \(2\,10^7\)).

In the absence of interaction \(J = 0\) we identify two extreme behaviors: for \(\theta = 0\) the particle remains confined in the neighborhood of its initial position; in contrast, for \(\theta = \pi/2\) the walker translates at speed 1 without spreading (the motion reduces to an amplitude swap to the left or to the right). Intermediate values give the usual ballistic spreading along the line of the Dirac quantum walk \cite{Asboth-2012qy}. In the interval \(\theta \in (\pi/2, \pi)\) the system's behavior symmetrically reverses. However, it should be noted that interactions may in principle break this symmetry.

In the interacting case we may classify the different regimes according to the particle motion (using \(p(x,t)\)), which can be localized or propagating, and to the spin dynamics (using \(\bm s(x,t)\)), which may exhibit oscillations, relaxation or be chaotic. In addition, each of these regimes may display a variety of entanglement behaviors that can be characterized by the von Neumann entropies \(S_l\) associated to each class of degrees of freedom, position, color and spin, and also by the entanglement of a connected or unconnected subset of spins \(S_A\).

\subsection{Weak coupling}

The motion of the walker for weak coupling \(J = 0.2\) is illustrated in Fig.~\ref{f:o54p}, where we represent in a logarithmic scale the probability \(\ln p(x,t)\) for the particle to be at step \(t\) at node \(x\). Between the case \(\theta = \pi/8\) (Fig.~\ref{f:o54p}a) and the case \(\theta = \pi/2\) (Fig.~\ref{f:o54p}b) we note that not only the propagation speed differs, with values of about \(1\) for \(\theta = \pi/2\) and about \(1/3\) for \(\theta = \pi/8\), but that the probability spreading is much stronger for small rotation angles. It is worth mentioning that, at variance to the case `x' (\(\theta = \pi/2\)), the initial condition `z' is a proper state of \(U(J, \pi/2)\) leading to a trivial translation. A slight dispersion of the position probability appears in the case \(\theta = \pi/2\) for larger times (visible for \(t \gtrsim 150\)). This is an interesting effect of the interaction, it induces a dispersion on the underlying Dirac quantum walk.

The difference in the spreading of the particle reflects in the spin dynamics. We show in Fig.~\ref{f:o54}ac, for the same parameters (\(\theta = \pi/8, \pi/2, J = 0.2\)), the mean spin vector \(\bm s(t)\) as a function of time and the von Neumann spin entropy \(S_l(t)\) (\(l = x,c,s\)). Both cases exhibit spin oscillations, yet for \(\theta = \pi/8\) an almost recurrence to the initial state appears (relaxation is very weak: the peaks are \(1, 0.98, 0.97\), at \(t = 0, 839, 1648\)), while for \(\theta = \pi/2\) oscillations are damped (relaxation is stronger: the peaks are \(1, 0.74, 0.39\), at \(t = 0, 834, 1679\)). We observe that in spite of the irregular and fine grained features of the particle motion, the mean magnetization is smooth and behaves as if it would obey to a simple dynamical system.

The oscillation period can be empirically estimated to be,
\begin{equation}
\label{e:oscT}
T \approx 4\pi |V|/J\,,
\end{equation}
which is \(T = 817\) for \(|V| = 13\) and \(J = 0.2\), and depends essentially on \(J\). However, the amplitude of the oscillations is strongly dependent on the rotation angle \(\theta\): for \(\pi/8\) the amplitudes are much smaller than for \(\pi/2\).

\begin{figure}[tb]
  \centering
  \includegraphics[width=0.51\textwidth]{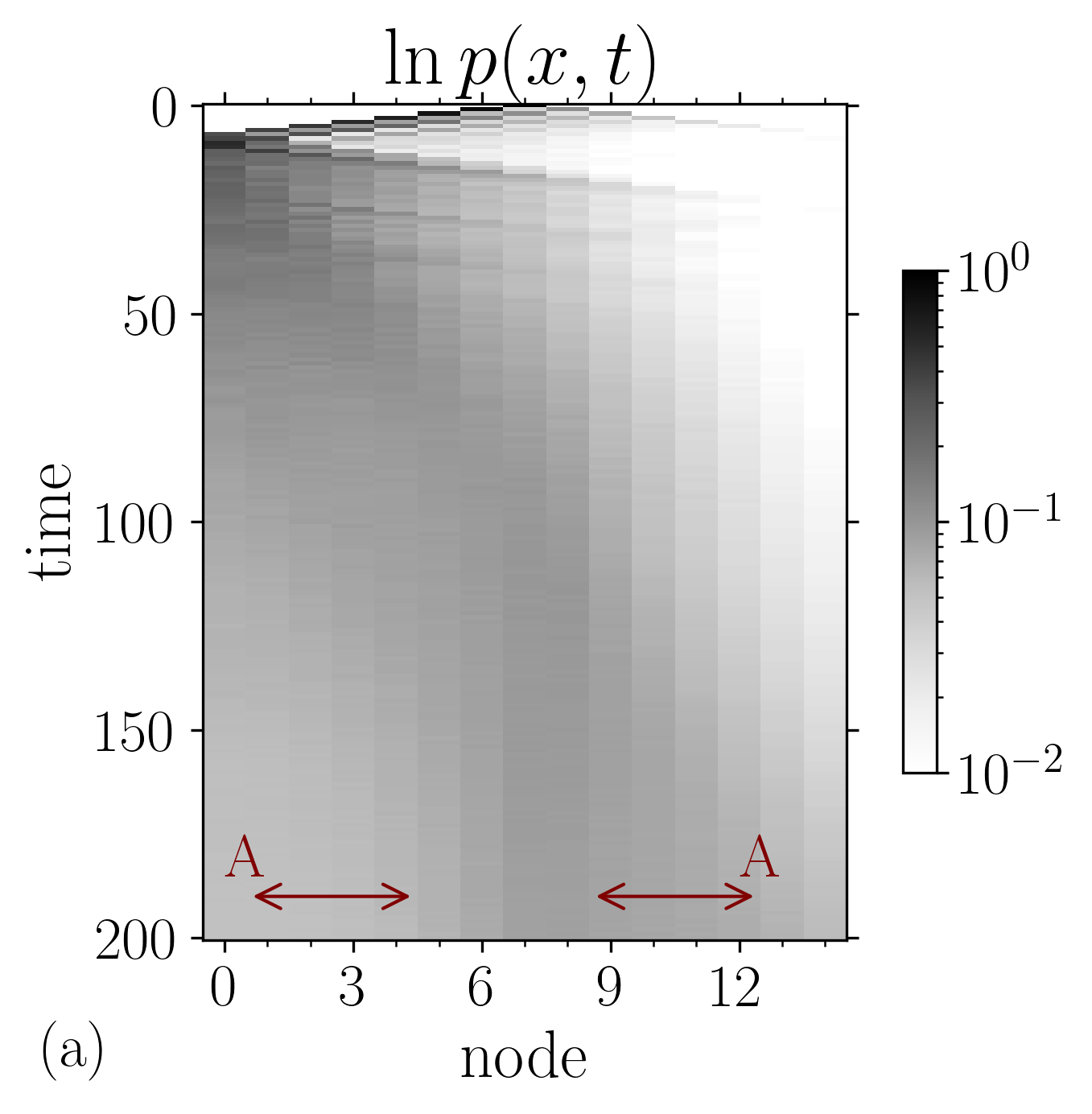}%
  \includegraphics[width=0.49\textwidth]{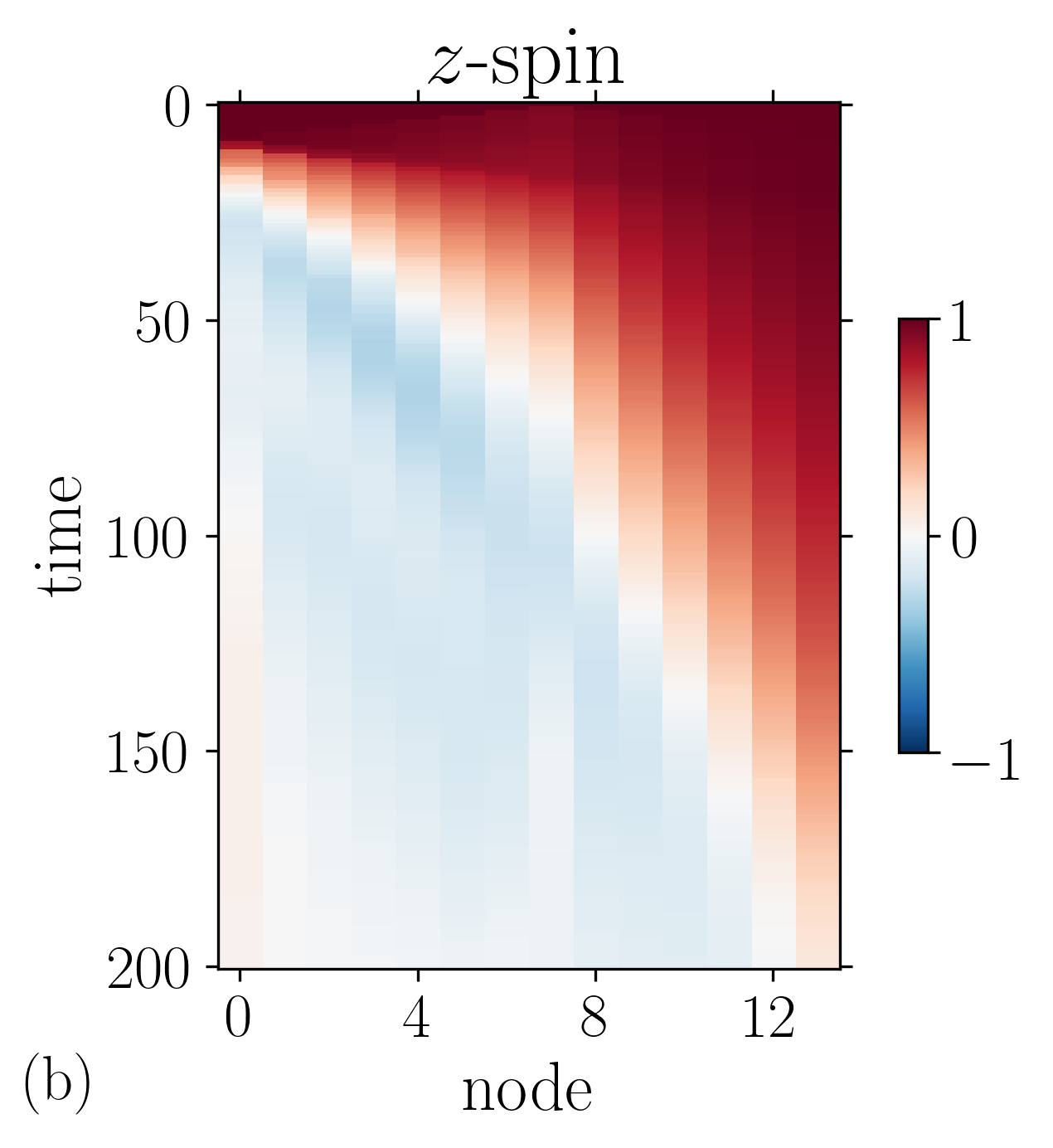}
  \caption{Simulation with closed boundary conditions `b'. Spatiotemporal particle and $z$-spin density. Parameters $(3\pi/8,1)$, `z'. The arrows in (a) show the set A of selected spins.
  \label{f:bz}}
\end{figure}

\begin{figure}[hbt]
  \centering
  \includegraphics[width=0.5\textwidth]{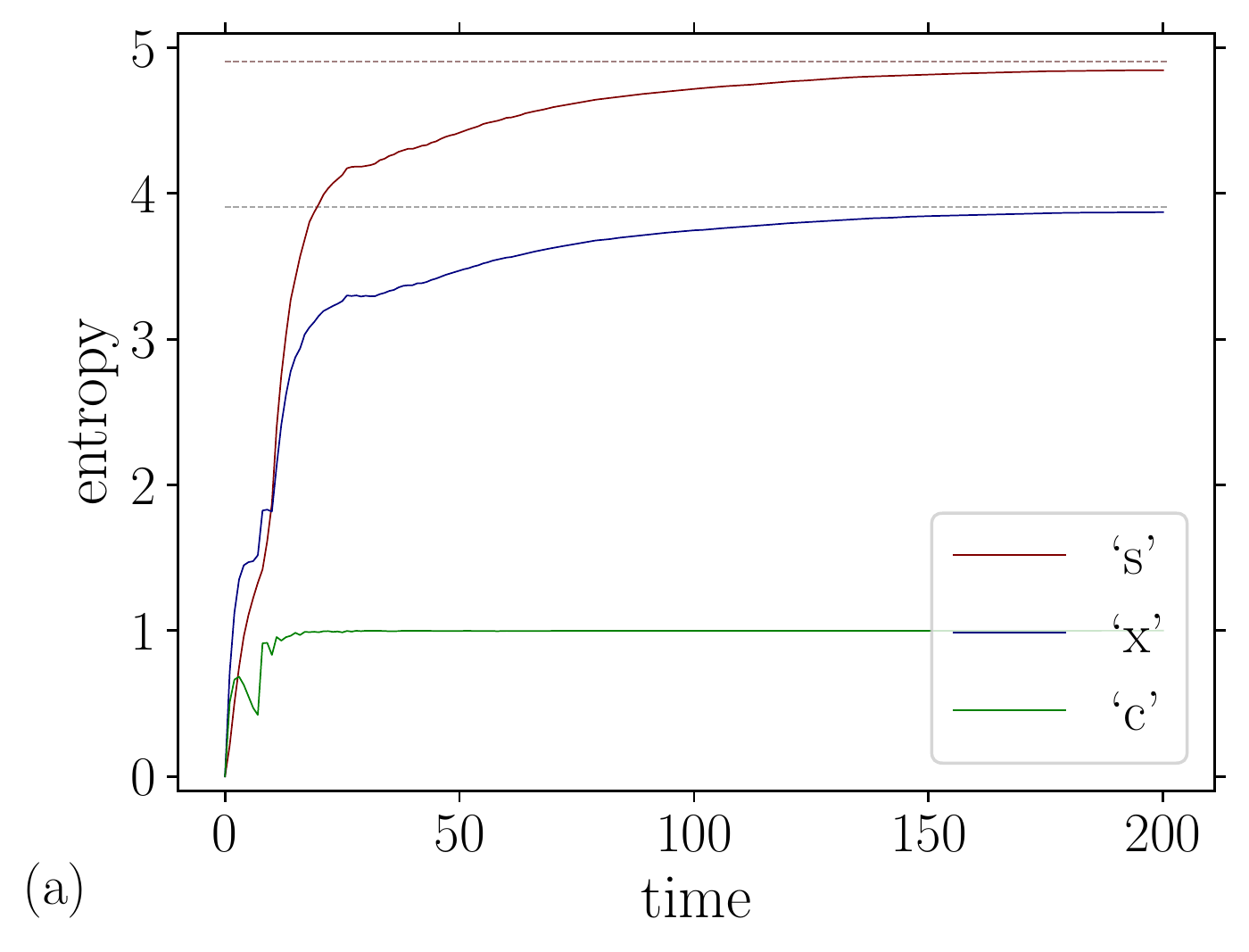}%
  \includegraphics[width=0.5\textwidth]{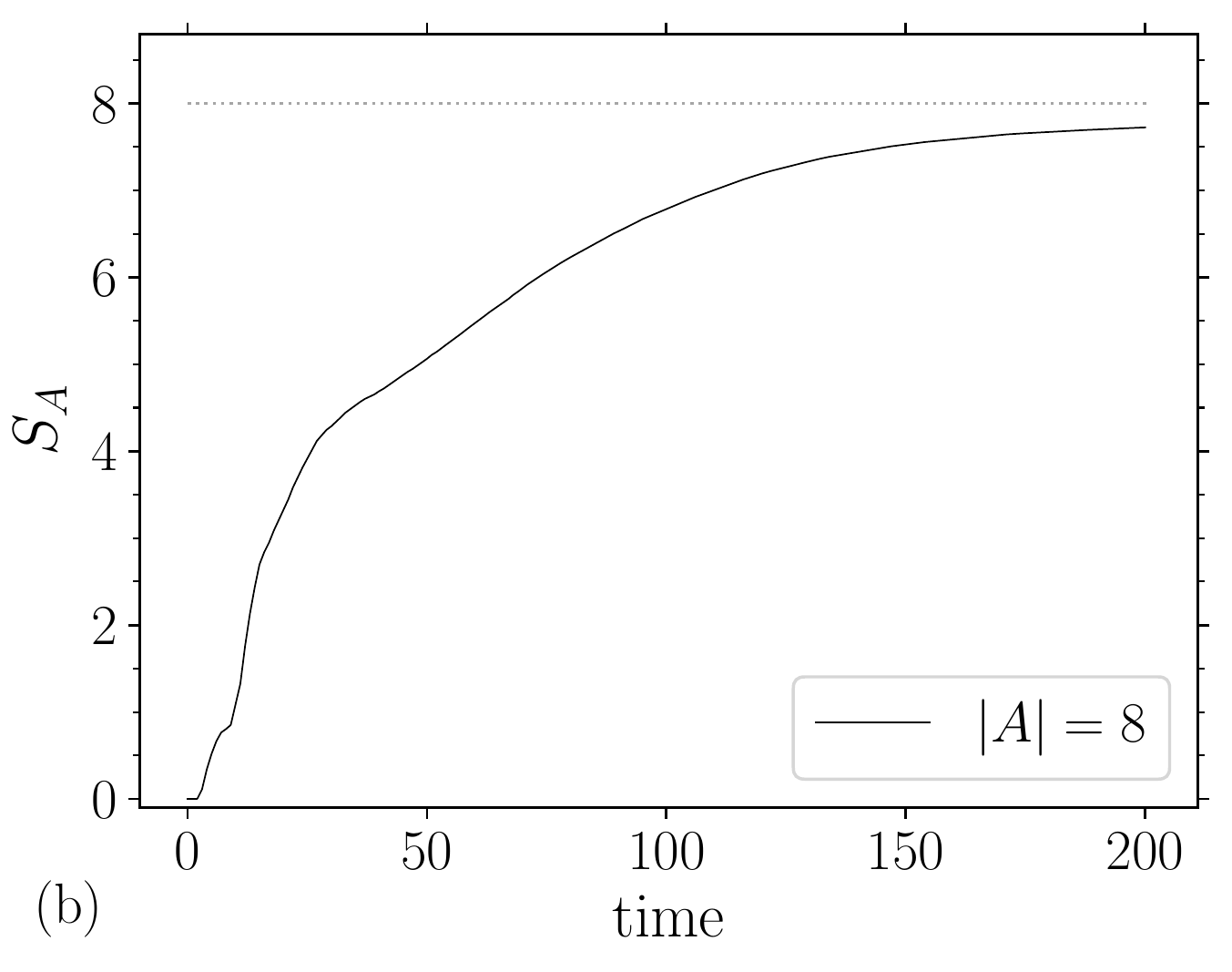}
  \caption{Entanglement with closed boundary conditions `b'. Parameters $(3\pi/8,1)$, `z'. (a) The $(xcs)$ entropies saturate to their maximum values. (b) Spin entropy of a disconnected set A, shown in Fig.~\protect\ref{f:bz}a.
  \label{f:bzA}}
\end{figure}

The motion and spin distributions for weak coupling display some qualitative differences depending on the values of the rotation angle, these differences also appear in the walk entanglement properties (Fig.~\ref{f:o54}bd). The recursive behavior found in the case \(\pi/8\) is also present in the entropy \(S_l(t)\). In the case \(\pi/2\) we have instead a steady increase of the entanglement towards saturation. The spin entropy is smaller than the position one \(S_x \gtrsim S_s\) and remains well under the maximum possible values
\[S_x < \log(|V|) \approx 3.7,\; S_s < \log(2\times |V|) \approx 4.7 \quad (|V| = 13)\]
for the smaller angle; while for the larger one, the spin and position entropies are larger, rather close to each other, and for long times we have \(S_x \lesssim S_s\). Another important qualitative difference we observe is the relatively large fluctuations of the node entropy \(S_x\) in the small angle case. The color entropy \(S_c\) generally reach its saturation value at one qubit, in a time which is of the order of the propagation time of the walker over the system length.

Summarizing this weak interacting case, we find that the walker propagates for a large range of rotation angles with increasing dispersion for small angles; the mean spin shows temporal oscillations with a period determined by the size of the system and the coupling constant; the entanglement manifests the recursive dynamics at small angles and, for larger angles, it is smooth and slowly increases to saturation at long times.

\subsection{Strong coupling and entanglement}

\begin{figure}[htb]
  \centering
  \includegraphics[width=0.33\textwidth]{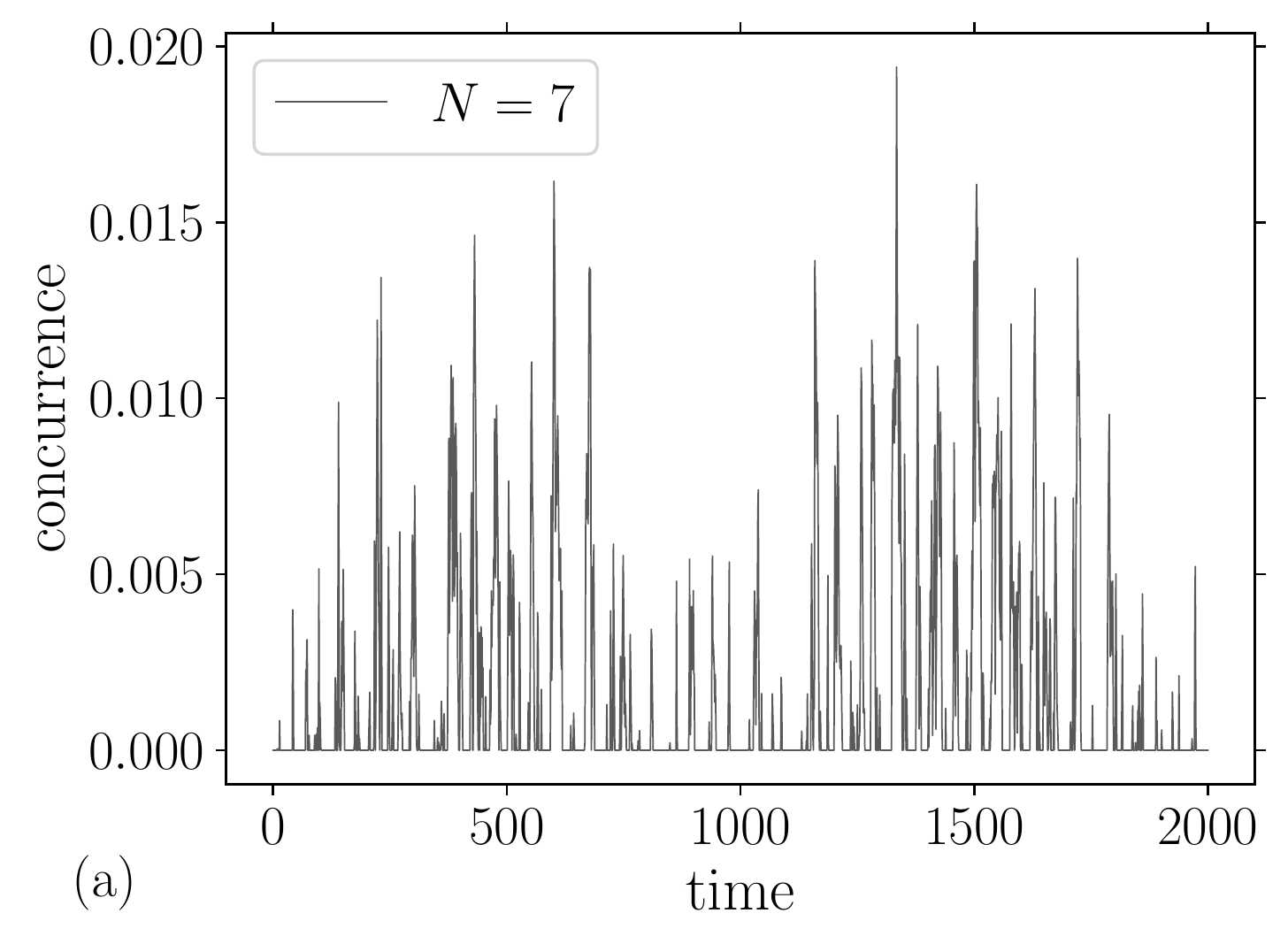}%
  \includegraphics[width=0.33\textwidth]{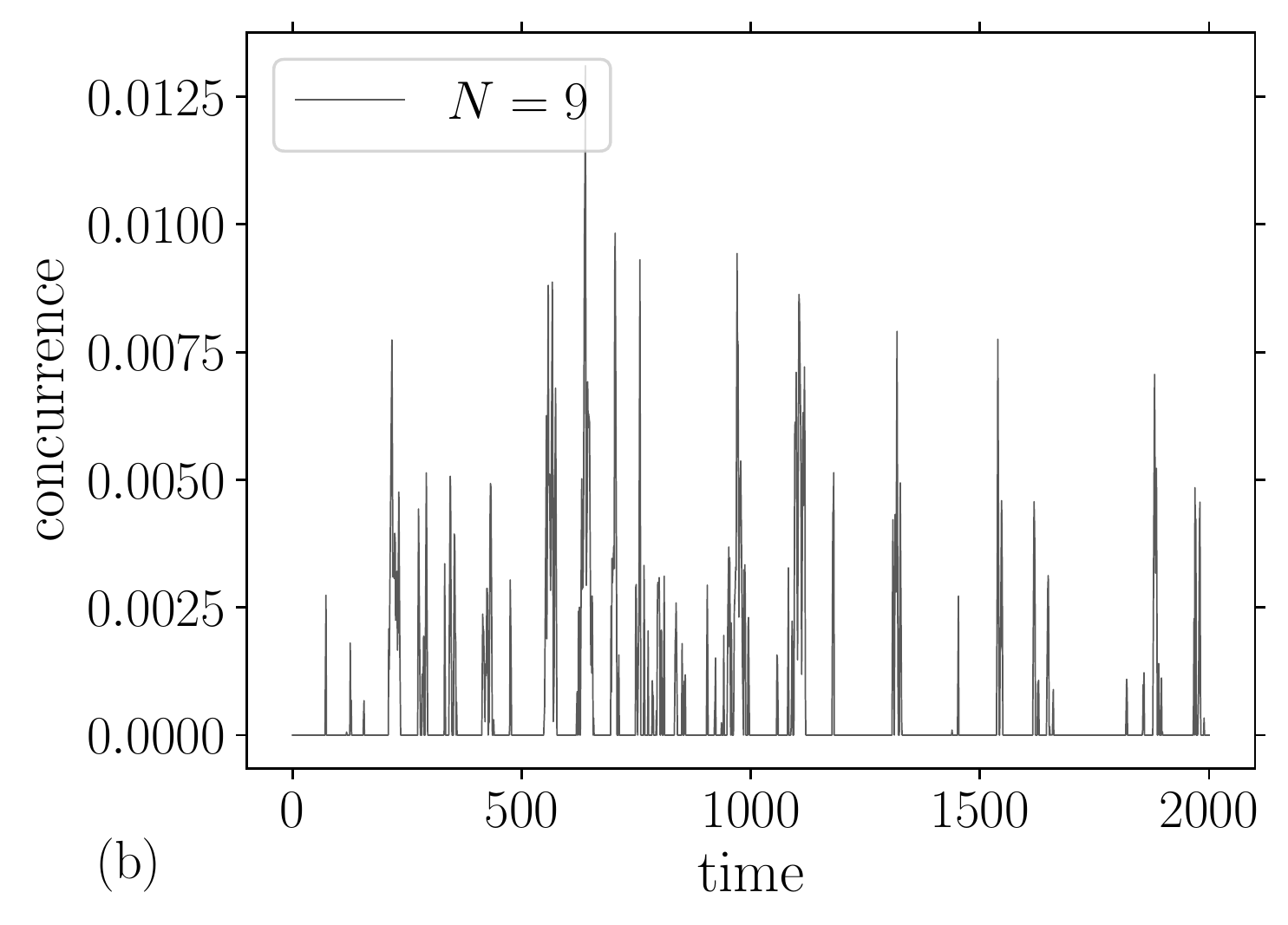}%
  \includegraphics[width=0.33\textwidth]{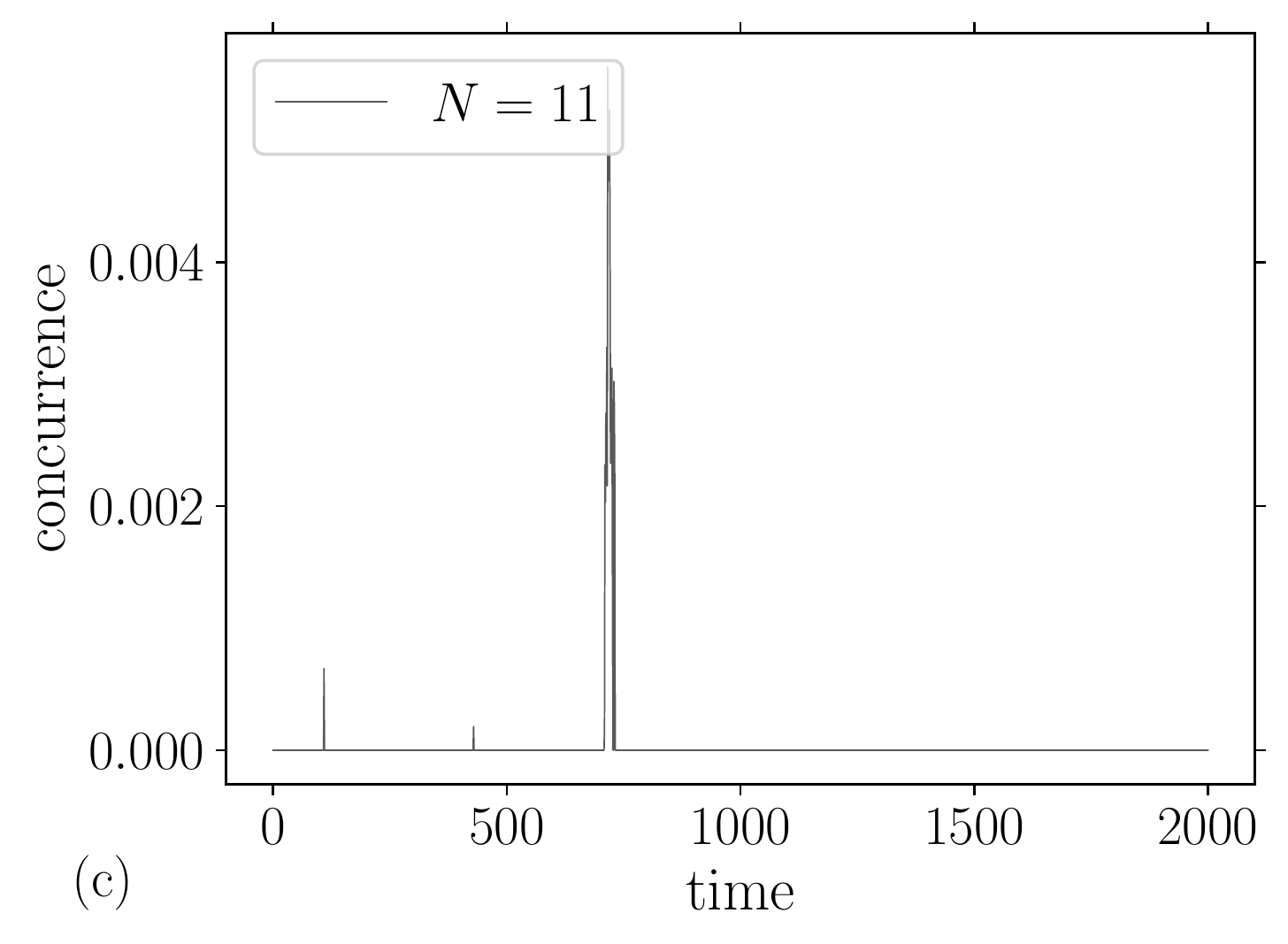}
  \caption{Concurrence as a function of the system size. Parameters $(1.5,1.2)$, `z'; $|V|=7,9,11$.
  \label{f:cN}}
\end{figure}

We turn now to the strong coupling regime and compare for the initial condition `z', two cases with fixed \(J = 1.2\) and angles \(\theta = 3\pi/8, \pi/8\). Both cases show a rapid spin relaxation (Fig.~\ref{f:zst}), with a smoother evolution in the large angle case \(3\pi/8\). The walker density of Fig.~\ref{f:zp} revels two distinct propagation regimes. Comparing the two cases we see a qualitatively new behavior appearing for \((\pi/8, 1.2)\) (Fig.~\ref{f:zp}c): the walker mainly stays in the neighborhood of its initial position in sharp contrast to the \((3\pi/8,1.2)\) case. We refer this situation as pertaining to a `localization' regime, at variance to the `propagation' regime of Figs.~\ref{f:o54p} and \ref{f:zp}a. Taking into account the ballistic motion of the underlying walker in absence of interaction, we may refer to this enhanced probability at the origin as interaction induced localization. A related phenomenon linking interaction and localization was found in the quantum walk of two coupled particles \cite{Verga-2018}.

\begin{figure}[tb]
  \centering
  \includegraphics[width=0.5\textwidth]{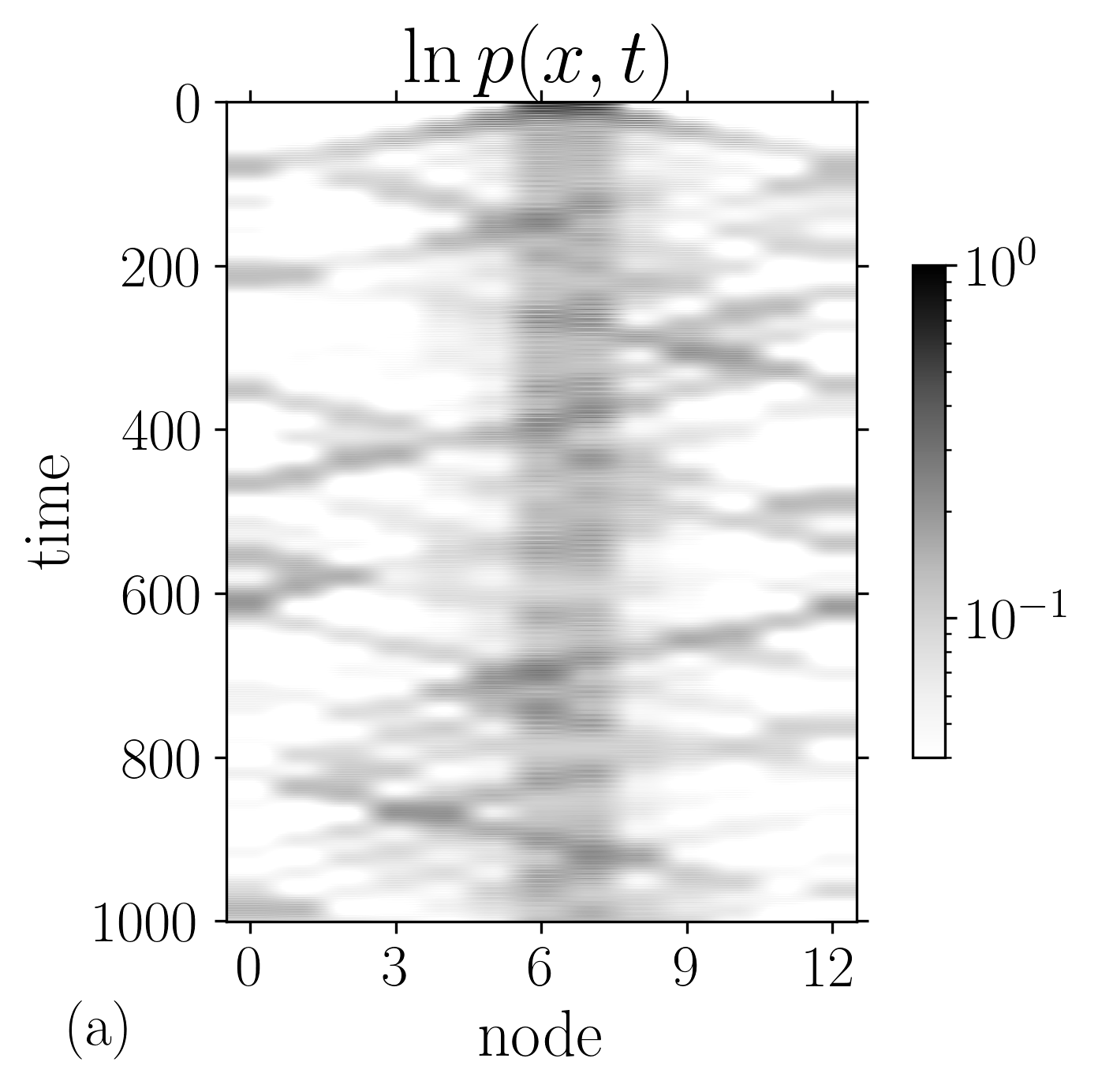}%
  \includegraphics[width=0.5\textwidth]{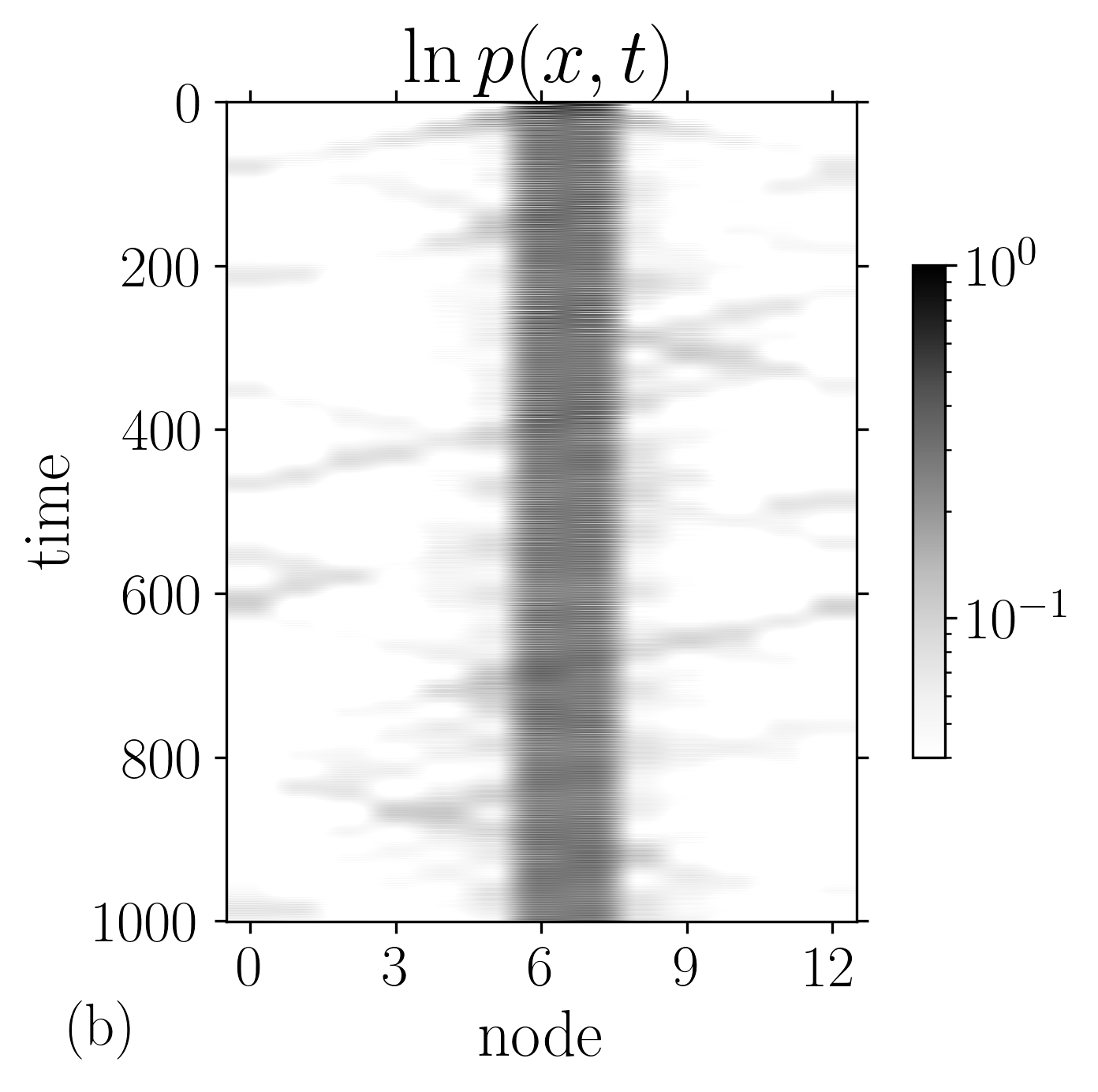}
  \caption{Irregular dynamics. Particle density for two different initial conditions: (a) `x', (b) `zx`. Parameters (0.1, 2). In addition to a central concentration the particle makes intermittent ballistic excursions, rarer in the `zx' case (b).
  \label{f:o3p}}
\end{figure}

\begin{figure}[tb]
  \centering
  \includegraphics[width=0.9\textwidth]{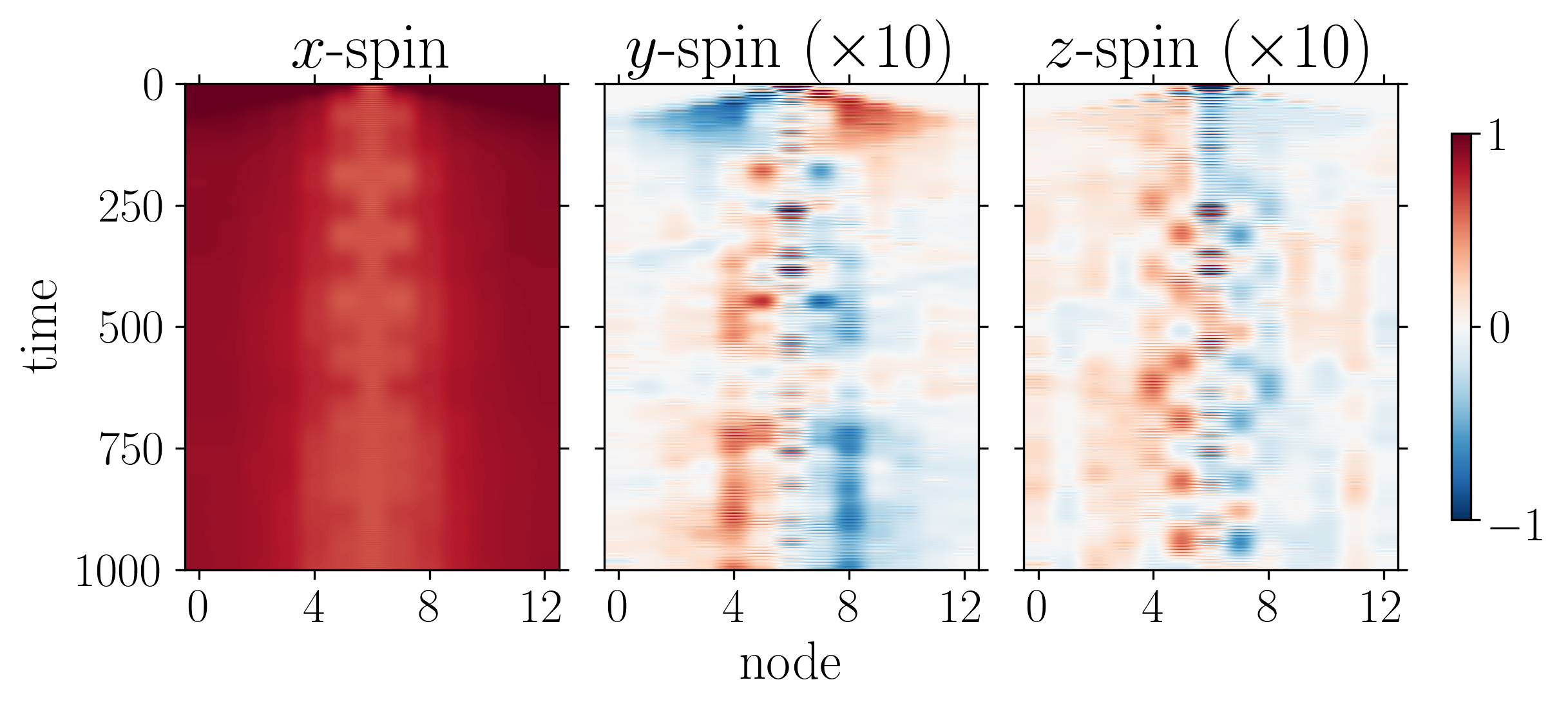}
  \caption{Spatiotemporal spin density. Parameters (0.1, 2) `x'. The ($y$,$z$) spin components are scaled by a factor 10. 
  \label{f:o3s}}
\end{figure}

\begin{figure}[tb]
  \centering
  \includegraphics[width=0.5\textwidth]{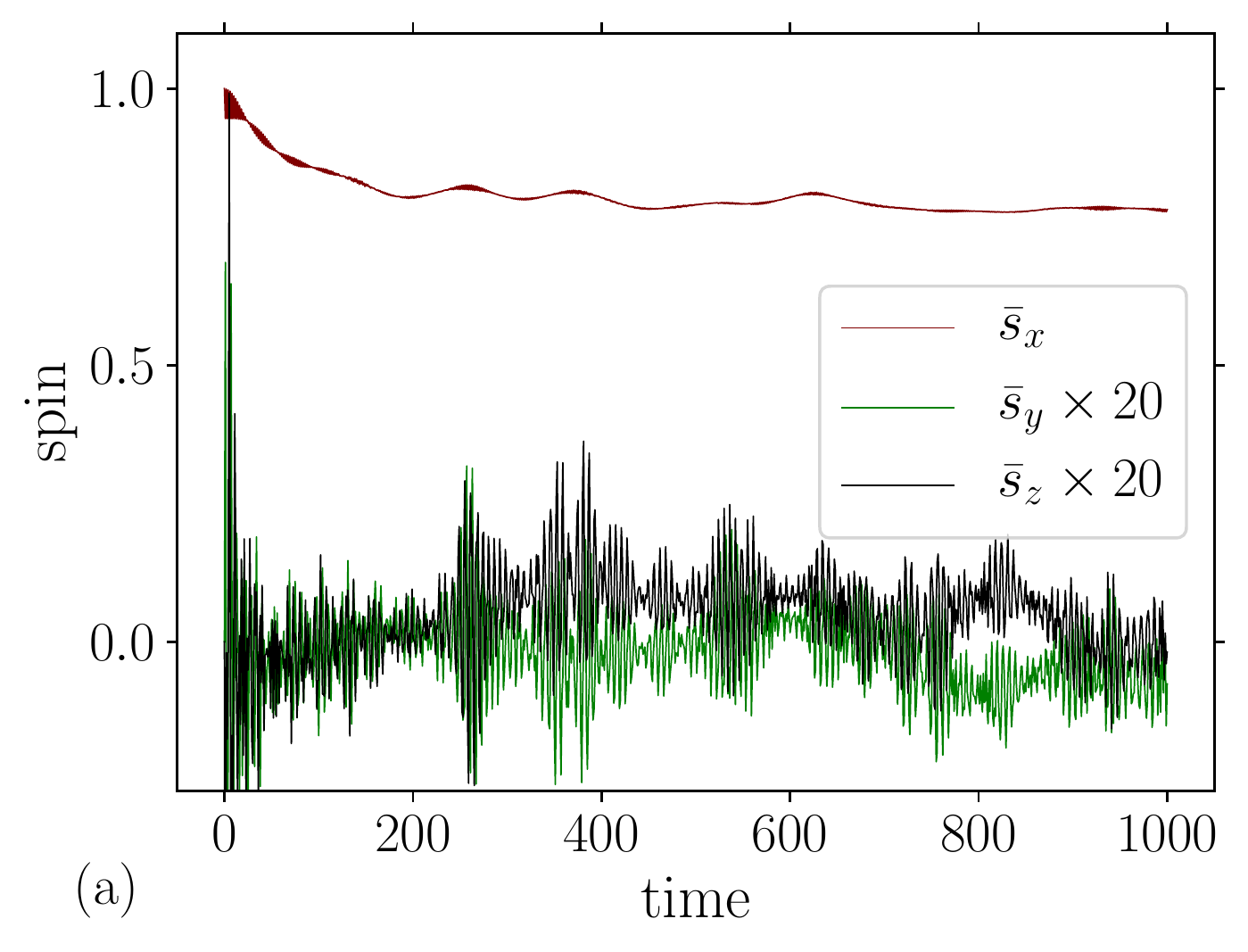}%
  \includegraphics[width=0.5\textwidth]{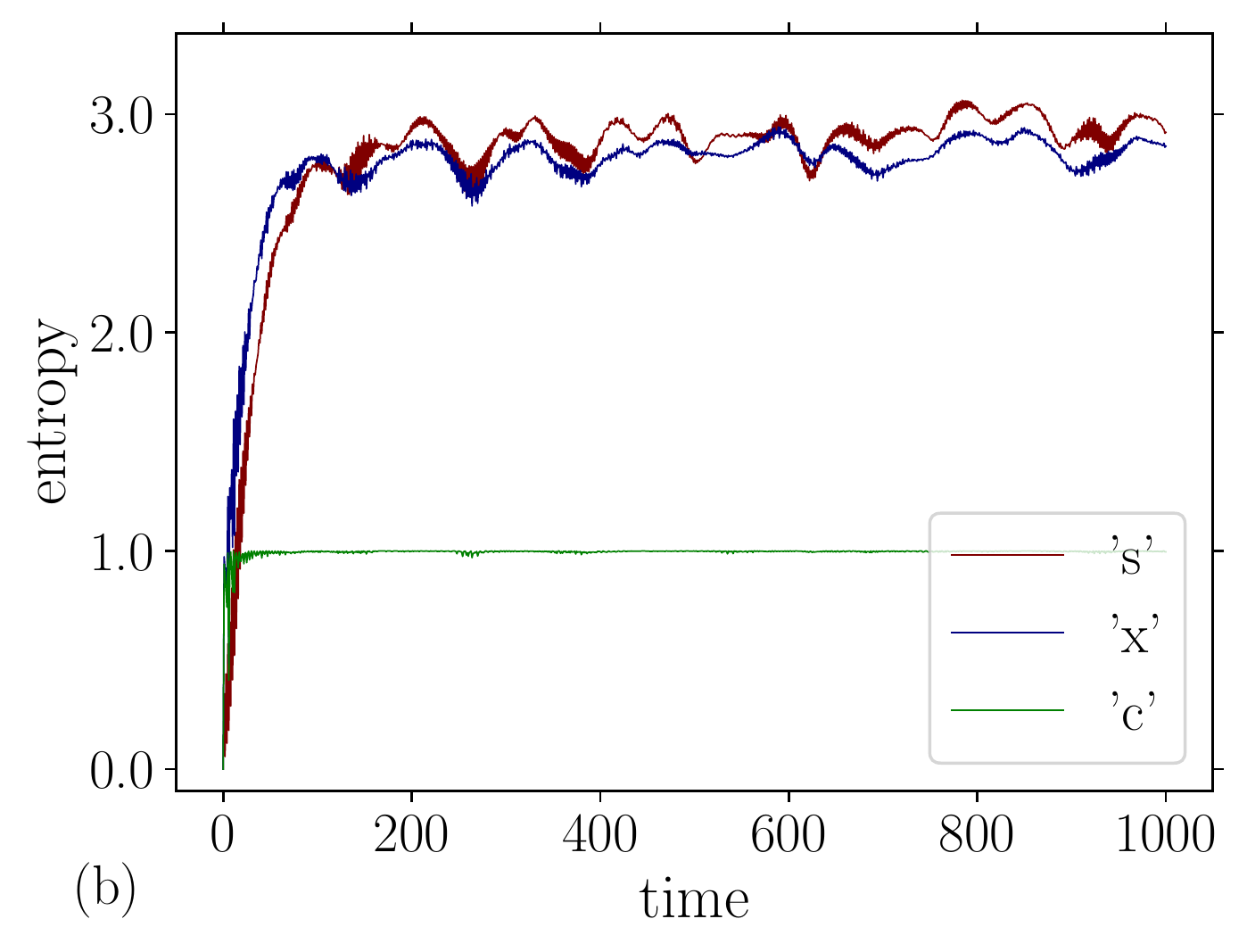}
  \caption{Averaged spin (a) and entanglement entropy (b), for strong coupling and small angle showing irregular dynamics. Parameters (0.1, 2), 'x' as in Figs.~\protect\ref{f:o3p}a and \protect\ref{f:o3s}. Particle and spin entropies follow the same pattern, remaining at a low level with respect to the random state value $\max S_x \approx 3.7$ (the coin entropy saturates quickly).
  \label{f:o3}}
\end{figure}

\begin{figure}[tb]
  \centering
  \includegraphics[width=0.45\textwidth]{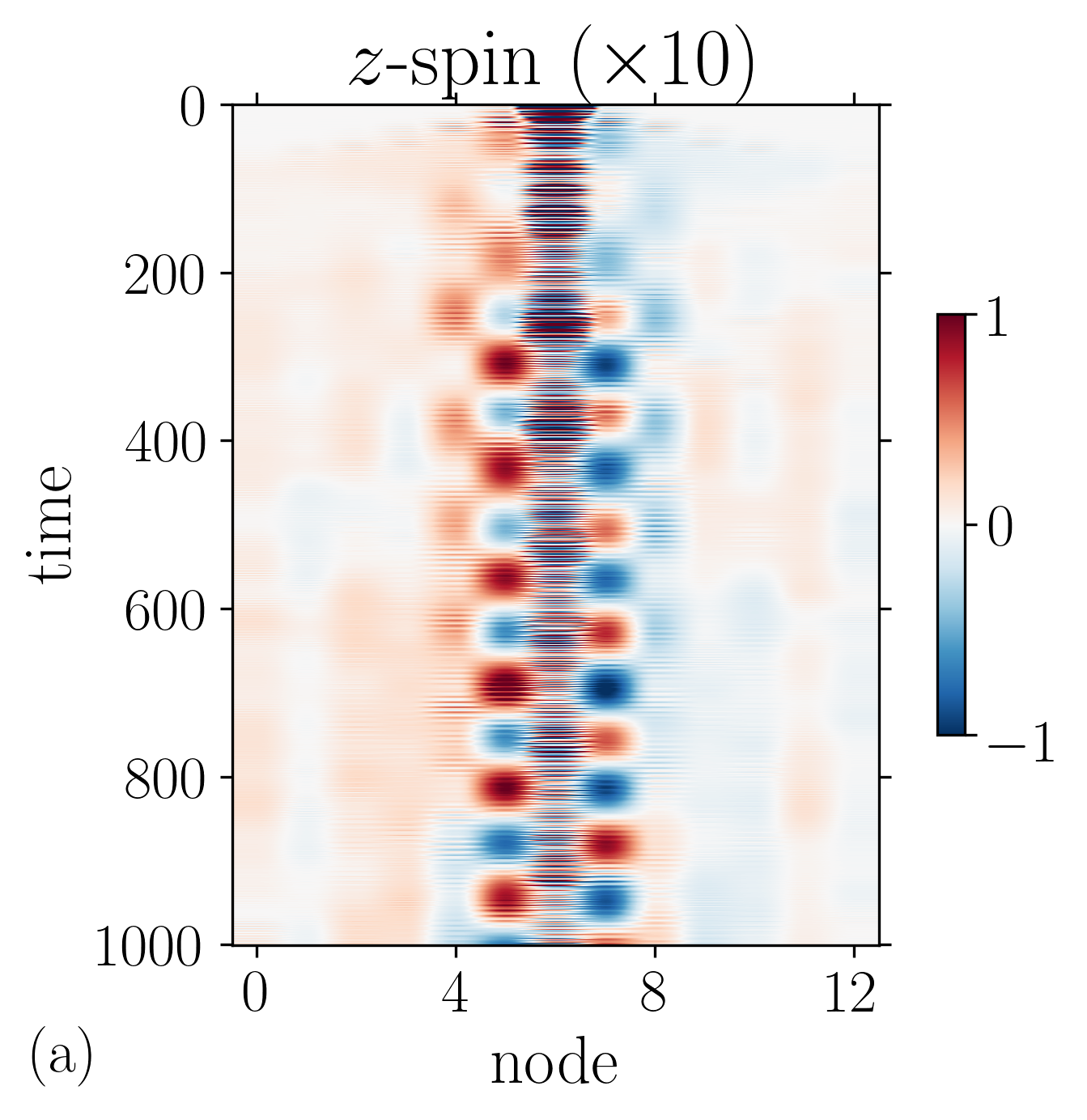}%
  \includegraphics[width=0.55\textwidth]{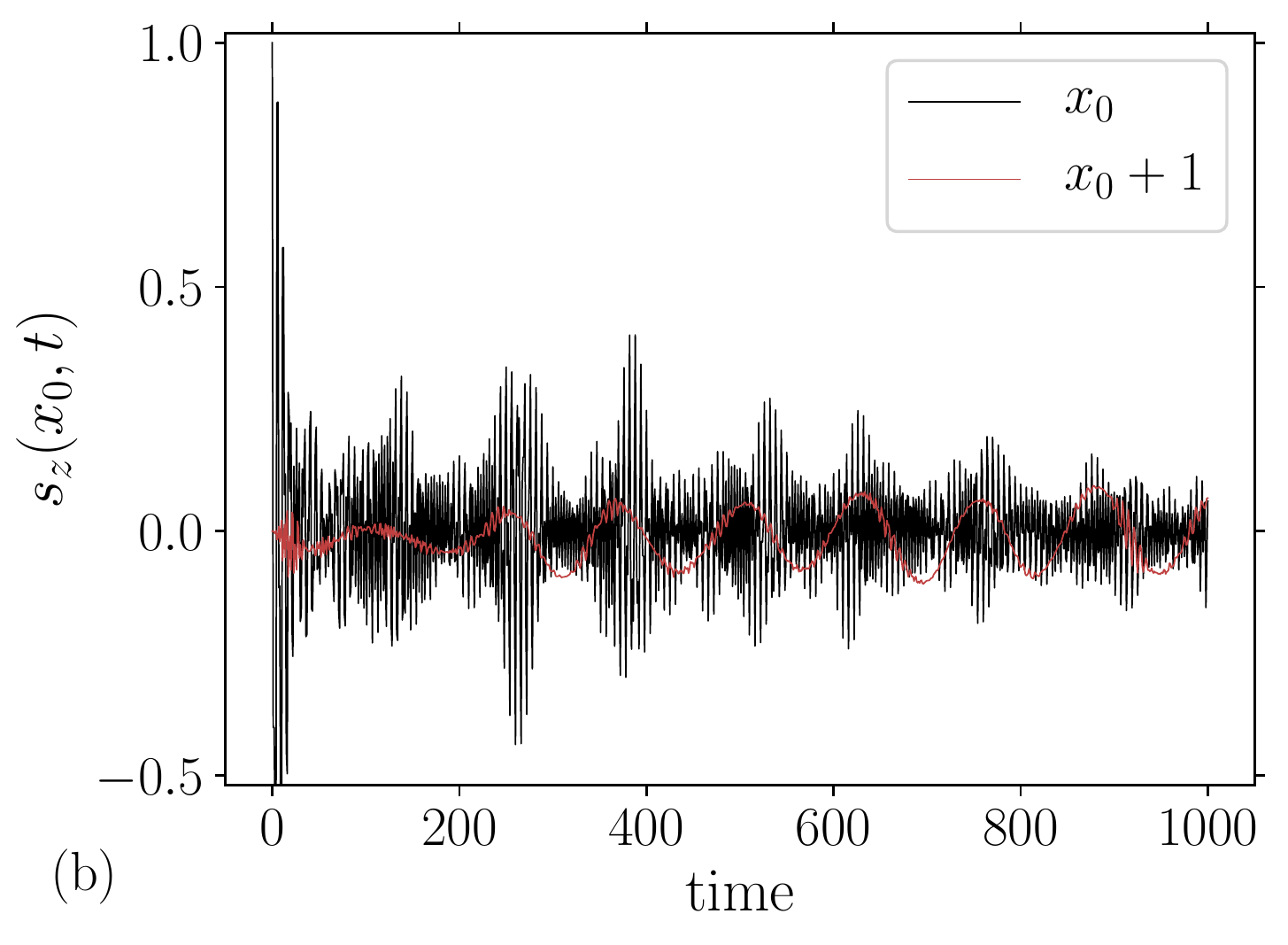}
  \caption{Spatiotemporal plot of the $z$ spin component (scaled) (a), and the corresponding temporal profile of node $x=x_0$. Parameters, (0.1, 2),'zx' as in Figs.~\protect\ref{f:o3p}b. The rapid variation at $x=x_0$ contrasts with the regular oscillations of the neighbor $x = x_0 + 1$, the red line in (b).
  \label{f:o2}}
\end{figure}

The entanglement entropy of a set of spins \(S_A(t)\) is presented in Figs.~\ref{f:zp}bd. The main distinction between the propagating and localized regimes is revealed by their respective growth laws. In the propagating case, we remark a linear stage for times \(t < 50\), after which follows a saturation regime where entanglement slowly increases. In the localization case, the linear stage disappears. In the case where the walker spreads ballistically approaching a uniform distribution over the line, the entropy sharply grows, while for the localized dynamics the growth of the entanglement is smoother. In contrast with normal exponential saturation, the spin set entropy of the localized dynamics can be fitted with a stretched exponential (the gray shaded curve in Fig.~\ref{f:zp}d):
\begin{equation}
\label{e:SAfit}
S_A(t) \sim 1 - \E^{-\nu t^\alpha}\,,
\end{equation}
where the exponent is \(\alpha = 0.5\), and \(\nu = 0.11\) a constant dependent in principle on the system's parameters \((\theta,J)\) (and also it might be slightly dependent on \(|V|\), in particular due to finite size effects). The fit is consistent with an initial stage \(S_A(t) \sim \sqrt{t}\), instead of the linear one \(S_A(t) \sim t\) observed in the propagation case.

One interesting question is whether spins spatially separated are entangled. We investigate this question in a system with closed (refelctive) boundaries `b', and using an initial state that propagates preferentially in one direction. We define a disconnected set of spins near the two borders:
\[A = \{1,2,3,4, 9,10,11,12\}, \quad |A| = 8,\quad |V| = 15\]
and measure \(S_A(t)\) using \eqref{e:SA}, in a system with \(15\) nodes. In Fig.~\ref{f:bz} we show together the particle and spin densities; the logarithmic scale of \(p(x,t)\) covers two orders of magnitude to reveal the correlation with the spin distribution; we see in particular, a propagating front where the spin reverses. The \((xcs)\) entanglement entropy, displayed in Fig.~\ref{f:bzA}a saturates to the random state values. The answer to the question is provided by the behavior of \(S_A\) shown in Fig.~\ref{f:bzA}b. We find that the separated spins entangle concomitantly with the progression of the mentioned flip spin front. Therefore, we may conclude that in addition to the global entanglement of the distinct degrees of freedom spanning the Hilbert space, entanglement also possesses a spatial organization that reflects in the possibility of distant spins entanglement through the interlacing created by the nonlocal smeared particle.

However, the state of two arbitrary selected spins do not form a Bell state as can be measured by the concurrence \eqref{e:Cee}. We find that whatever the pair of values \((\theta,J)\) the concurrence essentially vanishes. Indeed, a favorable situation for the formation of entanglement of two spins is when the state is near `z' eigenstate so that the particle can propagate almost freely \(\theta \approx \pi/2\) together with a strong coupling \(J > 1\); we show in Fig.~\ref{f:cN} the concurrence as a function of the number of sites. The increasing time interval between concurrence revivals can be interpreted as a progressive transition between a quasiperiodic and a chaotic dynamics, in analogy with the kicked top phenomenology \cite{Wang-2004fk}, which corresponds to the actual behavior of the system for the chosen parameters. In addition to this interpretation, the vanishing of the concurrence, that is to say the absence of bipartite entanglement correlated with an increase in the multipartite entropy \(S_A\), can be related to the presence of nonlocal entanglement, not necessarily associated with non regular dynamics; a scenario consistent with the mediated character of the spin-spin interaction. We discuss below the appearance of irregular behavior.

In summary, we have shown in Figures~\ref{f:o54p}-\ref{f:zp} the characteristic oscillations which are present for the full range of angles in the weak coupling limit \(J \ll 1\), and pointed out some of the qualitative differences observed between angles in the first and second quarters. We found that systems in the oscillation-relaxation regimes can be characterized by their distinct entanglement features, from recurrent to monotone growth towards saturation. At variance to the weak coupling case, for strong interaction oscillations are almost suppressed by a rapid relaxation to a saturation regime. We also observed that, depending on the point in the phase space \((\theta,J)\), two distinct particle density distributions arise, one uniform which can be associated with ballistic propagation, and another picked at the origin, which reveals localization driven by interaction. Finally we found evidence of entanglement between distant spins as unveiled by the entropy of a disconnected set.

\begin{figure}[thb]
  \centering
  \includegraphics[width=0.6\textwidth]{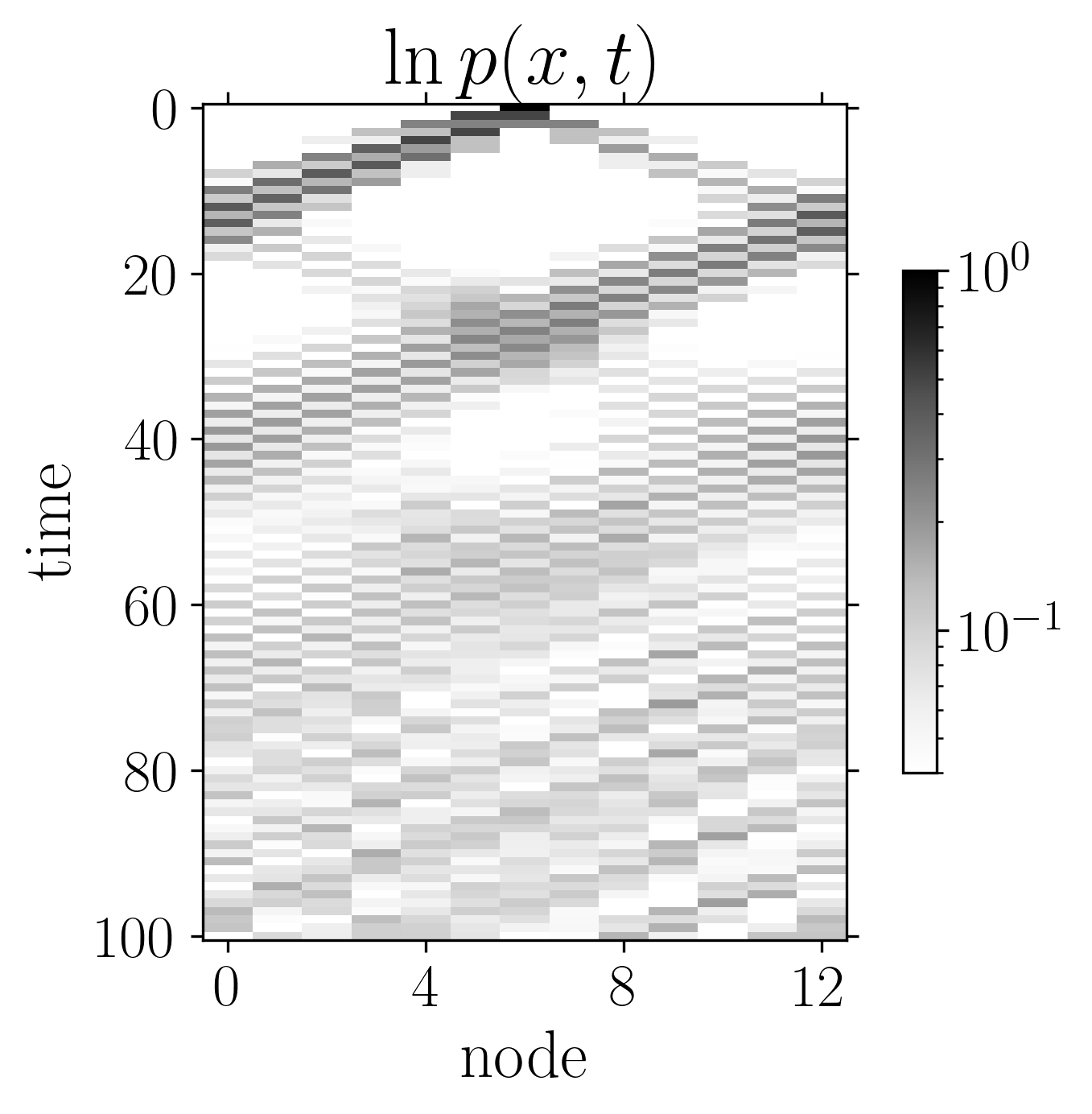}
  \caption{Spatiotemporal plot of the position probability showing irregular dynamics without localization. Parameters \((\pi/2, \pi)\), `x'.
  \label{f:PP_pL}}
\end{figure}

\begin{figure}[tb]
  \centering
  \includegraphics[width=0.33\textwidth]{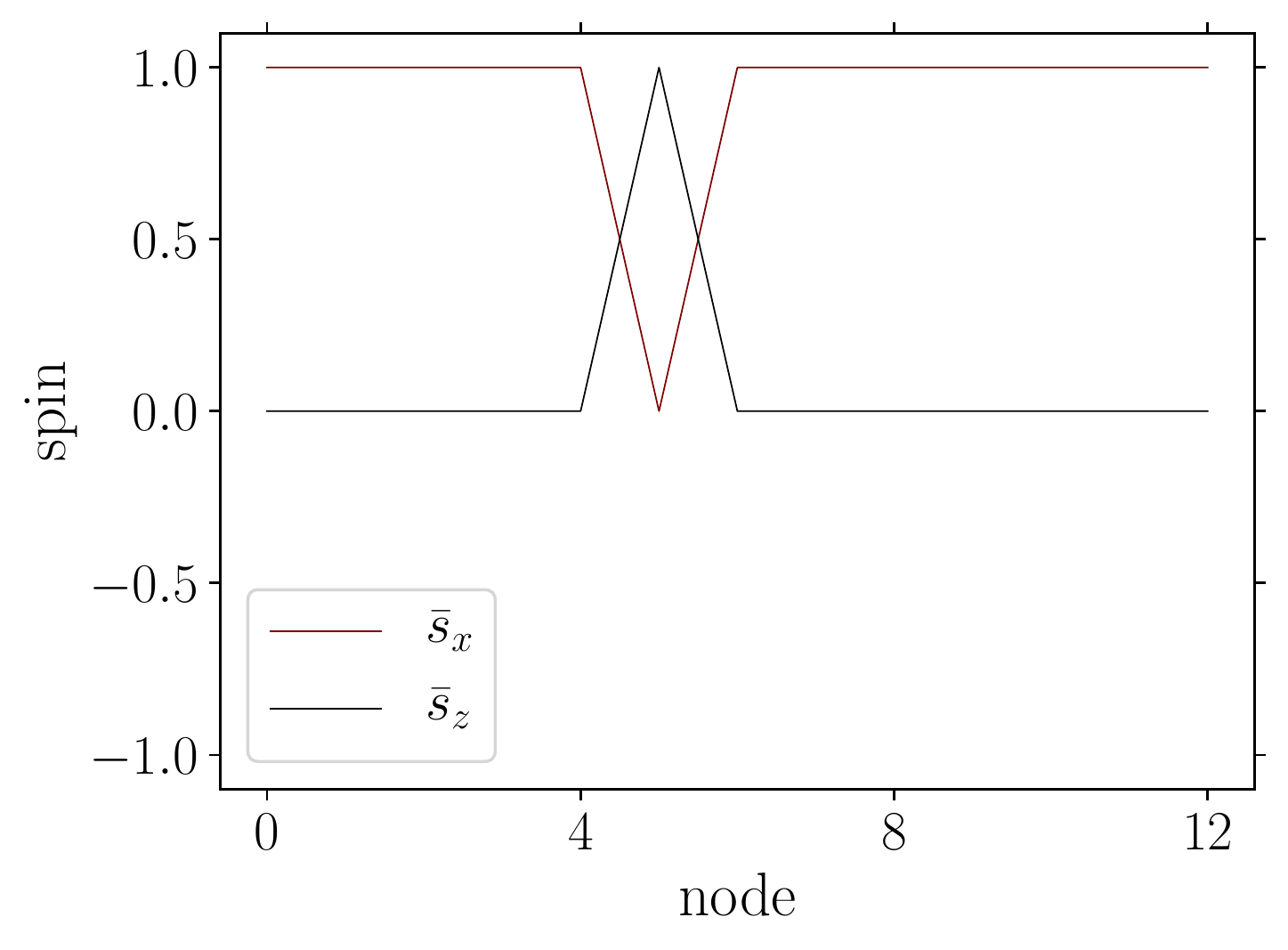}%
  \includegraphics[width=0.33\textwidth]{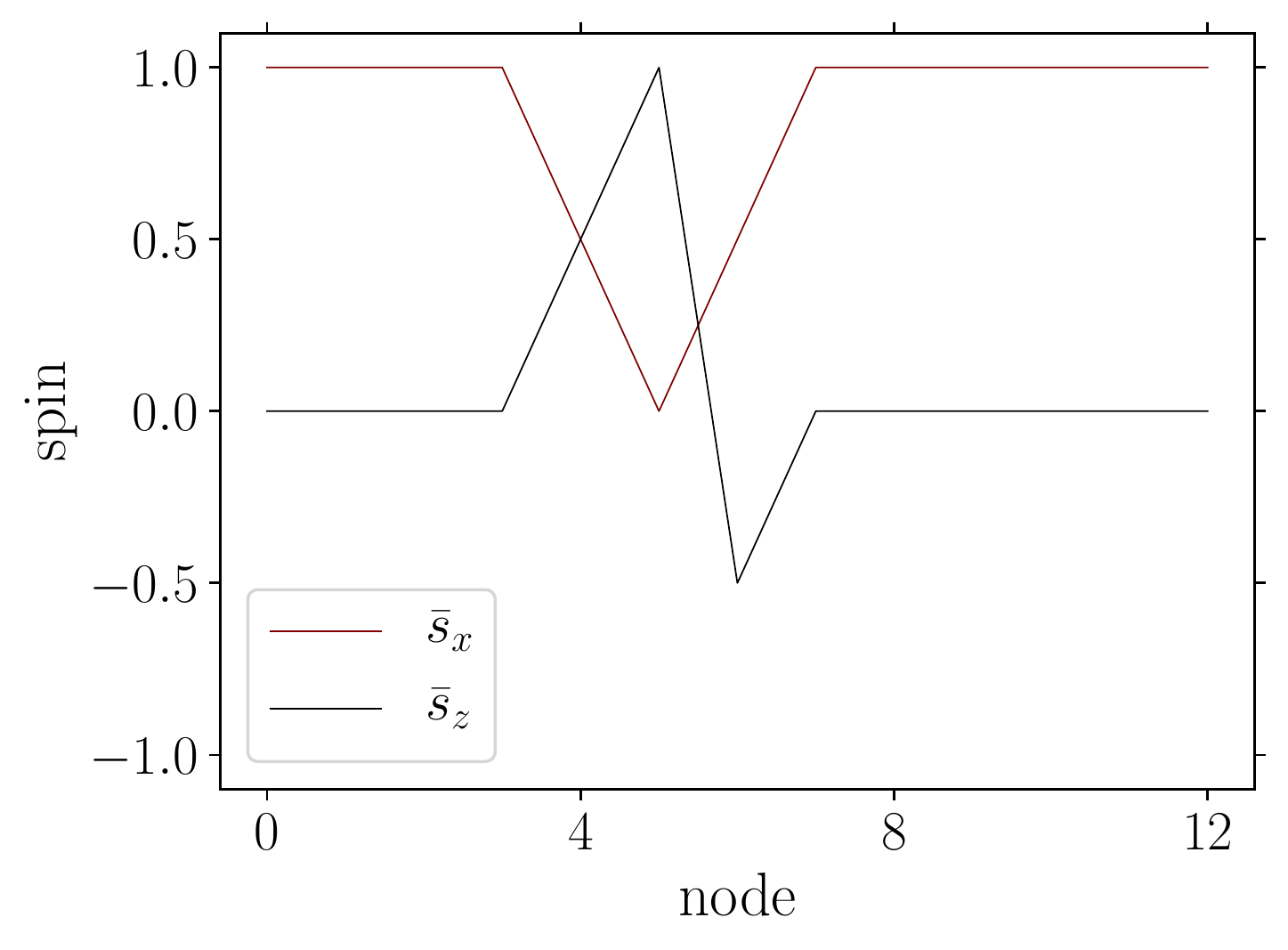}%
  \includegraphics[width=0.33\textwidth]{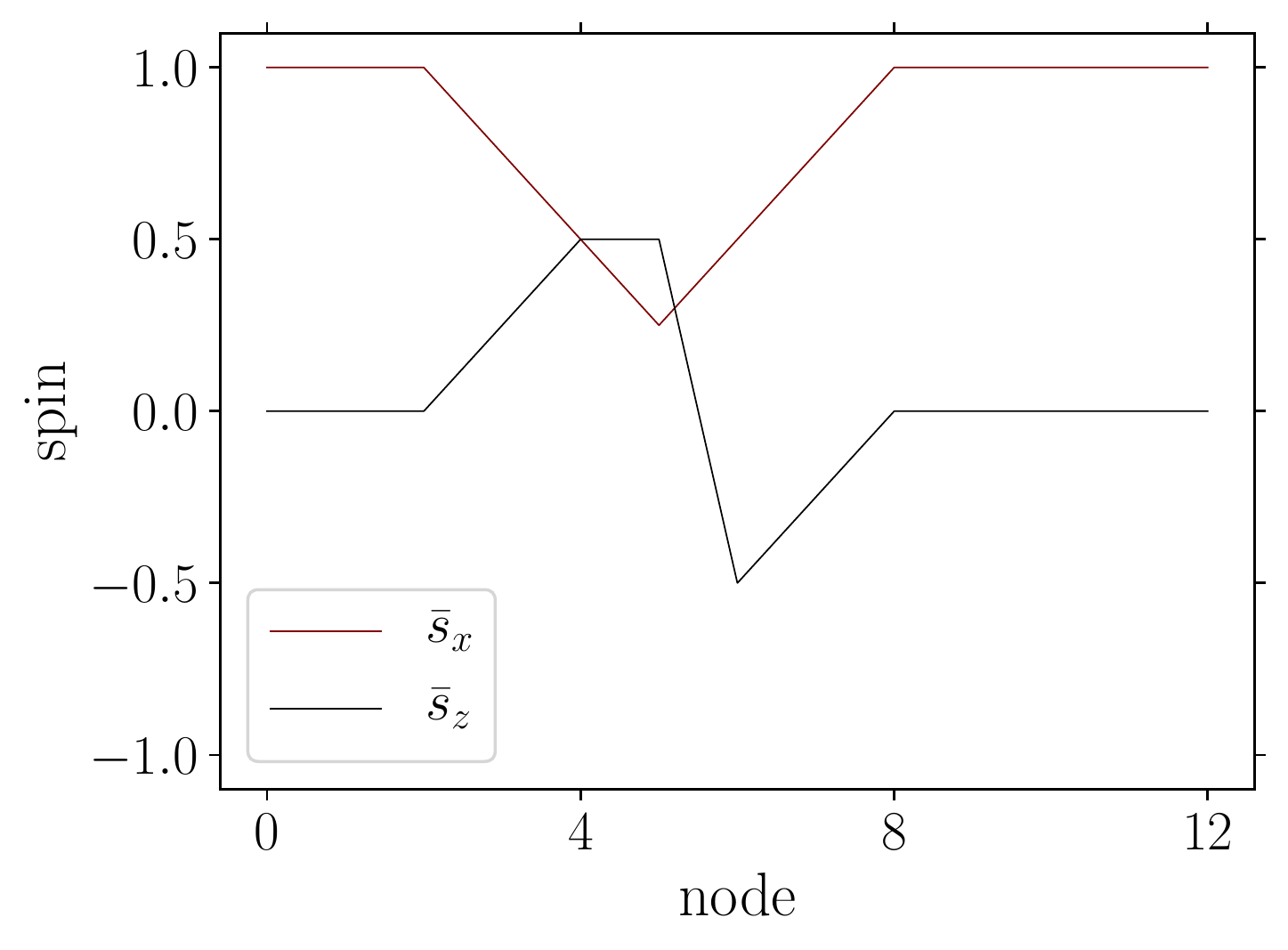}
  \caption{Three steps ($t=1,2,3$) of the quantum walk showing the changes in spin. Parameters \((\pi/2, \pi)\), `x' (as in Fig.~\protect\ref{f:PP_pL}).
  \label{f:PP_s1}}
\end{figure}

\begin{figure}[tb]
  \centering
  \includegraphics[width=0.97\textwidth]{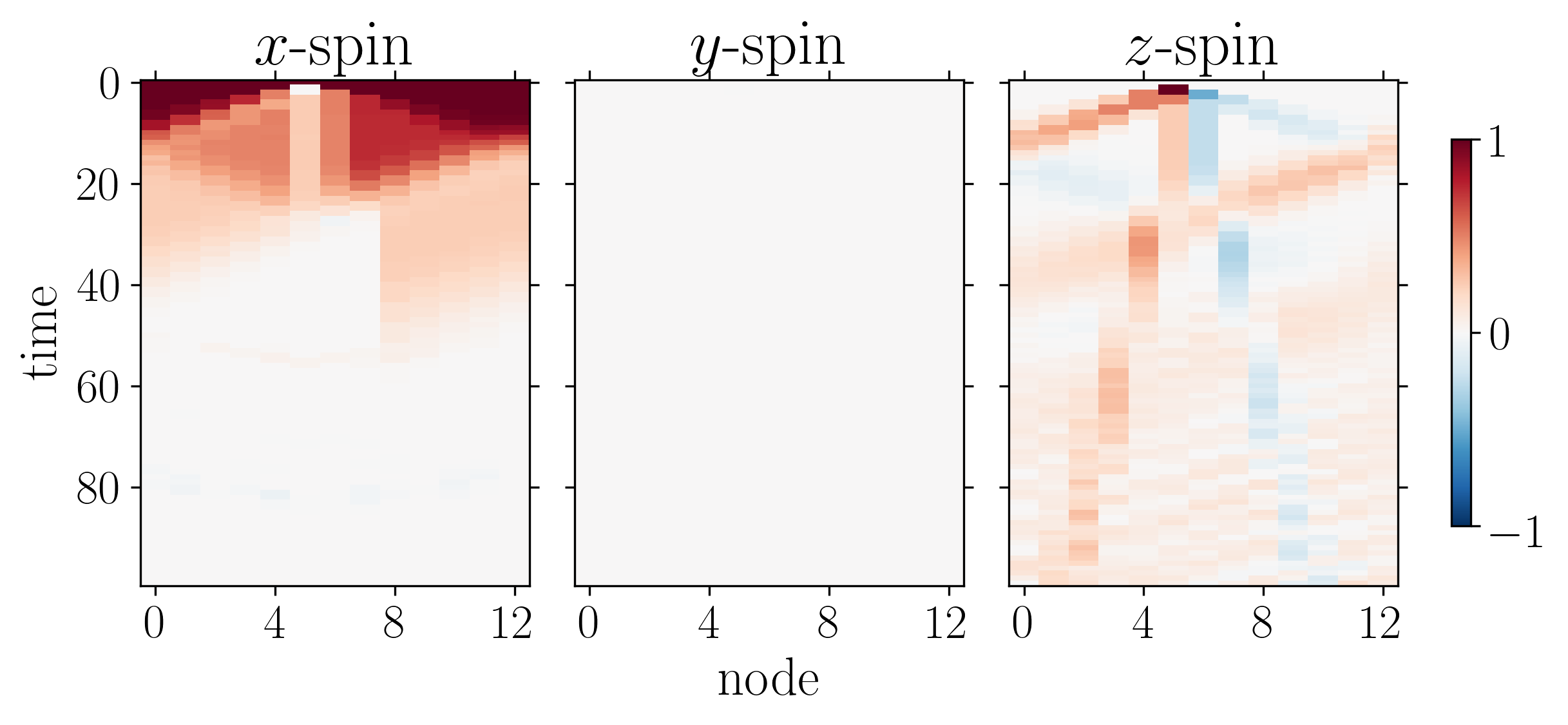}
  \caption{Spatiotemporal diagram of the spin $\bm s(x,t)$. Parameters \((\pi/2, \pi)\), `x'. 
  \label{f:PP_s}}
\end{figure}

\begin{figure}[tb]
  \centering
  \includegraphics[width=0.7\textwidth]{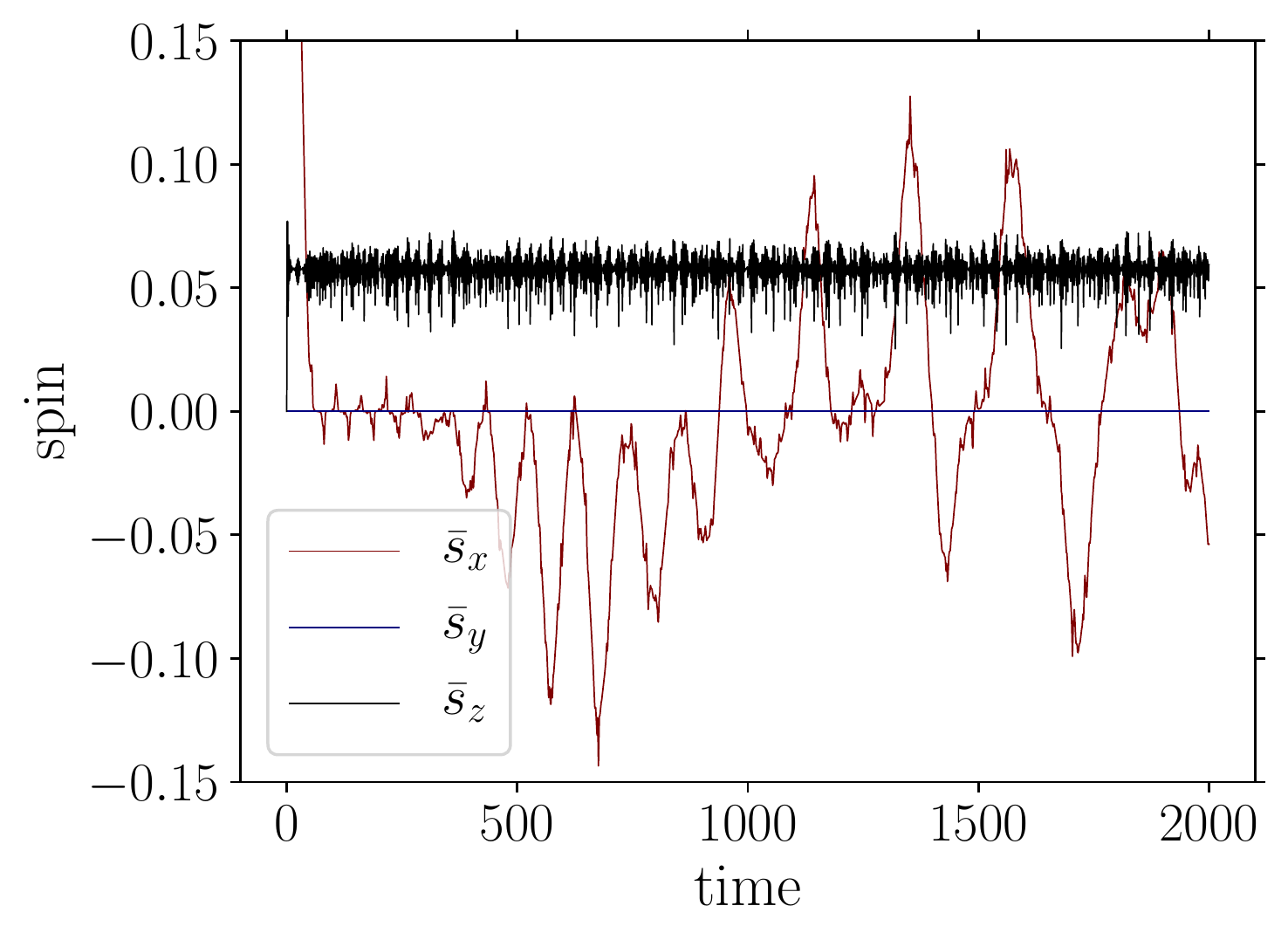}
  \caption{Long time evolution of the averaged spin as a function of time, showing the irregular dynamics in the propagating regime. Parameters \((\pi/2, \pi)\), `x'(c.f. Figs.~\protect\ref{f:PP_pL} and \protect\ref{f:PP_s}).
  \label{f:PP_st}}
\end{figure}

\subsection{Chaos}

In addition to oscillations and relaxation, the interacting walk exhibits irregular dynamics, characterized by a complex behavior of the observables. We fix the point \((0.1,2)\) in the \(\theta \approx 0\) region of the parameter phase space, with initial conditions `x' and `zx', which differ in the spin state of the edge \(e = (x_0, x_0+1)\). A remarquable pattern shows up in Fig.~\ref{f:o3p}. The particle density is essentially concentrated at the origin, however, intermittently ballistic excursions spark throughout the line. We note that the localized spin initial state `zx' leads to a stronger particle localization, underlining the close link between entanglement features and spin distribution. The spatiotemporal spin dynamics (Fig.~\ref{f:o3s}) shows the emergence, from an initial homogeneous spin state perturbed by the motion of the particle starting from one node, of structures with a large scale range. The intermittent behavior is clearly seen in Fig.~\ref{f:o3} where we plot the mean spin vector as a function of time (we zoomed into the \(y\) and \(z\) spin components), and the entanglement entropies. It is worth noting the close relationship between the spin and particle entanglements, meaning that the system is far from spin entanglement saturation (as would be the case for an almost random quantum state).

For the `zx' initial state, it is interesting to follow the evolution of the spin up located at the central bond (Fig.~\ref{f:o2}). The irregular motion of the central spin contrasts with the smoother oscillatory behavior of its neighbors. This is an interesting illustration of the spectrum of scales created by the iteration of \(U(\theta,J)\), in particular the appearance of large spatiotemporal scales.

We discuss now a case `x' with parameters \((\pi/2, \pi)\), in the large angle strong coupling region, to illustrate irregular dynamics without localization. Figure~\ref{f:PP_pL} displays the particle density, which corresponds to a ballistically propagating and spreading particle. The step by step motion of the walker modifies the spin state, as illustrated in Fig.~\ref{f:PP_s1}: a spin wake follows the particle progressively transforming the initial homogeneous \(\ket{+}\) state, enriching it with new amplitudes and exciting the other spin components. The initial spatiotemporal spin evolution is presented in Fig.~\ref{f:PP_s}; it is interesting to note the strong correlation of the particle motion in Fig.~\ref{f:PP_pL} with the behavior of the $z$ spin component. At long times this simple initial pattern becomes a complex motion as reflected by the mean spin shown in Fig.~\ref{f:PP_st}.

In brief, we observed that in some regions of the parameter space irregular dynamics of the particle and magnetization are possible. The chaotic like behavior of the observables occurs  for both localized and propagating states. One characteristics of the `chaotic' regime is its relatively small entanglement entropy, with the spin entropy sticking to the particle one.

\subsection{Topology}

\begin{figure}[tb]
  \centering
  \includegraphics[width=0.33\textwidth]{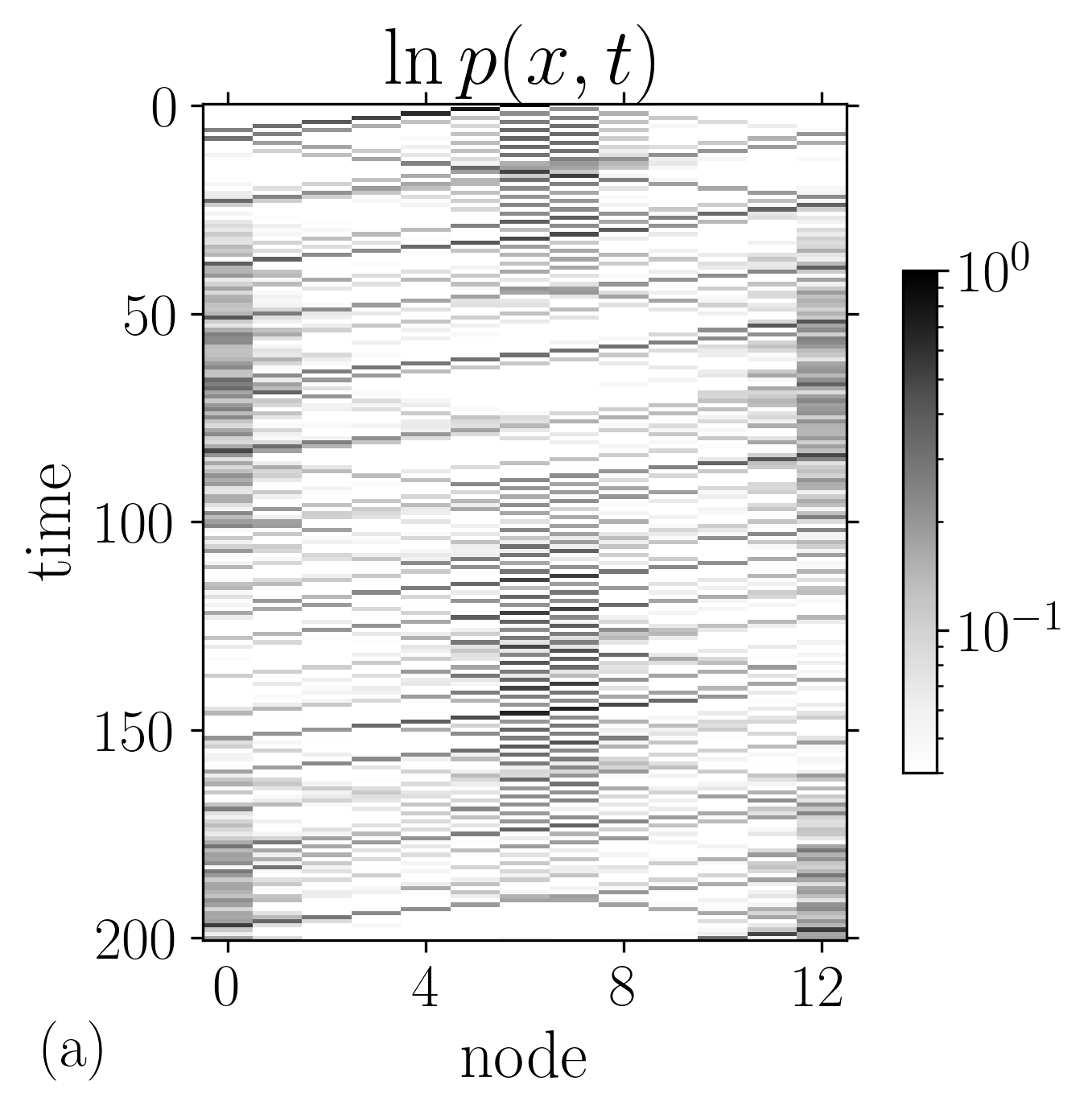}%
  \includegraphics[width=0.33\textwidth]{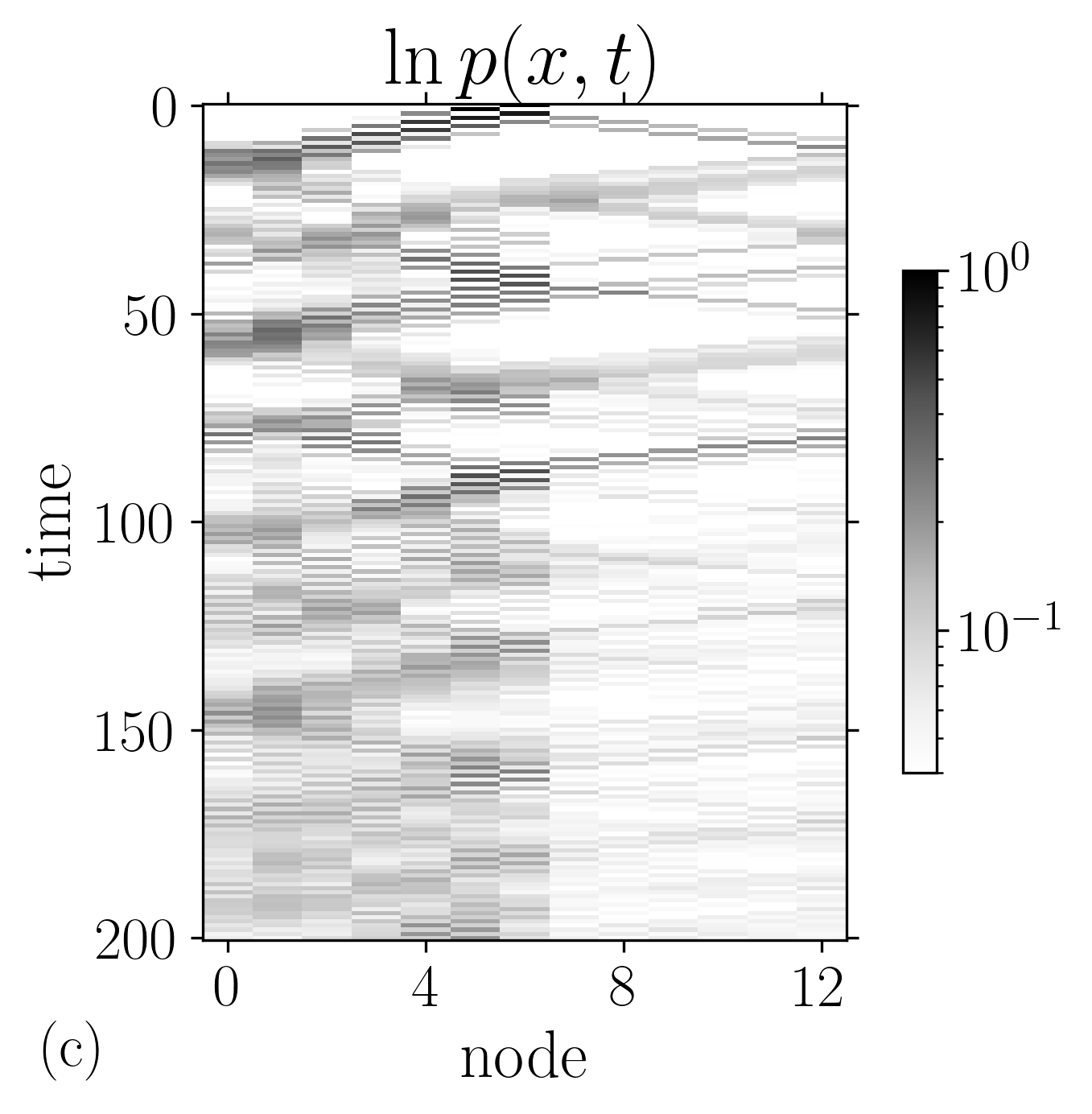}%
  \includegraphics[width=0.33\textwidth]{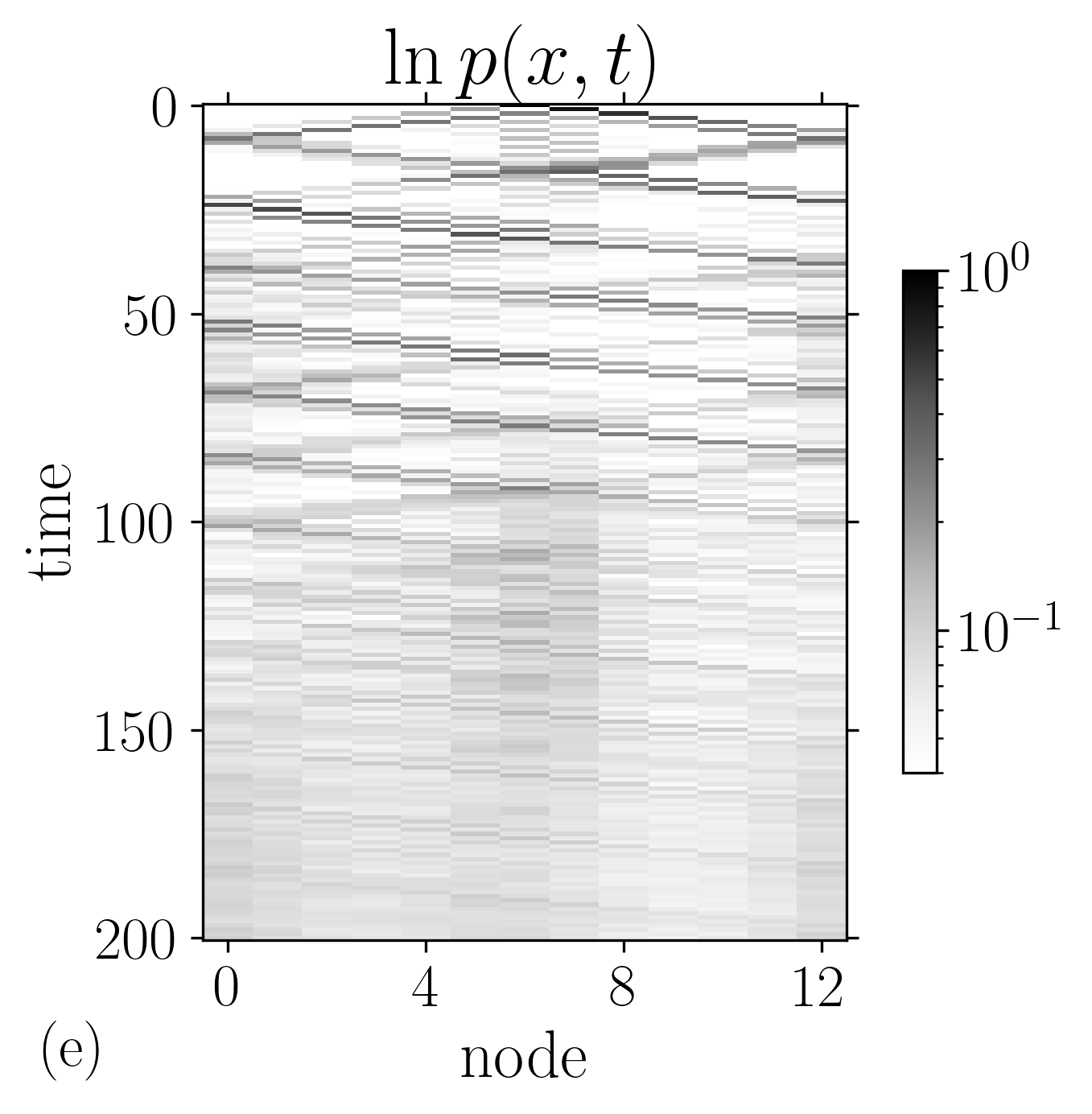}\\
  \includegraphics[width=0.33\textwidth]{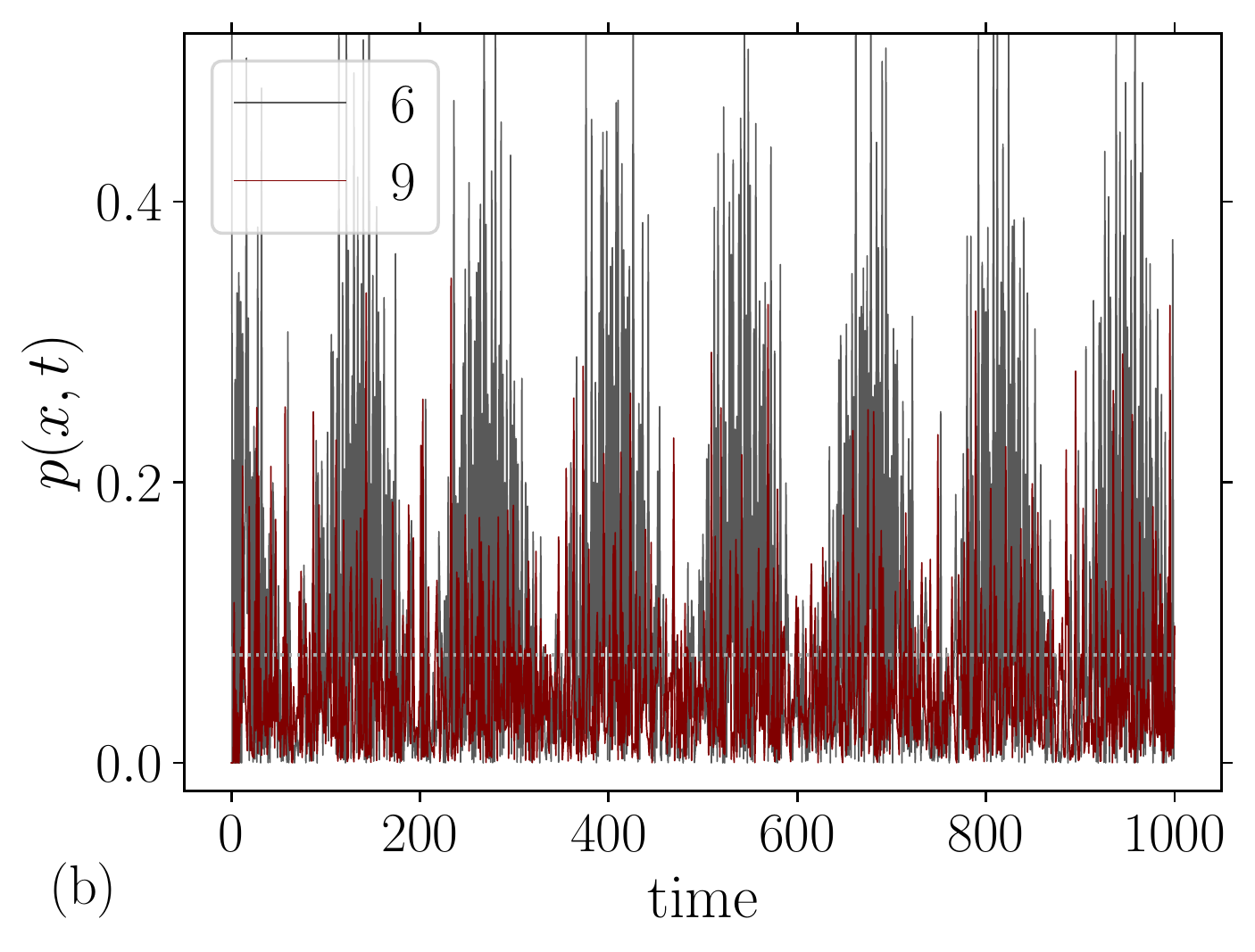}%
  \includegraphics[width=0.33\textwidth]{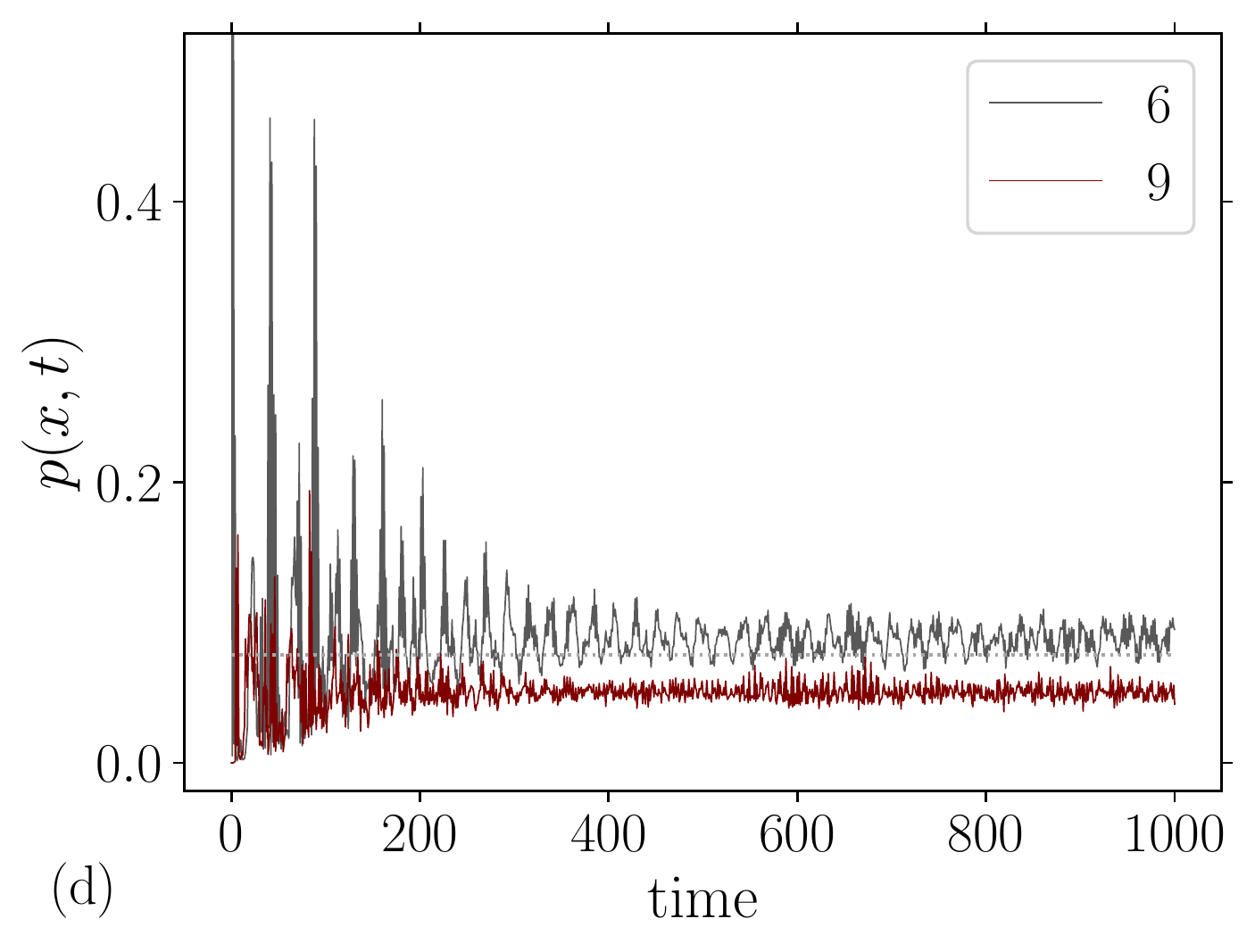}%
  \includegraphics[width=0.33\textwidth]{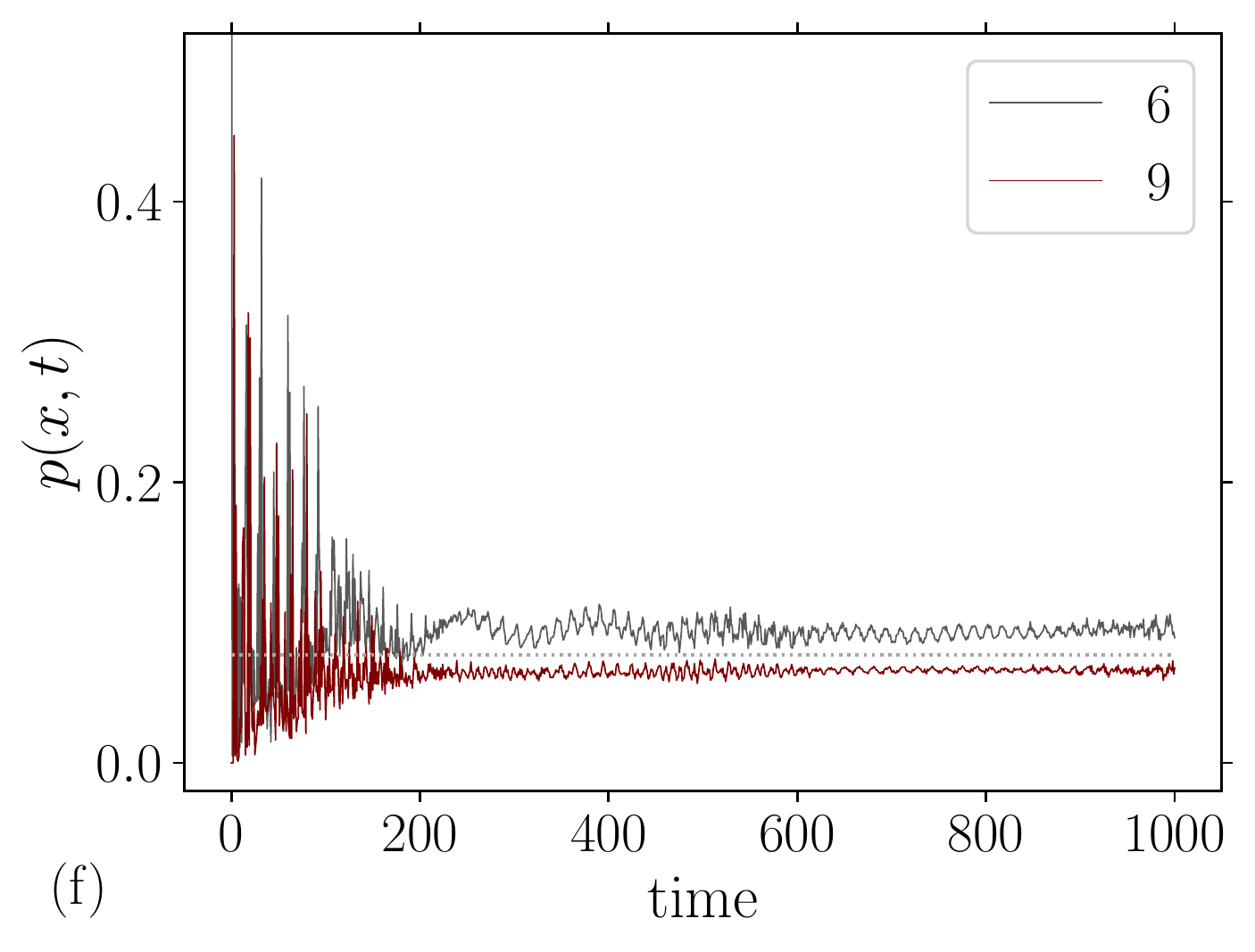}
  \caption{Topology `t'. Columns represent different interfaces and interaction strengths: (a-b) $J=0$, (c-d, e-f) $J=0.2$; (a-b, e-f) topological interface, $\theta(x<6) = 1.1$, $\theta(x \ge 6) = 2.1$; (c-d) normal interface $\theta(x<6) = 0.5$, $\theta(x \ge 6) = 1.2$. Interfaces are at $x=0$ and $x=6$, due to the periodic boundary condition. The top row shows the position distribution in logarithmic scale $\ln p(x,t)$; the bottom row the time dependency of the probability $p(x,t)$ for the central node $x=6$ and a ``body'' node $x=9$. Initial condition `x'. The thin gray line in (b,d,f) corresponds to the uniform probability.
  \label{f:tp}}
\end{figure}

A hallmark property of quantum free walks is that they can be used as a laboratory to investigate topological phases and the bulk-boundary correspondence, as we mentioned in the introduction. Less investigated is the interplay between topology and interaction in quantum walks. For instance, one may think that tuning the coupling constant \(J\), it would be possible to change the topology of a given phase. The Dirac walk, as we shall discuss in the next section, has two topological phases for \(\theta < \pi/2\) and \(\theta > \pi/2\), corresponding to each side of the parameter boundary \(\theta = \pi/2\) for which the spectral gap of its associated effective Hamiltonian closes \cite{Kitagawa-2010jk}. However, the physical difference between the two phases is somewhat arbitrary because of the existence of a simple relationship relating \(\theta < \pi/2\) and \(\theta > \pi/2\) (c.f. \cite{Asboth-2012qy} for a discussion about the chiral symmetry of the Dirac walk). In the absence of interaction the walk is governed by \(W(\theta)=MR(\theta)\), hence a change \(\theta \rightarrow \pi/2 - \theta\) corresponds to a change in the rotation direction, or equivalently to a change \(Y \rightarrow -Y\), up to a constant unitary transformation, which do not modify the behavior of the observables. This is similar to what happen in other context, with the Dirac equation as already noted by Shen et al. \cite{Shen-2011kx}. The Dirac equation do support edge localized states protected by the mass gap, however it is not a satisfactory model of a topological insulator: the symmetry between positive and negative energy states must be broken in this case to obtain a genuine model of distinct topological phases. In our case the presence of the interaction, whose unitary operator do not commute with color rotations, breaks the symmetry between the two angle sectors. We may expect new effects to appear due to interaction and related to the existence of edge states at the interface separating the phases.

\begin{figure}[tb]
  \centering
  \includegraphics[width=0.33\textwidth]{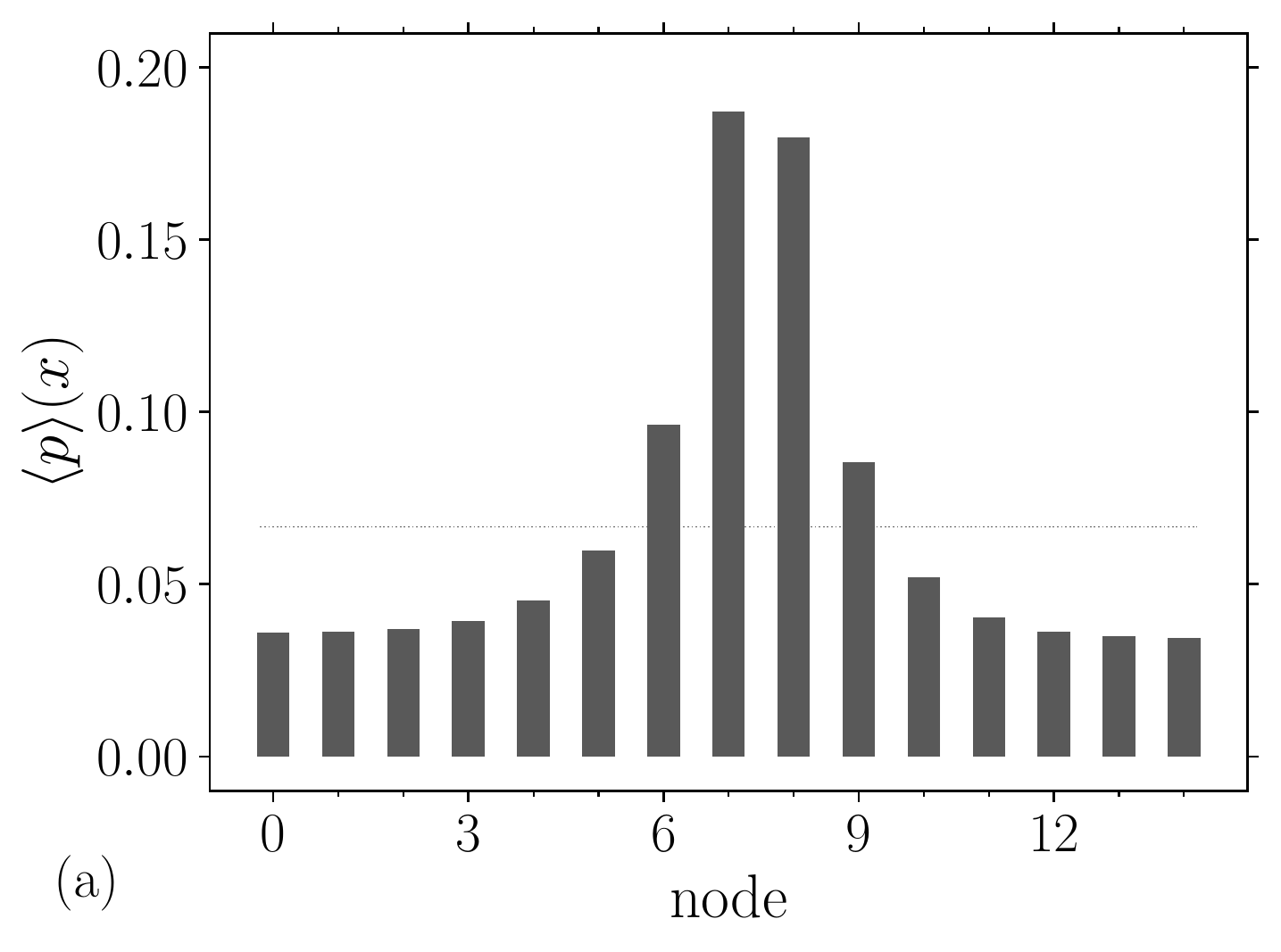}%
  \includegraphics[width=0.33\textwidth]{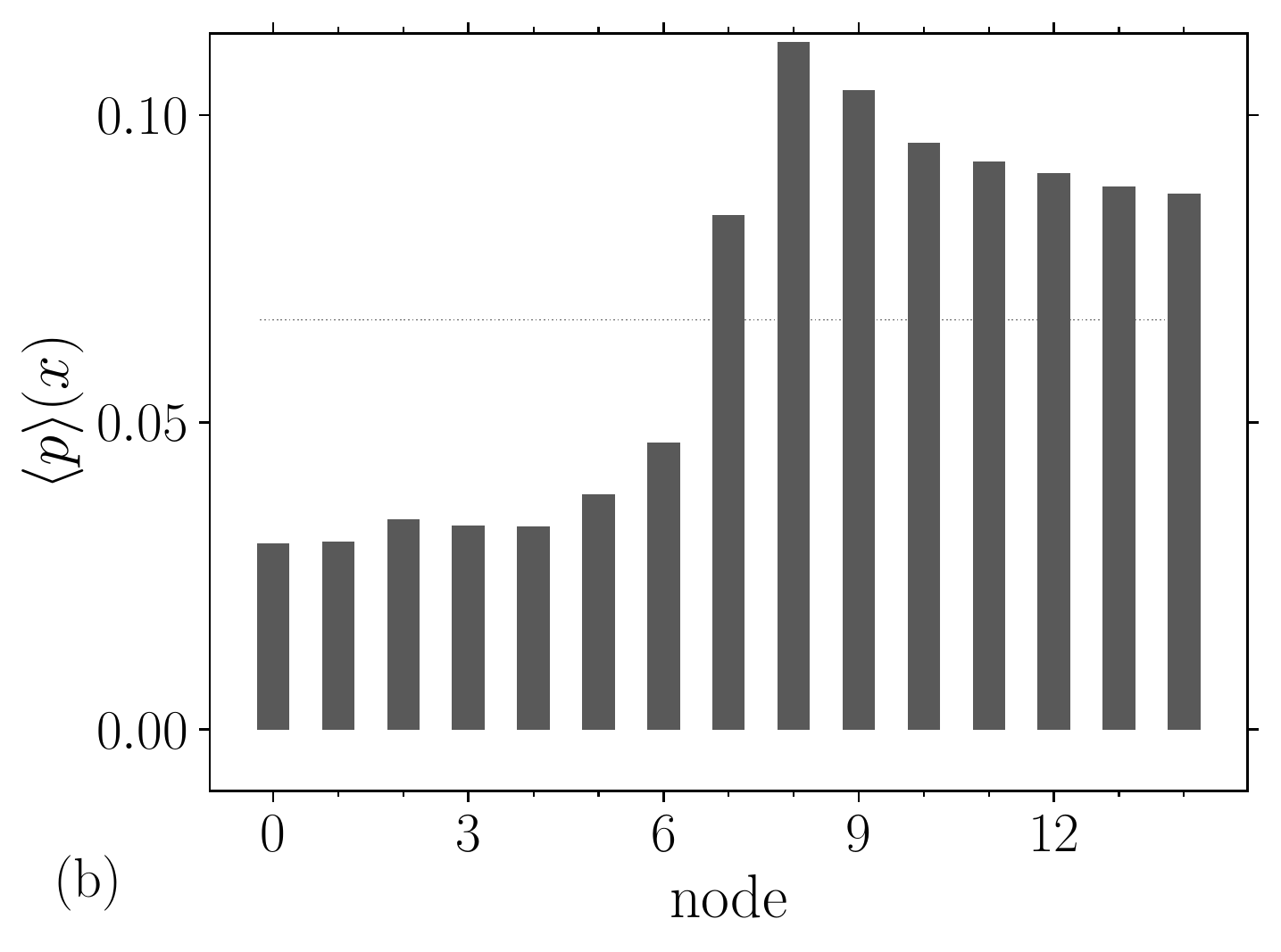}%
  \includegraphics[width=0.33\textwidth]{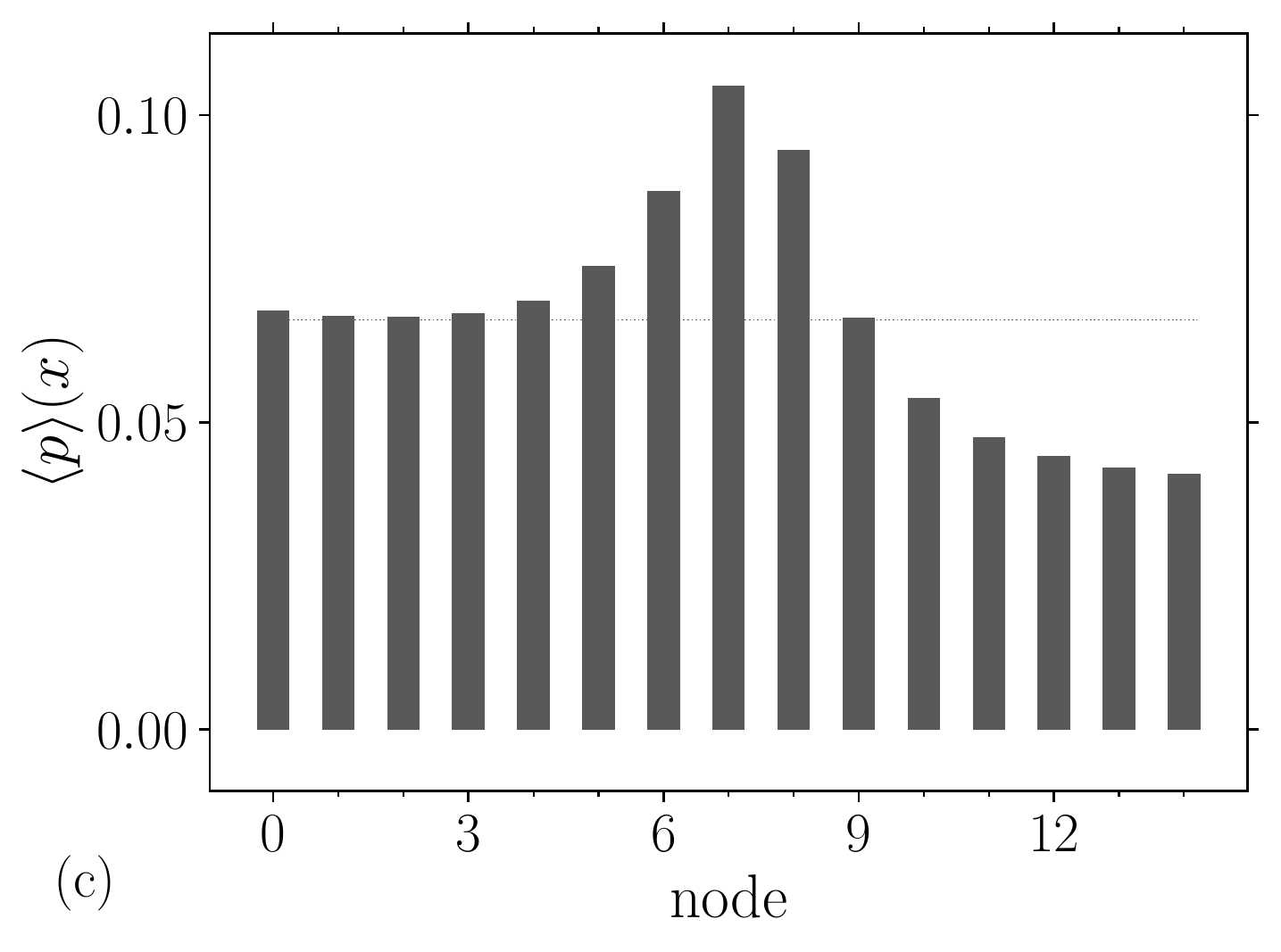}
  \caption{Topology `t'. Particle density averaged over time. The light dashed line is the uniform level distribution. Parameters `bz' (closed boundary conditions); (a) free walk $J=0$, $\theta_-=1.1$ $\theta_+=2.1$; (b) trivial interface $J=05$, $\theta_-=0.5$ $\theta_+=1.2$, (c) topologically nontrivial interface $J=0.5$, $\theta_-=1.1$ $\theta_+=2.1$. Interface at $x=7$ ($|V|=15$).
  \label{f:tbh}}
\end{figure}

\begin{figure}[tb]
  \centering
  \includegraphics[width=1\textwidth]{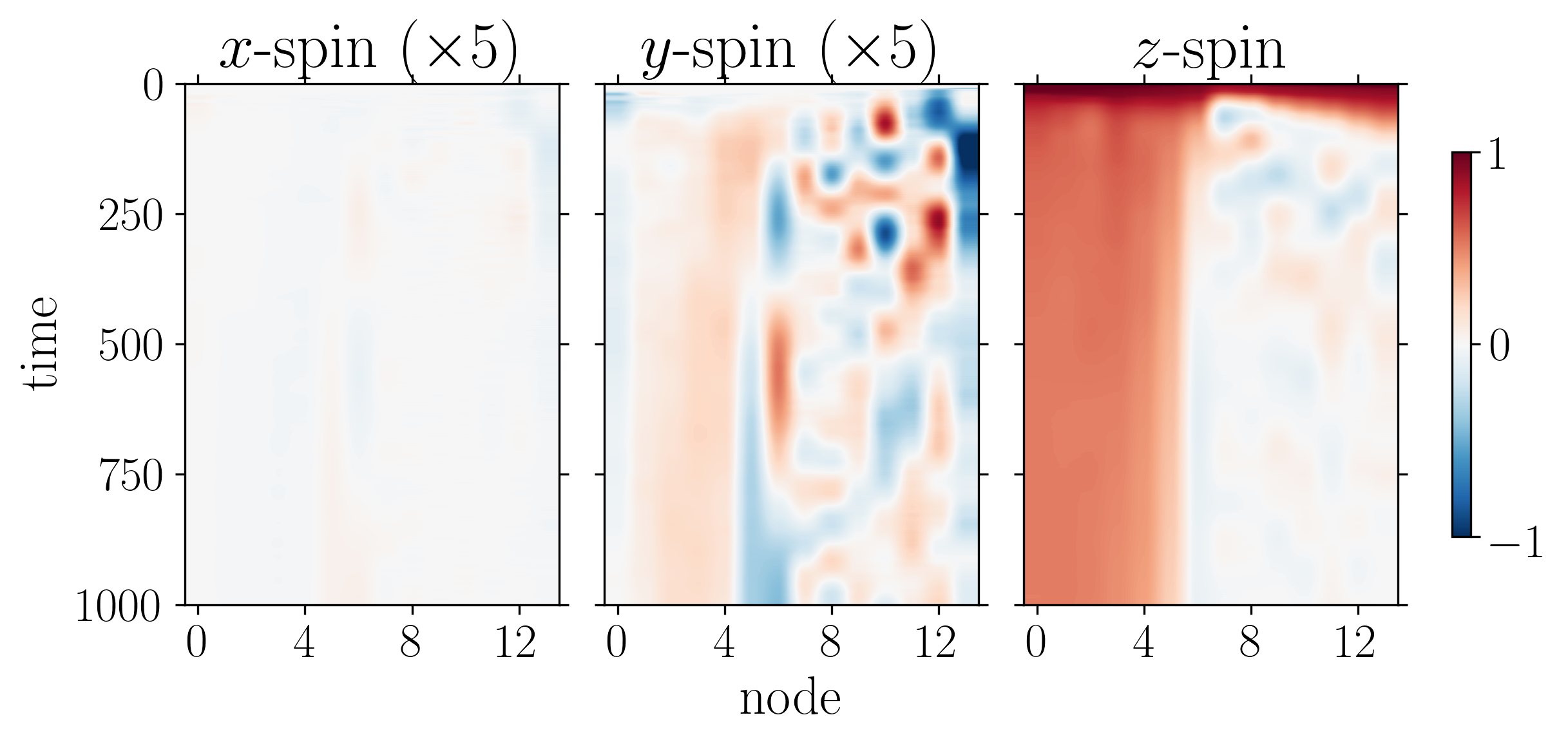}\\
  \includegraphics[width=1\textwidth]{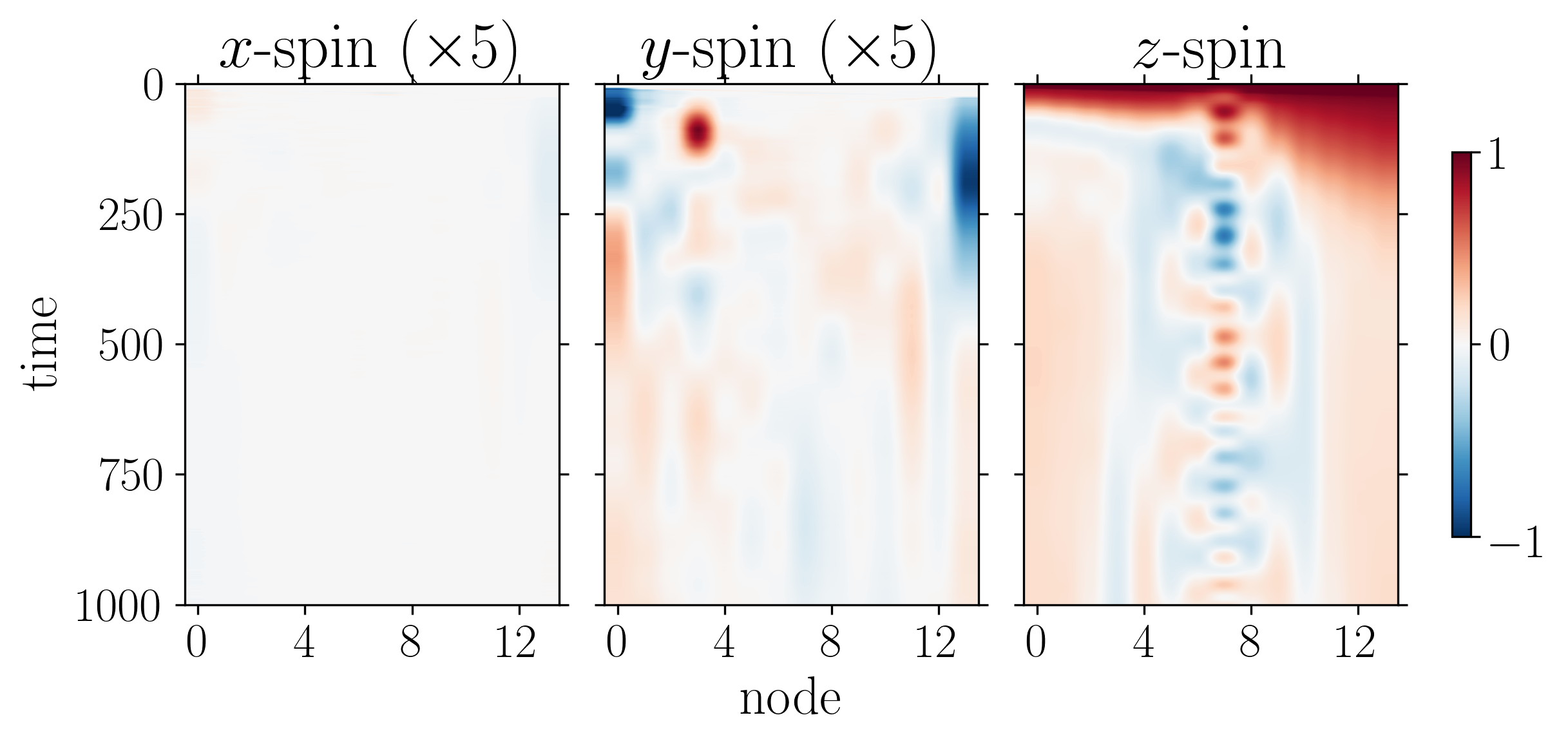}
  \caption{Topology `t'. Spin density evolution. Top, trivial interacting case; bottom, nontrivial case. Parameters as in (b) and (c) of Fig.~\protect\ref{f:tbh}.
  \label{f:tbs}}
\end{figure}

As a consequence, to investigate the topological properties of the interacting walk, we generalize the color operator \(C\) to take into account the change in the value of \(\theta\) at an interface defined at \(x = x_0\). In practice, we introduce two operators \(R(\theta)\) differing only in the value of the angle: \(\theta_-\) and \(\theta_+\) for \(x < x_0\) and \(x > x_0\), respectively. In the case of periodic boundary conditions, a second interface at \(x=0\) is also present. In Fig.~\ref{f:tp} we compare three cases distinguished by their topological properties, as defined by the associated noninteracting case. We choose for the topological interface \(\theta_-=1.1\), \(\theta_+=2.1\), which are at similar distance to the critical value \(\theta = \pi/2\), and for the trivial one \(\theta_-=0.5\), \(\theta_+=1.2\). The free case \(J=0\) with nontrivial interface, shows almost recurrences corresponding to probability concentrations alternating between the two interfaces. This behavior reflects in the temporal evolution of the probability at specific nodes, as shown in the second row of the figure, where \(p(x,t)\) is represented for a central and a bulk node, \(x=6,9\). The interacting trivial case \(J=0.2\) (c-d), shows normal scattering at the discontinuities of the coin parameter \(\theta\), but no edge states. The interacting non trivial case \(J=0.2\) (e-f), shows the convergence towards a probability at the origin larger than the equidistribution (thin dotted line); recurrences and alternation between the interfaces disappear and a probability concentration forms. The level of the probability at the interface \(x=6\), slightly larger than the mean in the present case, grows with the value of the coupling constant, which is consistent with the presence of an edge state. A more precise diagnostic is provided by the probability distribution averaged over time, which we discuss next.

In Fig.~\ref{f:tbh} we present the time averaged particle density for three cases illustrating the free nontrivial (a), interacting trivial (b) and interacting nontrivial (c) situations. In Fig.~\ref{f:tbs} the spin density comparing the trivial (top) and nontrivial (bottom) interacting cases. At variance to the previous case, the initial condition is `z' and the boundary is closed, which allow us to focus in the dynamics around the unique interface. The particle distribution shows that in the free case there is a symmetry between the two phases, symmetry which is broken by the interaction in a case having the same angle parameters (Fig.~\ref{f:tbh}c). Both (a) and (c) possess edge states, where the particle probability concentrates. In the interacting trivial interface case (b), a strong asymmetry arises between the two regions; this asymmetry is also present, albeit smoother, in the nontrivial case (c). The discontinuity found in the interacting trivial case can be related to different propagation properties at small and large angles: for \(\theta = 0.5\) the particle tends to localize, while for \(\theta = 1.2\) it moves ballistically with a small dispersion. In (c) the asymmetry between the two sides of the interface, absent in the equivalent free case, is related to the preferring direction of the particle motion (towards the left). In the free case, this tendency is compensated by the alternation between the to sides of transmission and reflection which cancel when averaged over time: with \(J >0\) this symmetry is broken.

\begin{figure}[htpb]
  \centering
  \includegraphics[width=0.5\textwidth]{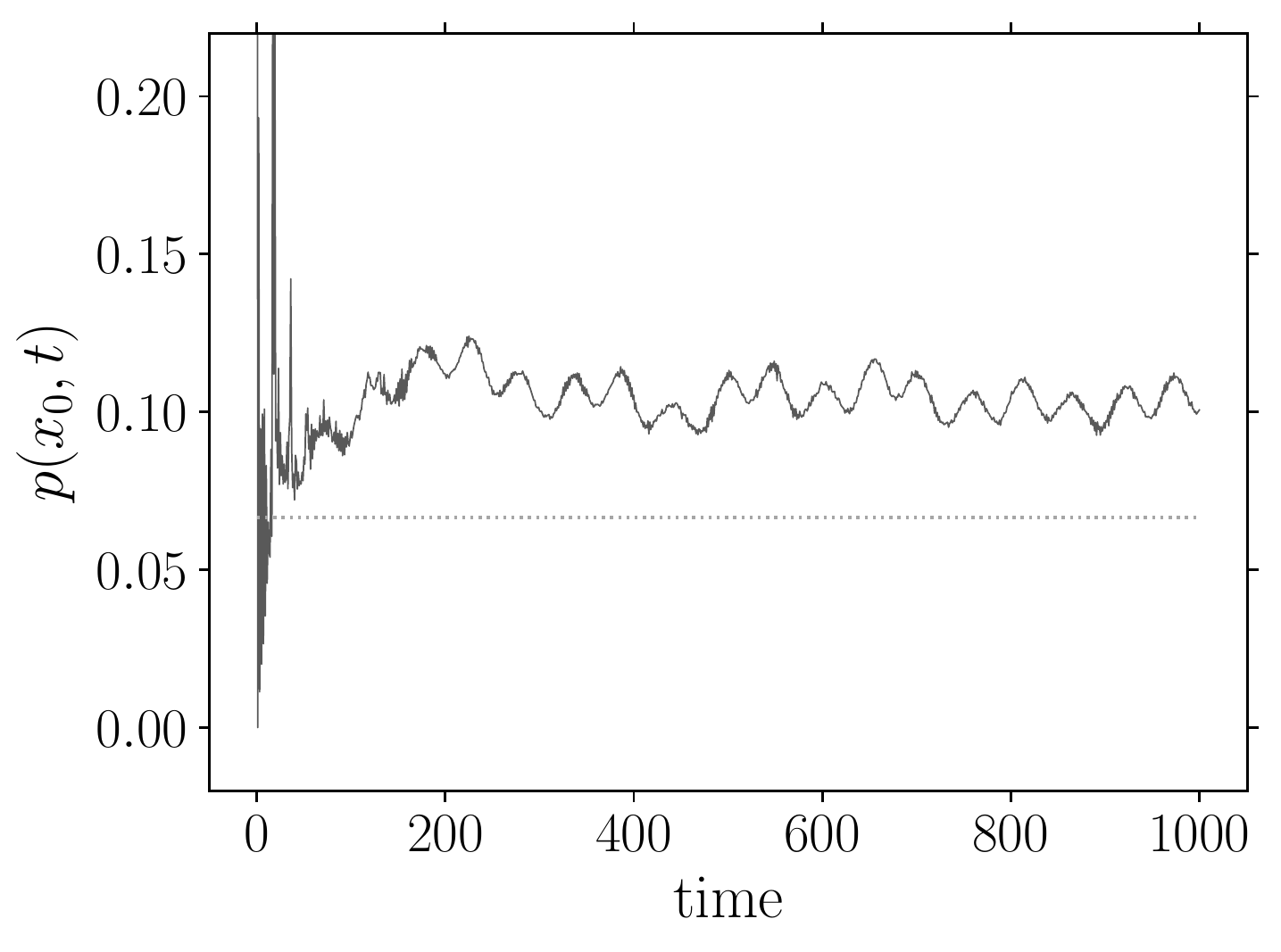}%
  \includegraphics[width=0.5\textwidth]{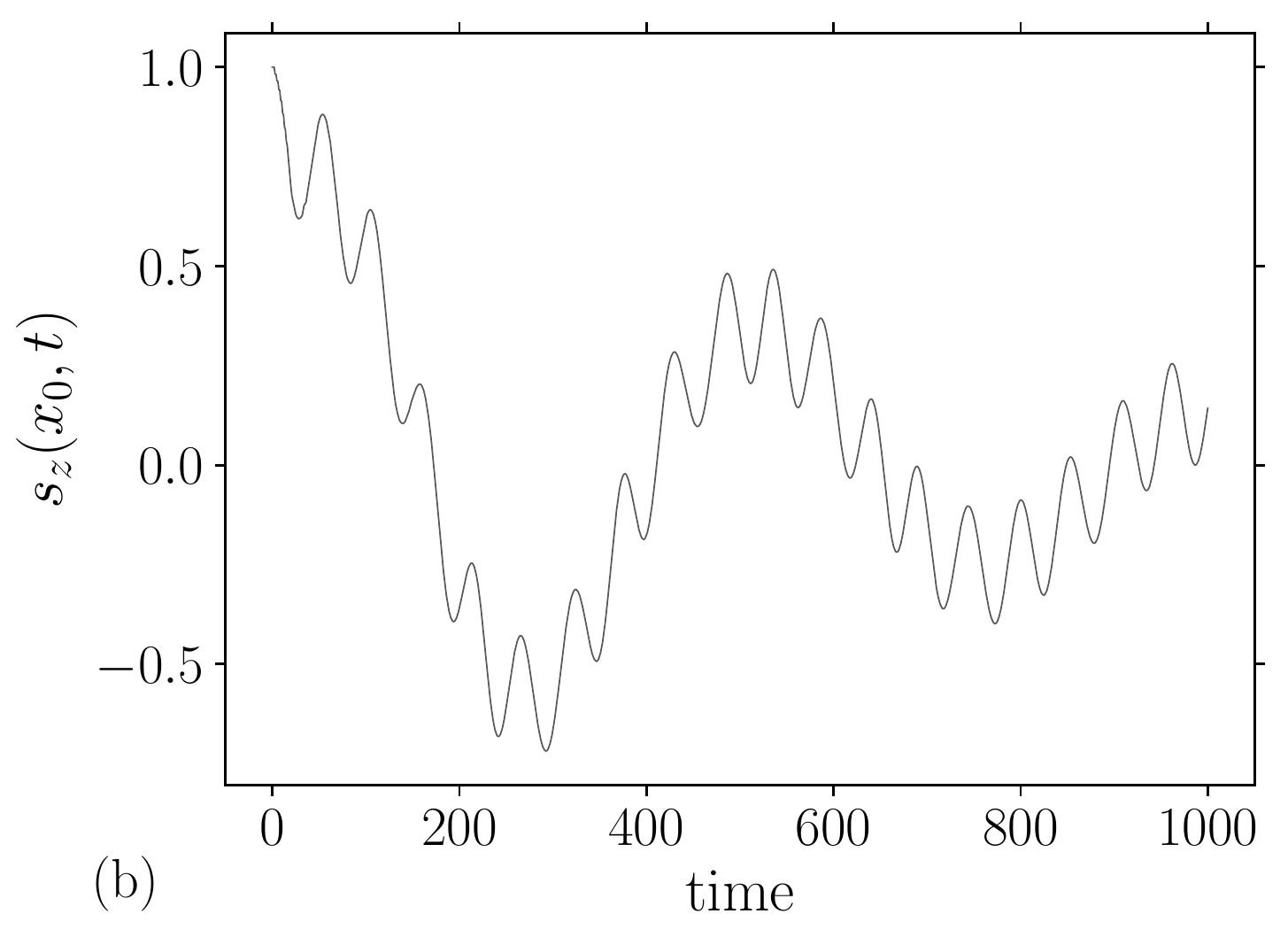}
  \caption{Topology `t'. Particle probability (a) at the origin $x_0=7$, and spin at the adjacent link (b); parameters as in the previous Fig.~\ref{f:tbs} (bottom panel). The thin line in (a) is the uniform distribution value.
  \label{f:tb0}}
\end{figure}

The role of the particle-spin interaction is apparent in the behavior of the spin density \(\bm s(x,t)\) at the interface (Fig.~\ref{f:tbs}). In the trivial case (top) a simple discontinuity in the \(z\) spin component is present at the interface, while in the nontrivial case (bottom) a rich quasiperiodic dynamics develops, within an overall symmetric distribution. The initial uniformly up polarized state relaxes towards two different values in the topologically trivial case, on both sides of the discontinuity of the rotation angle. In the localized region (on the left) the spin fades out smoothly, in contrast to the fluctuating behavior in the propagating region, in which at long times the mean spin tends to vanish. In the nontrivial case, both regions are propagating and the spin density evolves similarly, the main effect is on the interface, where a faster scale dynamics sets up (but slower than the free particle return time to the origin). The oscillations of the central spin are correlated with the oscillations of the particle density at the same position (the two nodes defining the corresponding link), as we can see in Fig.~\ref{f:tb0}, where we observe the correlation arising after a short initial transient. In fact, at the interface the particle executes a zig-zag motion between the neighbors at \(x_0\) and \(x_0+1\), the lattice link where the spin resides (Fig.~\ref{f:tb0}b).

The main point about the topological properties of the system is its ability to create at the interface between topologically different regions, edge states where the particle concentrates and where a distinct spin dynamics develops. The edge states are robust under wide changes of the phase parameters, as well as to the initial condition (whose overlap with this state should be nevertheless significant).

\begin{figure}[b]
  \centering
  \includegraphics[width=0.5\textwidth]{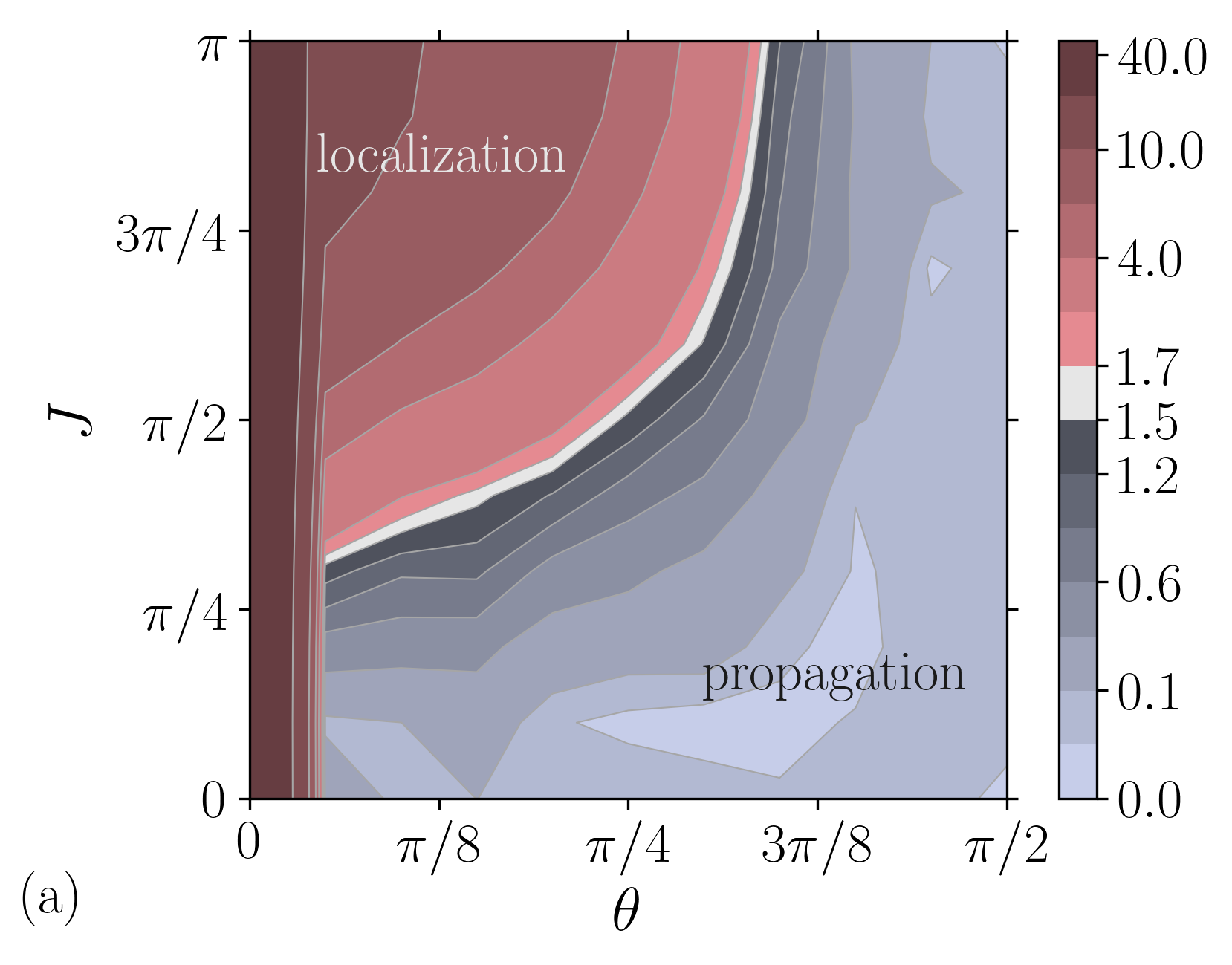}%
  \includegraphics[width=0.5\textwidth]{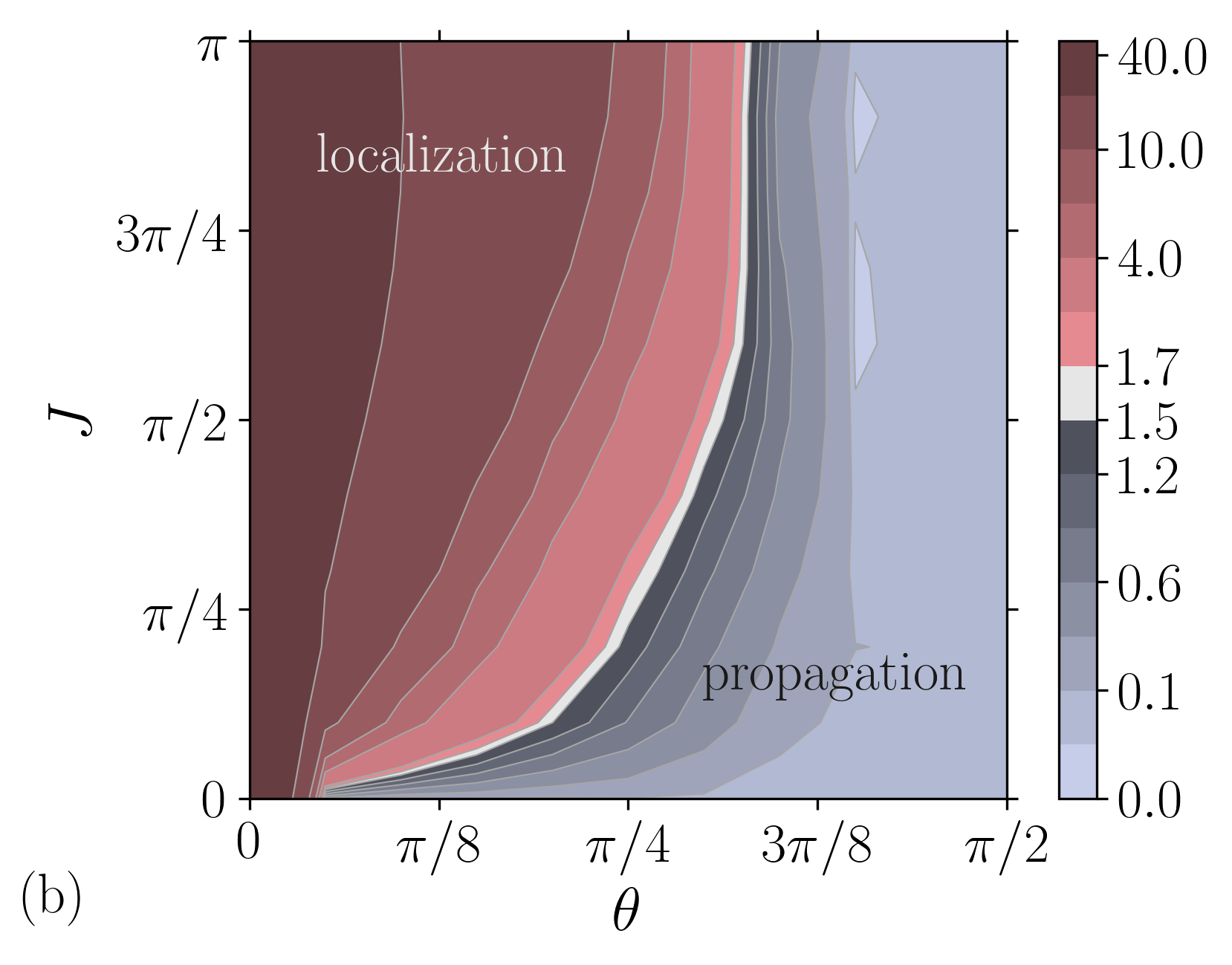}
  \caption{Phase space. (a) Initial state `x', (b) `z'. We use the Kolmogorov-Smirnov distance to the uniform position distribution to determine the localization degree; the threshold $D_\text{KS} = 1.5/100$ roughly corresponds to $D_\text{KS} = D_\text{KS}(\pi/4, \pi/2)$ for the initial condition `x'. Note that the scale of colors is not linear. (The scale of $D_\text{KS}$ is magnified by a factor 100.)
  \label{f:PSxz}}
\end{figure}

\subsection{Parameter phase space}

\begin{figure}[tb]
  \centering
  \includegraphics[width=0.5\textwidth]{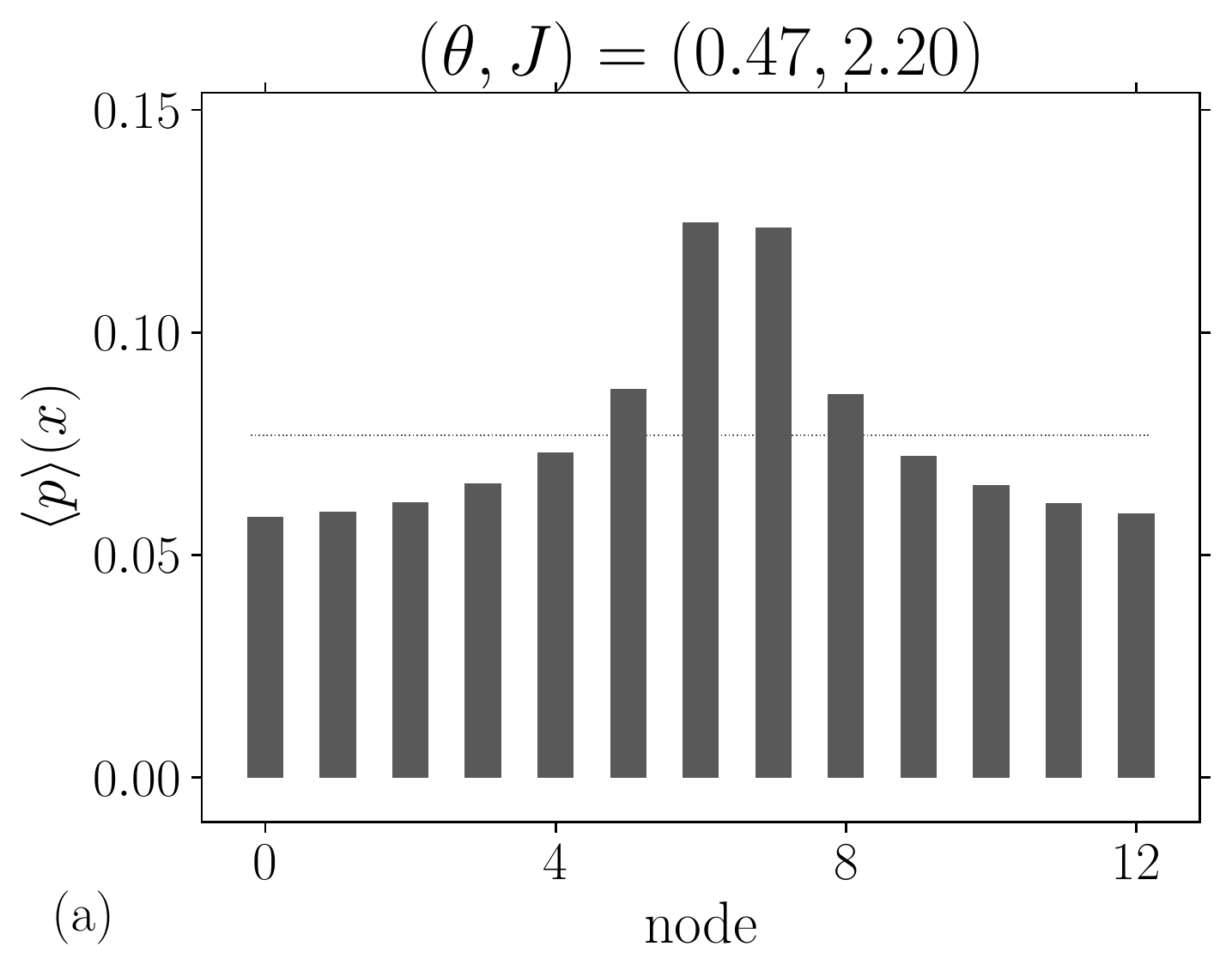}%
  \includegraphics[width=0.5\textwidth]{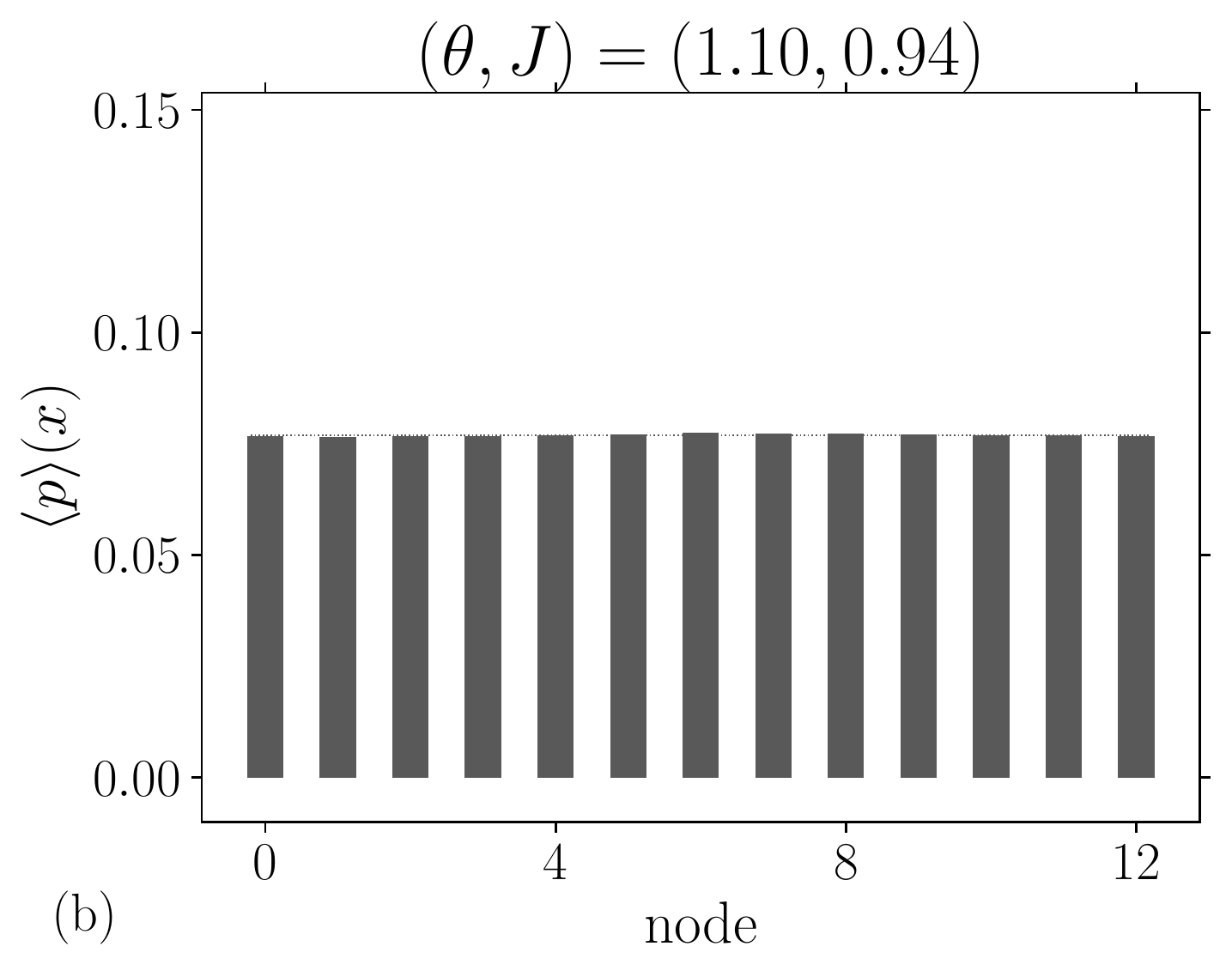}
  \caption{Position distribution averaged over time showing typical localization (a) and propagation (b) situations. The thin line gives the uniform value; `x'.
  \label{f:PSp}}
\end{figure}

\begin{figure}[tb]
  \centering
  \includegraphics[width=0.5\textwidth]{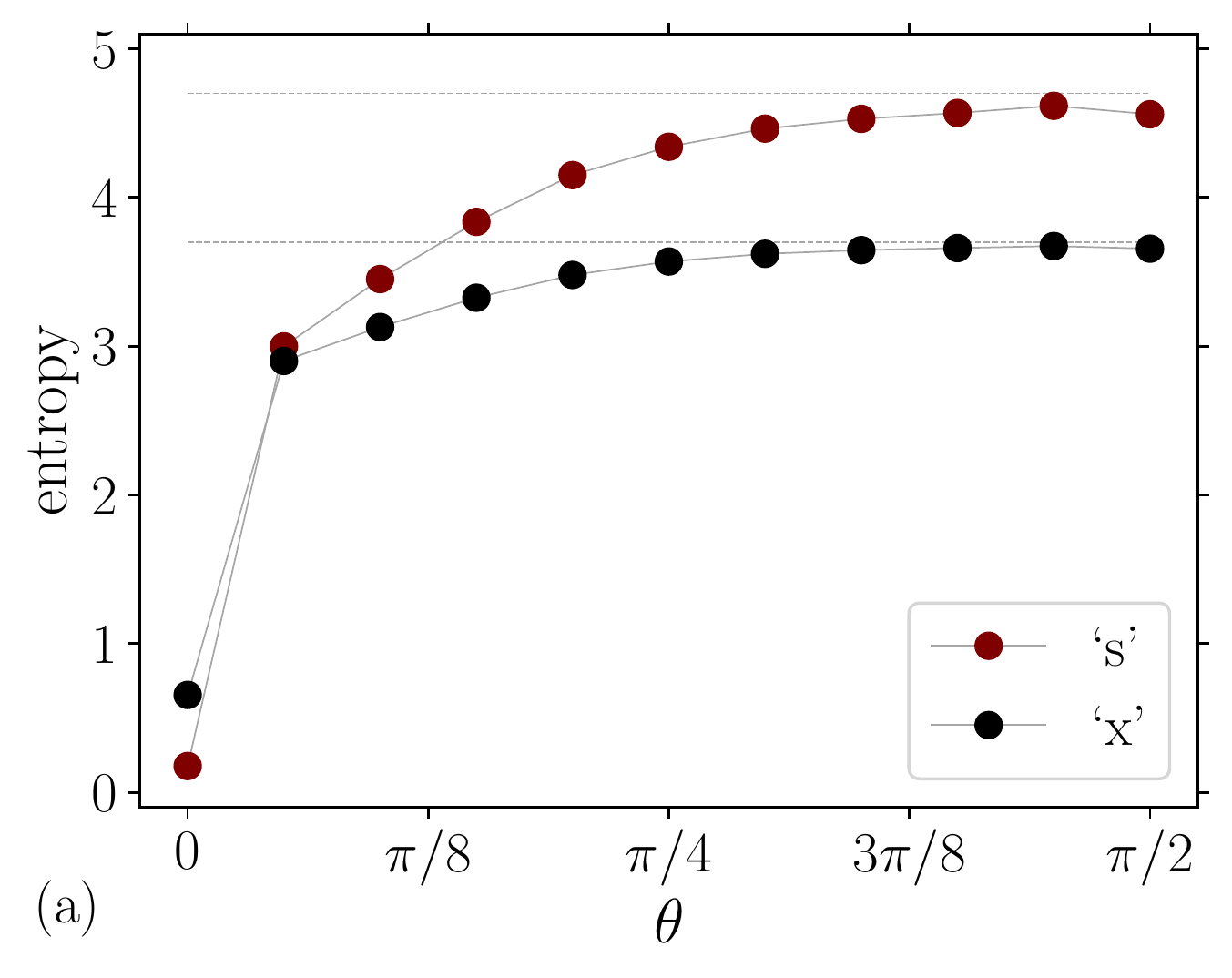}%
  \includegraphics[width=0.5\textwidth]{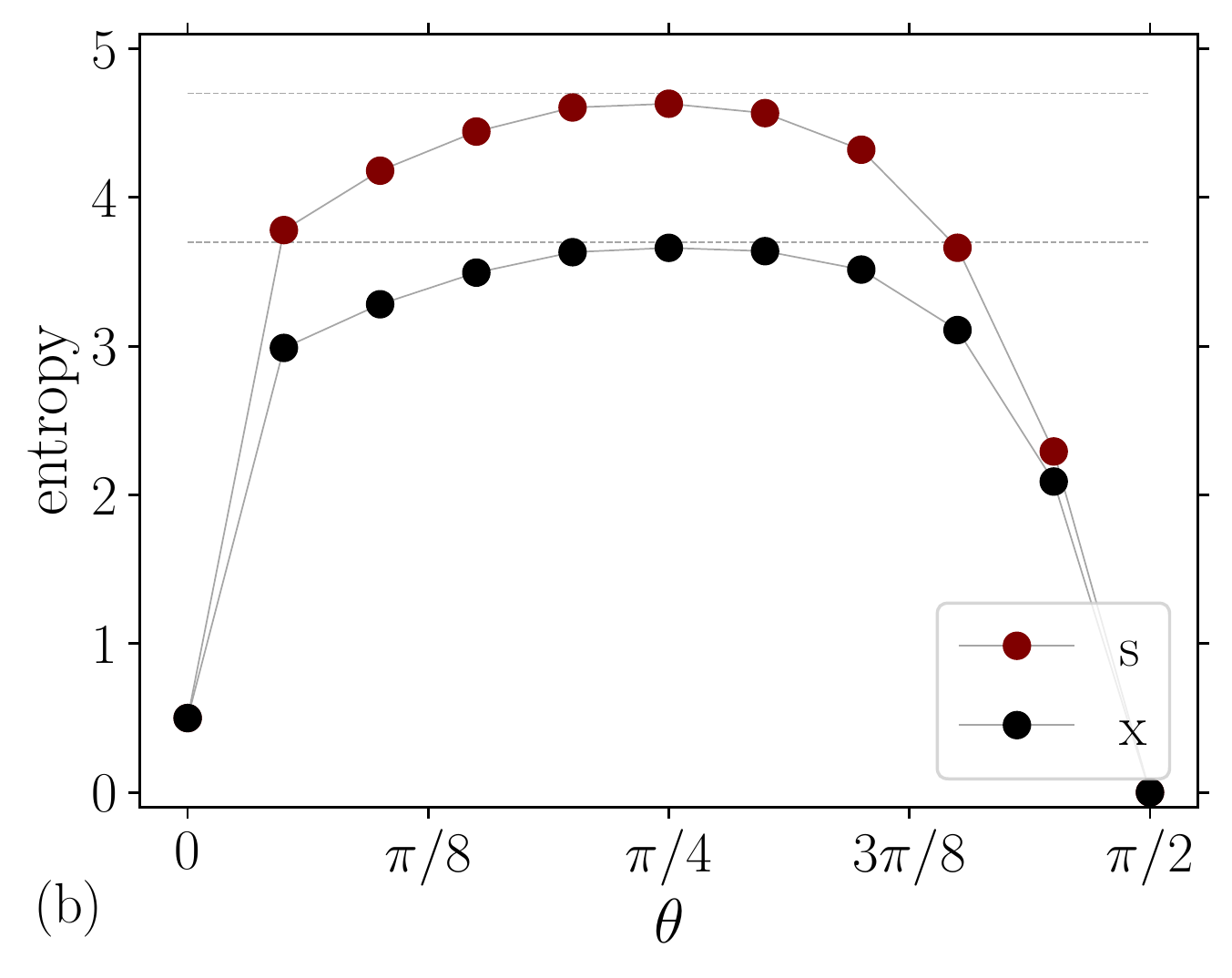}\\
  \includegraphics[width=0.5\textwidth]{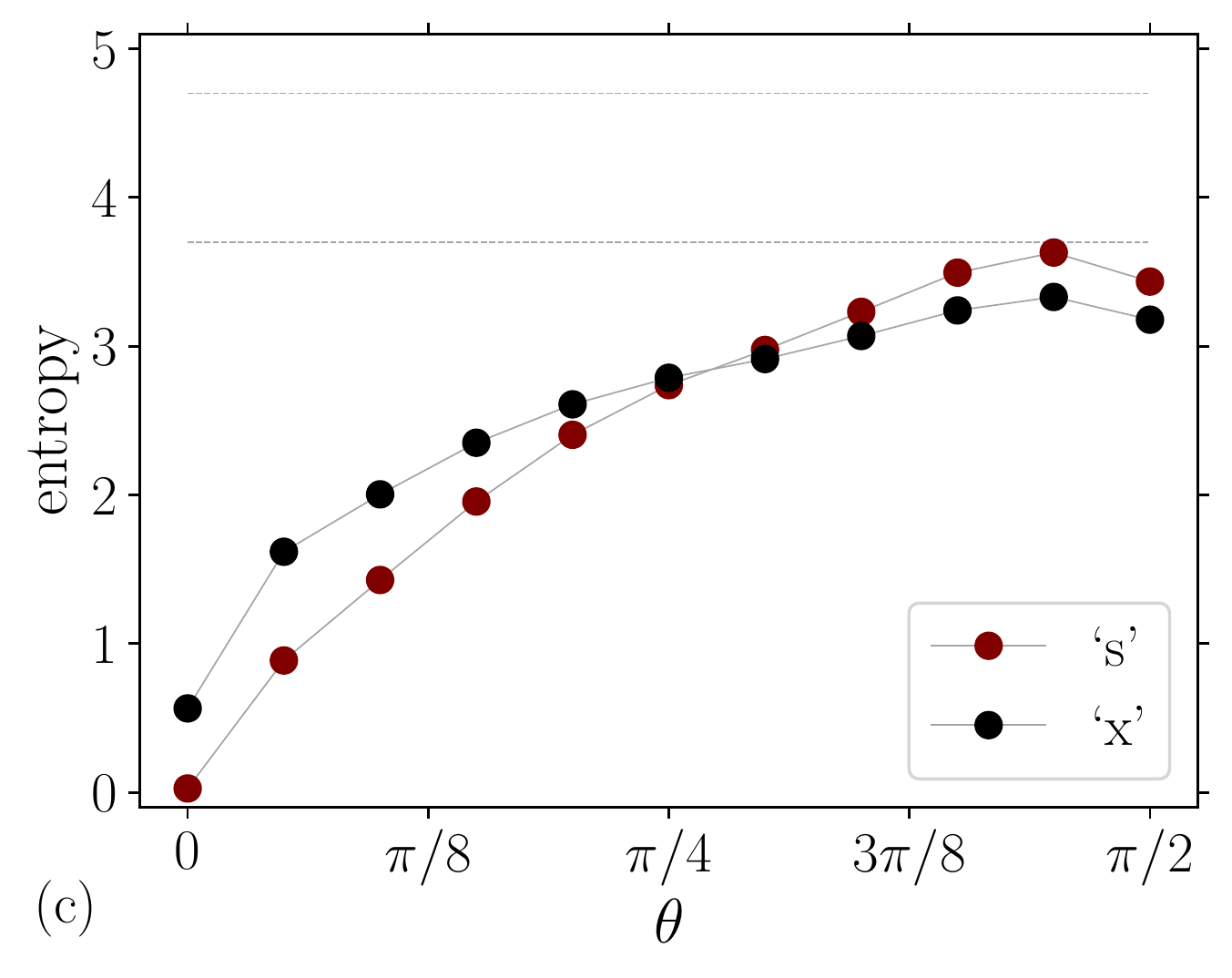}%
  \includegraphics[width=0.5\textwidth]{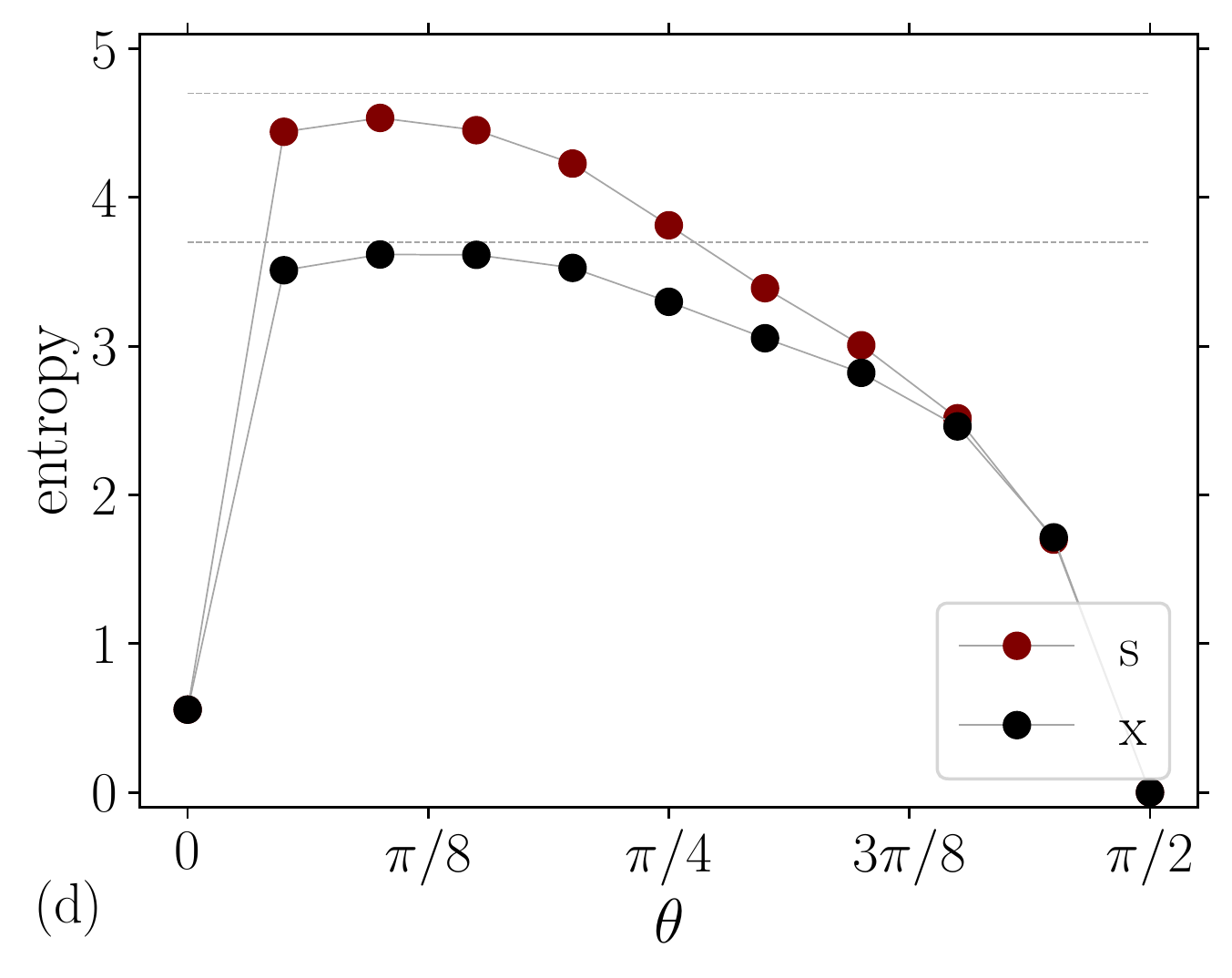}
  \caption{Entanglement as a function of the rotation angle $\theta$ for two initial conditions, `x' in the first column (a,c), and `z' in the second (b,d), and two values of the coupling $J = \pi/2, \pi/10$, first row (a,b), and second row (c,d), respectively.
  \label{f:VNxz}}
\end{figure}

\begin{figure}[tb]
  \centering
  \includegraphics[width=0.5\textwidth]{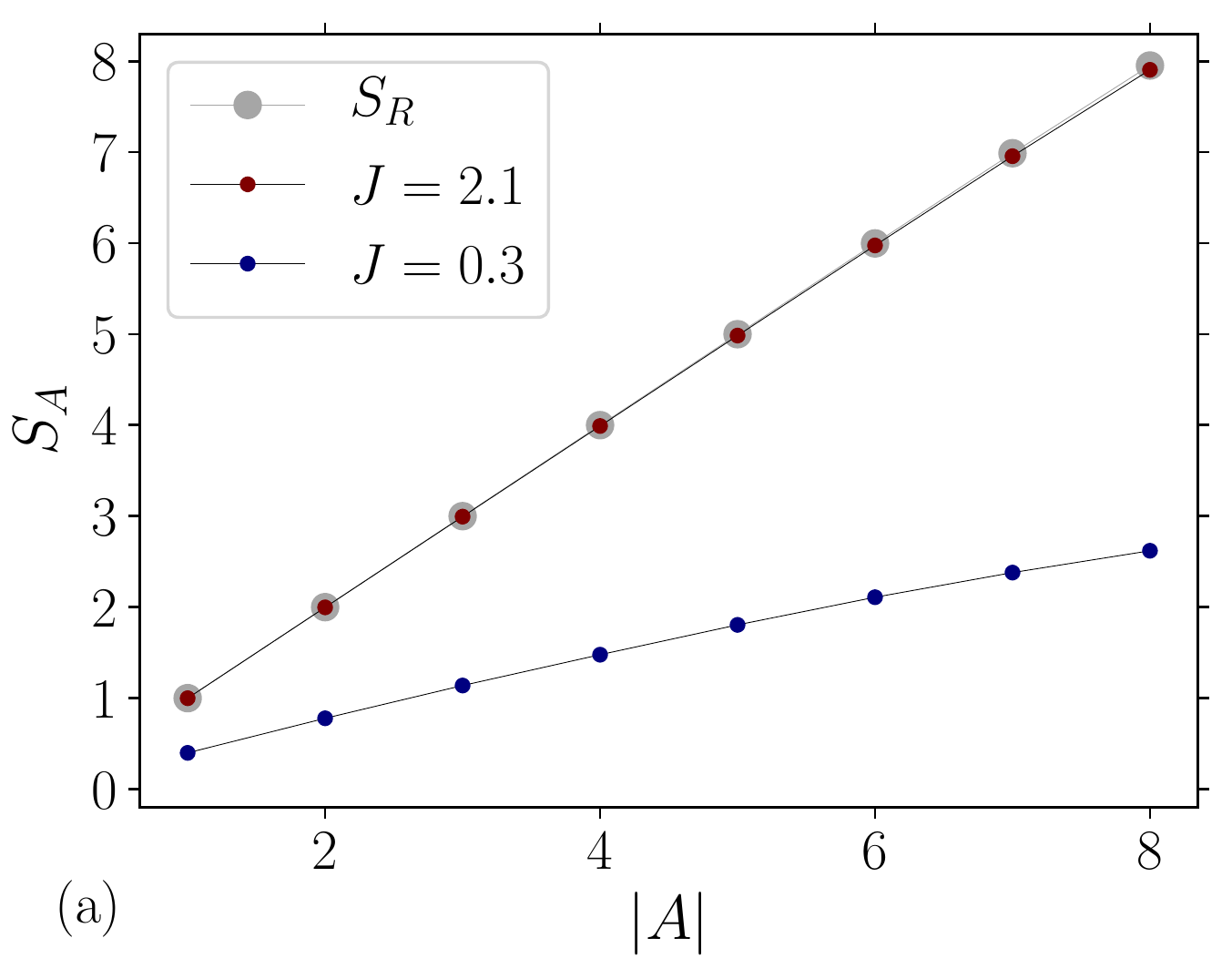}%
  \includegraphics[width=0.5\textwidth]{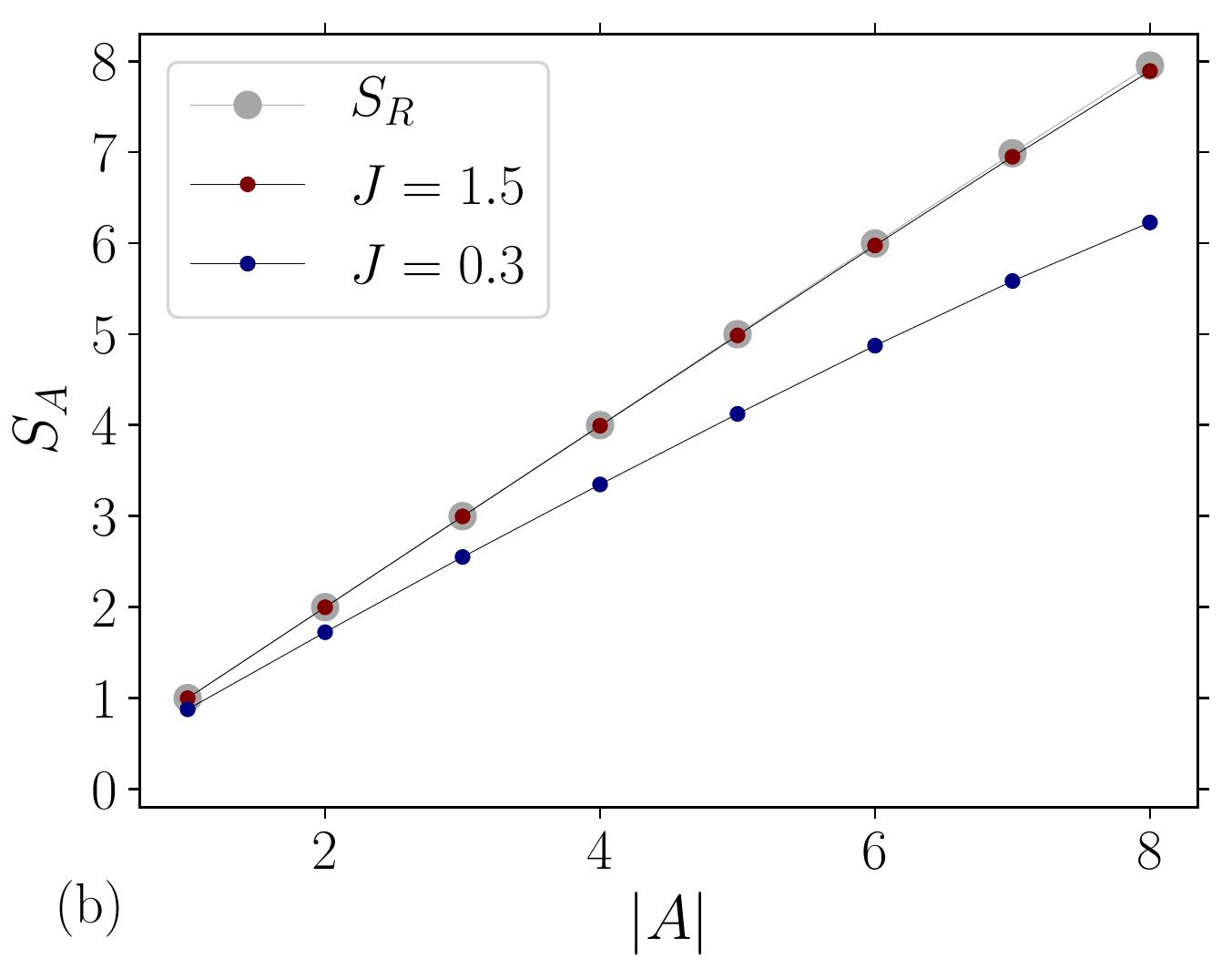}
  \caption{Entanglement entropy $S_A$ of a set $A$ of spins as a function of its size. $S_A$ is compared to the entropy of a pure random state $S_R$. Parameters: (a) $(0.9\, \pi/2, 2\pi/3)$ (red), and $(\pi/4,\pi/10)$ (blue), `x'; (b) $(\pi/4, \pi/2)$ (red), and \((\pi/4,\pi/10)\) (blue), `z'. The system's size is $|V| = 15$. For strong coupling the spin entanglement reaches the random state value $S_R$ (gray).
  \label{f:SA}}
\end{figure}

In order to organize the information about the different physical regimes we described, it is desirable to define appropriated diagnostics amenable at distinguishing between motion regimes and entanglement characteristics. We focus first on one basic distinction, which determines many other properties of the system, the ability of the walker to propagate or to be localized. To this goal, we use as localization test the spatial distribution of the particle at long times: a uniform distribution means a propagation dominated regime, while a picked distribution is an indicator of localization. To measure the difference between the actual distribution and the uniform one we use the Kolmogorov-Smirnov distance. It is defined by the equation:
\begin{equation}
\label{e:DKS}
D_\text{KS} = \max_x \left| F_x[\braket{p}] - u_x \right|\,,
\end{equation}
where \(F\) is the actual distribution (in the probability language: a monotone increasing function) associated to the time averaged position probability density \(p(x,t)\) and \(u\) the uniform distribution over the line (a simple linear function).

The parameter phase space \((\theta,J)\) is represented in Fig.~\ref{f:PSxz} for two initial states (`x' and `z'). Two examples of the particle density are given in Fig.~\ref{f:PSp}, where we find typical localized and uniform densities. The transition between localization and propagation regimes (the white level) is rather smooth, at least in this small size system \(|V| = 13\). However, within the localized region of the phase space, the position density increases rapidly with the interaction, given large values of \(D_\text{KS}\), while in the propagation part of the phase space the particle density is much more homogeneous (note the change of color scale between the two phases). Localized states occupy a larger part of the parameter space in the case of the `z' initial condition than in the case of the `x' initial condition, for which the propagation region is more extended. The interaction \eqref{e:VxyJ} acts very differently on up and plus states; for instance, when applied to a spin up in a uniform position state, it gives a state that only change the \(z\) spin component, at variance, when applied to a \(\ket{+}\) state all spin components are affected. We discuss in the next section the origin of this dynamical anisotropy. In a spatially uniform state the homogeneous spin up state, but not the plus state, is an eigenvector of \(U\). The restricted spin dynamics in the `z' case facilitates then localization of the walk.

To complement this view of the parameter phase space based on the particle motion, it is interesting to explore the entanglement as a function of \((\theta,J)\). This is given in Fig.~\ref{f:VNxz}, where the von Neumann entropy of the \((xcs)\) degrees of freedom is depicted. The left column corresponds to `x', and the right one to `z'. The first row corresponds to \(J=\pi/2\), and the second one to \(J=\pi/10\). We observe that, for strong coupling, there is a tendency of a weaker entanglement between the particle and spin degrees of freedom in the localized region for the `x' initial state, and a stronger entanglement in the propagation region; for `z', the entanglement is maximum at the transition between the two regimes. In the case of a weaker coupling, entanglement is globally smaller, with comparable values of the position and spin entropies. The small angle part of the `z' \(J=\pi/10\) is biased by finite size effects, even if the decreasing entanglement with the angle is qualitatively correct (c.f. Fig.~\ref{f:zp}). The vanishing of the entanglement for \(\theta = \pi/2\), `z', is simply a consequence that this initial state is a proper state of \(U\). By extension, the dynamics in the propagation region of `z', is regular showing smooth slightly damped oscillation for \(J \lesssim 1\), and relaxation for larger \(J\).

In the region of strong entanglement, we verified that the spin entropy of a set of spins is proportional to its size. We show \(S_A\) as a function of \(|A|\) in Fig.~\ref{f:SA} (red dots), where we compare it to the Don Page entropy \cite{Page-1993nr} of a pure random state (gray dots),
\begin{equation}
\label{e:SR}
S_R = \log(D_A) - \frac{D_A^2}{2D_V \ln2}, \quad D_A = 2^{|A|},\, D_V = 2 |V| 2^{|V|}
\end{equation}
which, for our range of parameters is close to \(S_R \approx |A|\). We find for \(J>1\) that the state of maximum spin entanglement is an extensive variable, well described by a random pure global state. However, random maximally entangled states disappear for weaker couplings, leading instead to smaller entanglement, which follows a sublinear increase law with the subsystem size, as can be appreciated in Fig.~\ref{f:SA}, where we also show \(S_A\) for \(J=0.3\) (blue dots, the system size is \(|V|=15\)).

Before closing this section on the phenomenology, let us mention that the initial spin entangled state `e' imposes a constraint on the absolute value of the mean spin \(\bar{s}\), which is zero at \(t=0\), very different from the other initial states (\(\bar{s}(0) =1\)). One observes the emergence of persistent irregular oscillations, whose period is also roughly given by \(\sim |V|/J\) and an amplitude generally smaller than the mean particle density, with a complex spatial distribution. At strong coupling in the propagation region, the spin polarization sticks to the particle propagation. At variance to the initial product states, no relaxation towards a stationary state is found.

\section{Discussion}

\subsection{Free walk}

We consider first the free walk \(J=0\). With periodic boundary conditions, the evolution operator reduces to a translation invariant walk operator \(U(\theta, 0 ) = W(\theta) = M R(\theta)\). Therefore, the evolution operator \(W\) can be diagonalized in Fourier space. We write, using \eqref{e:M1d}, the motion operator,
\[M = \sum_k \ket{k}\bra{k} \otimes M(k) \,,\]
where \(k \in (-\pi,\pi)\) is the quasimomentum, and
\begin{equation}
  \label{e:Mk}
  M(k) = \E^{-\I k X} = \begin{pmatrix} 0 & \E^{-\I k} \\ \E^{\I k} & 0 \end{pmatrix} \,.
\end{equation}
Then, after multiplying by \(R(\theta)\), the walk operator (in Fourier space) is given by the matrix,
\begin{equation}
  \label{e:Wk}
  W(k,\theta) = \begin{pmatrix}
    \E^{-\I k} \sin\theta & \E^{-\I k} \cos\theta \\
    \E^{ \I k} \cos\theta & -\E^{\I k} \sin\theta
  \end{pmatrix} = \E^{-\I H_0(k,\theta) + \I \pi/2}\,,
\end{equation}
where we introduced the free effective Hamiltonian,
\begin{equation}
  \label{e:H0k}
  H_0(k, \theta) = E(k, \theta) \, \hat{\bm d}(k, \theta) \cdot \bm \tau \,,
\end{equation}
with
\begin{equation}
  \label{e:En}
  \cos E(k,\theta) = -\sin k \sin \theta\,, \quad \hat{\bm d}(k, \theta) = \frac{1}{\sin E(k, \theta)}
  \begin{pmatrix}\cos k \cos \theta \\
    \sin k \cos \theta \\
    \cos k \sin \theta
  \end{pmatrix} \,,
\end{equation}
where \(E(k,\theta)\) is the quasienergy spectrum, and \(\hat{\bm d}\) a momentum like unit vector (in analogy with a spin-orbit coupling). The free spectrum displays two bands separated by a gap which is maximum for \(\theta = 0, \pi\), and closes for \(\theta = \pi/2\). At \(\theta = \pi/2\) the spectrum is similar to the massless Dirac cone.

\begin{figure}[tb]
  \centering
  \includegraphics[width=0.5\textwidth]{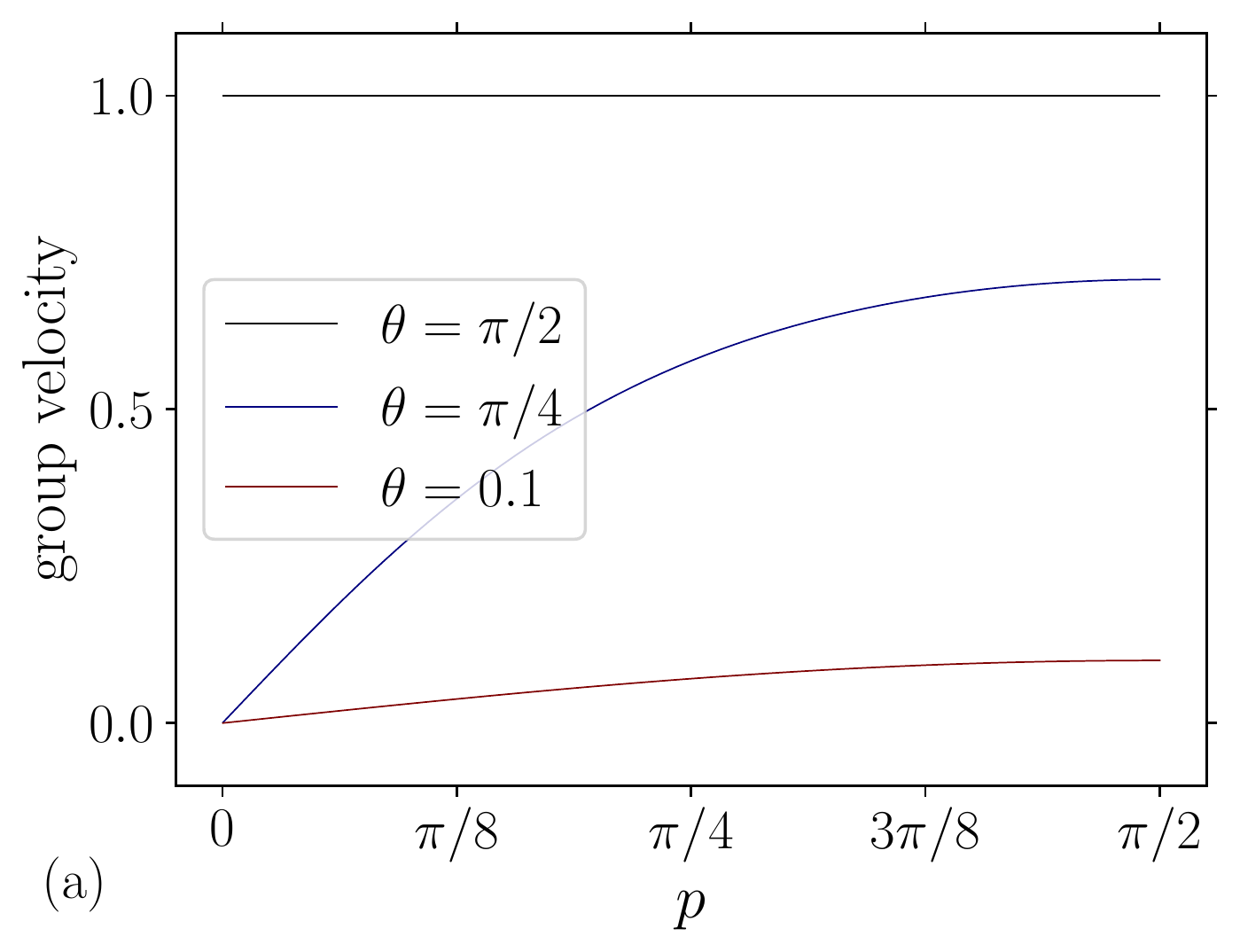}%
  \includegraphics[width=0.5\textwidth]{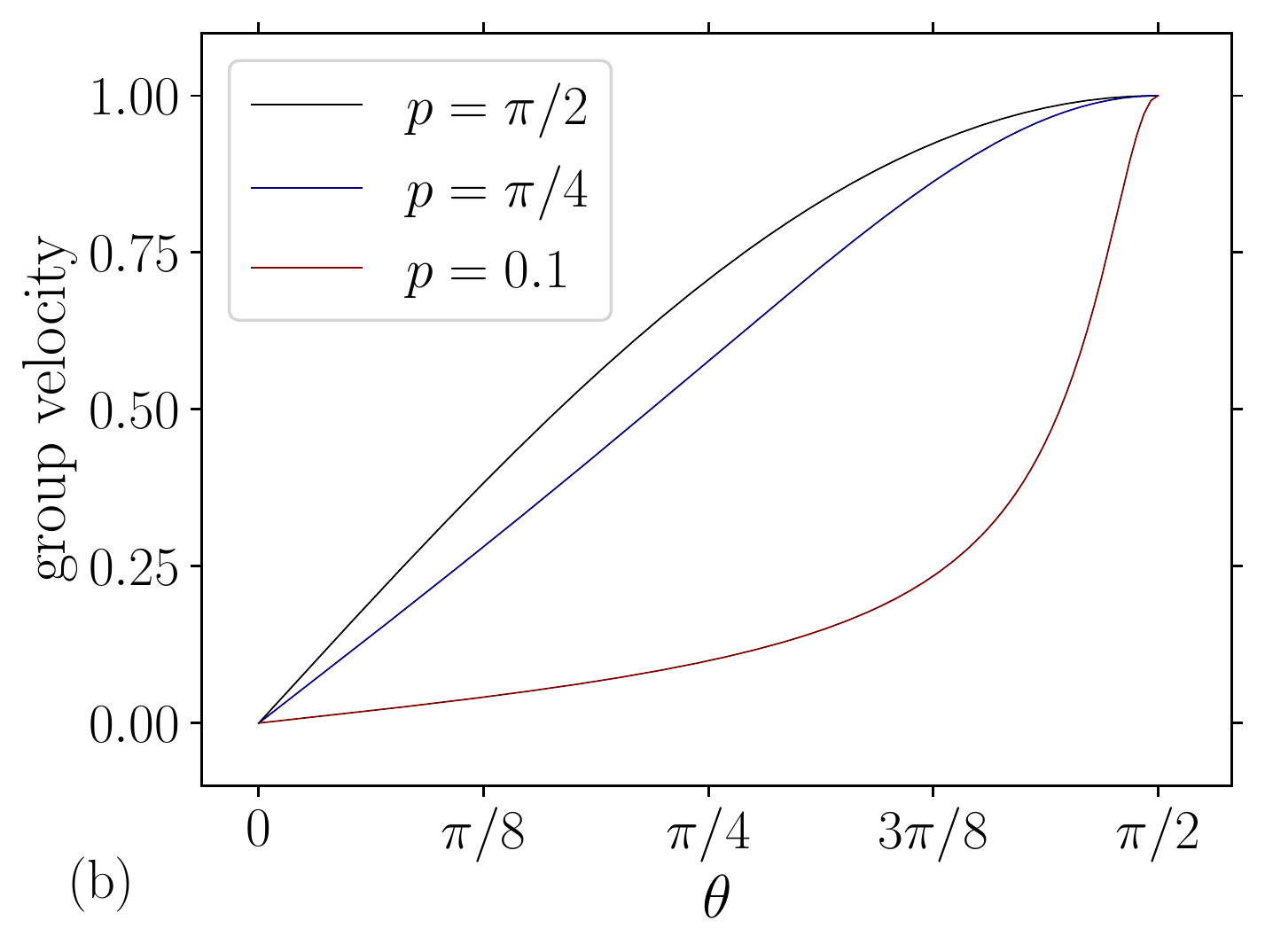}
  \caption{Group velocity as a function of $p=k-\pi/2$ (a) and $\theta$ (b).
  \label{f:vg}}
\end{figure}

In Fig.~\ref{f:vg} we represent the group velocity as a function of the shifted wavenumber \(p = k - \pi/2\) and the angle \(\theta\),
\begin{equation}
\label{e:vgE}
v_g(p,\theta) =  \frac{\sin p \sin \theta}{\sqrt{1 - \cos^2p \sin^2 \theta}}\,.
\end{equation}
The group velocity vanishes for \(\theta = 0\) and tends to a constant \(v_g=1\) for \(\theta = \pi/2\); between these two extreme values the wave propagation is dispersive; as a function of \(\theta\), it is an increasing function. These two regimes qualitatively follow the behavior observed in the interacting case for small \(J\), as shown in Figs.~\ref{f:o54p}, in which we found recurrent oscillations in the case of a slowly moving walker and damped oscillations with high entanglement in the case of a propagating walker. This picture is also consistent with the observation of localized states for angles smaller than \(\theta \sim \pi/4\). Nevertheless, in order to understand the rich phenomenology of the system we must take into account the interaction. 

\subsection{Homogeneous system}

We consider first the simplest situation in which the particle and spin distributions are initially homogeneous, implying that they remain homogeneous because of translation invariance. Under homogeneous conditions, the spatial dynamics becomes trivial and the corresponding motion operator essentially decouples from the other degrees of freedom. The consequence is that we can reduce \(U\) to the color-spin coupling and characterize the particle by its momentum \(p\). Moreover, for large times and most parameters (in the homogeneous case), we observe a smooth temporal evolution of the spin density \(\bm s(t)\), which is only function of the time. Therefore, it is natural to assume a mean field separation of scales, and split the evolution operator:
\begin{equation}
  \label{e:VW}
  U_t(\Delta t) = V_t(J) W(p, \theta) = \E^{-\I H_J(t) \Delta t} \E^{-\I H_0 \Delta t} \approx \E^{-\I H(t) \Delta t} \,,
\end{equation}
where \(W = M(p)R(\theta)\),
\begin{equation}
\label{e:homH}
H = H_0(p, \theta) + H_J(t), \quad H_J(t) = - \frac{J}{4} \bm s(t) \cdot \bm \tau
\end{equation}
and \(H_0\) is given by \eqref{e:H0k}. Note that \(U\) is here the one time step operator, acting at time \(t\) in which we replaced the operator \(H_J\) of \eqref{e:VJ} by a particle operator, reducing the problem to a selfconsistent determination of the mean field \(\bm s(t)\): we do not assume the splitting \eqref{e:VW} holds for all times, but only over one step on which we must \emph{update} the mean field \(\bm s(t)\). Actually, in the continuous time limit the Heisenberg equation for the time evolution of the expected value \(\bm s(t) = \braket{\bm \sigma^{\otimes |V|}} \equiv \braket{\bm \sigma}\) is,
\begin{equation}
\label{e:sigmat}
\frac{\D}{\D t} \braket{\bm \sigma} = \frac{J}{2} \braket{\bm \sigma \times \bm \tau}\,;
\end{equation}
the approximation consists in replacing \(\bm \sigma\) by the mean field, on the right hand side:
\begin{equation}
\label{e:LLmf}
\frac{\D}{\D t} \bm s(t) = \frac{J}{2} \bm s(t) \times \braket{\bm \tau}\,,
\end{equation}
where \(\braket{\bm \tau}\) is computed from an appropriated evaluation of the quantum state evolved with \eqref{e:VW}. Equation \eqref{e:LLmf} is a simple form of the Landau-Lifshitz equation, in which the magnetization precession is driven by the exchange of torque with the walker \cite{Ralph-2008ly,Sayad-2016a}:
\begin{equation}
\label{e:LLeq}
\dot{\bm s} = \frac{J}{2} \bm s \times \braket{\bm \tau}, \quad \braket{\bm \tau}(t) = \braket{\psi(t)|\bm \tau|\psi(t)}
\end{equation}
where \(\ket{\psi(t)}\) is the system's state at time \(t\), which must be deduced from the Schrödinger equation with Hamiltonian \eqref{e:homH}, and \(\braket{\bm \tau}\) is the effective particle `magnetic momentum'. In brief, for homogeneous systems, the large time scale spin dynamics is supposed to be driven by the itinerant particle spin and satisfies the Landau-Lifshitz equation; the interaction is selfconsistently computed solving the corresponding Schrödinger equation at each time step.

We start with the lowest order solution (in powers of \(J\)), in which we can solve the free particle dynamics, and apply the corresponding state to the computation of the effective torque. We determine first the eigenstates. To diagonalize the Dirac walk Hamiltonian,
\[
  H_0 = \bm d(p,\theta) \cdot \bm \tau\,,
\]
it is convenient to parametrize the vector \(\bm d = E \hat{\bm d}\) \eqref{e:En} that actually we might consider be arbitrary, by the spherical angles \((\alpha, \beta)\):
\[
  \bm d = |\bm d| (\sin\alpha \cos\beta, \sin \alpha \sin \beta, \cos \alpha)\,,
\]
functions of \((p,\theta)\); with this definition the wave function of a free particle (in the positive energy band of \(H_0\)) is,
\begin{equation}
\label{e:psi}
\psi_\PL(x,t) = \frac{1}{\sqrt{|V|}}\chi_\PL(p, \theta) \E^{-\I E_+t + \I p x}\,,
\end{equation}
where
\begin{equation}
\label{e:Epm}
  E_\pm = \pm |\bm d|
\end{equation}
are the positive and negative energy bands (c.f. \eqref{e:En}), the spinor of positive energy is,
\[
  \chi_\PL = \begin{pmatrix} \cos \frac{\alpha}{2} \\ \sin \frac{\alpha}{2} \, \E^{\I \beta} \end{pmatrix}\,,
\]
and similar expressions for the negative energy wave function.

We can now apply the solution \eqref{e:psi} of the free Dirac walk to the computation of \(\braket{\bm \tau}\). We readily obtain:
\[
  \braket{\bm \tau} = \psi_\PL^\dagger \bm \tau \psi_\PL = \frac{1}{|V|} \frac{\bm d}{|\bm d|}\,,
\]
which leads to the equation
\begin{equation}
\label{e:LLhom}
\dot{\bm s} = - \frac{J}{2|V|}\bm s \times \frac{\bm d}{|\bm d|}\,.
\end{equation}
(the minus sign corresponds to our choice of the rotation direction, see \eqref{e:Epm}). This is an interesting result, it predicts that in the homogeneous case the fixed spins will oscillate with a period of \(T = 4\pi |V|/J\), which we observed in \eqref{e:oscT}. However, this result supposes decoupling between the particle and spins, as implied by the neglect of correlations in \eqref{e:LLeq} and the free particle state approximation. In particular, it cannot account for the relaxation of the spin, note that the Landau-Lifshitz equation preserves the norm of \(\bm s(x,t)\).

\begin{figure}[tb]
  \centering
  \includegraphics[width=0.5\textwidth]{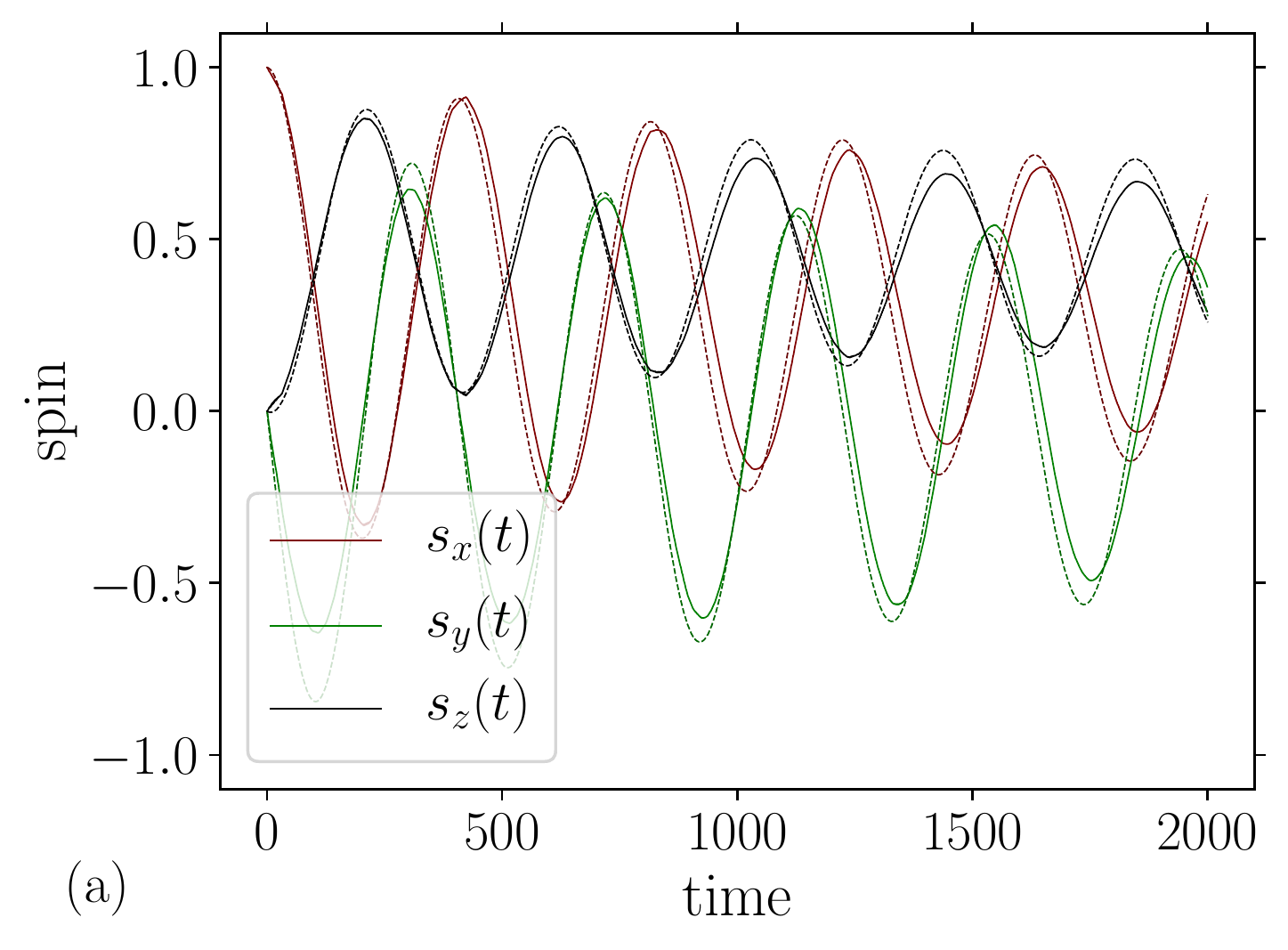}%
  \includegraphics[width=0.5\textwidth]{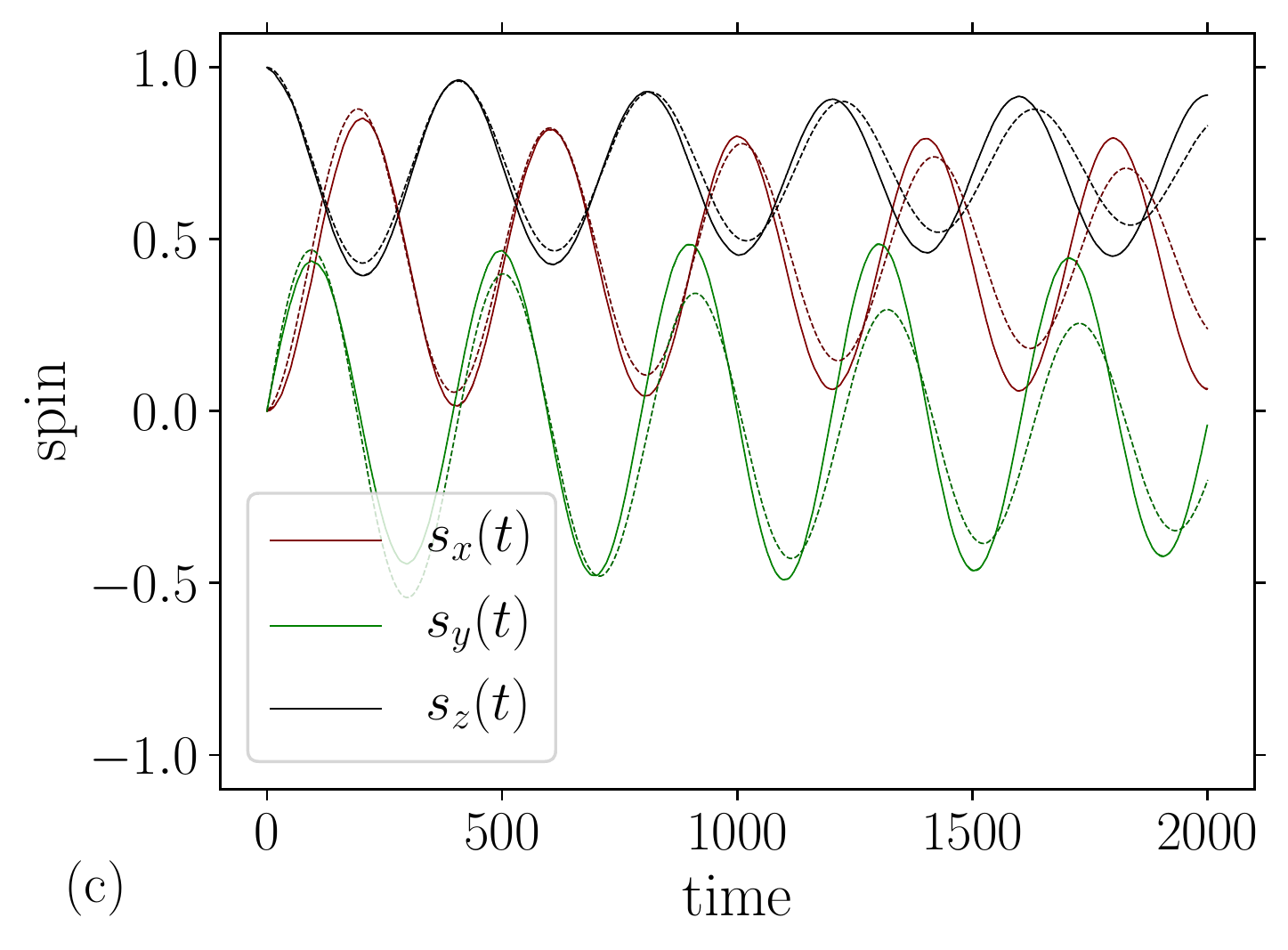}\\
  \includegraphics[width=0.5\textwidth]{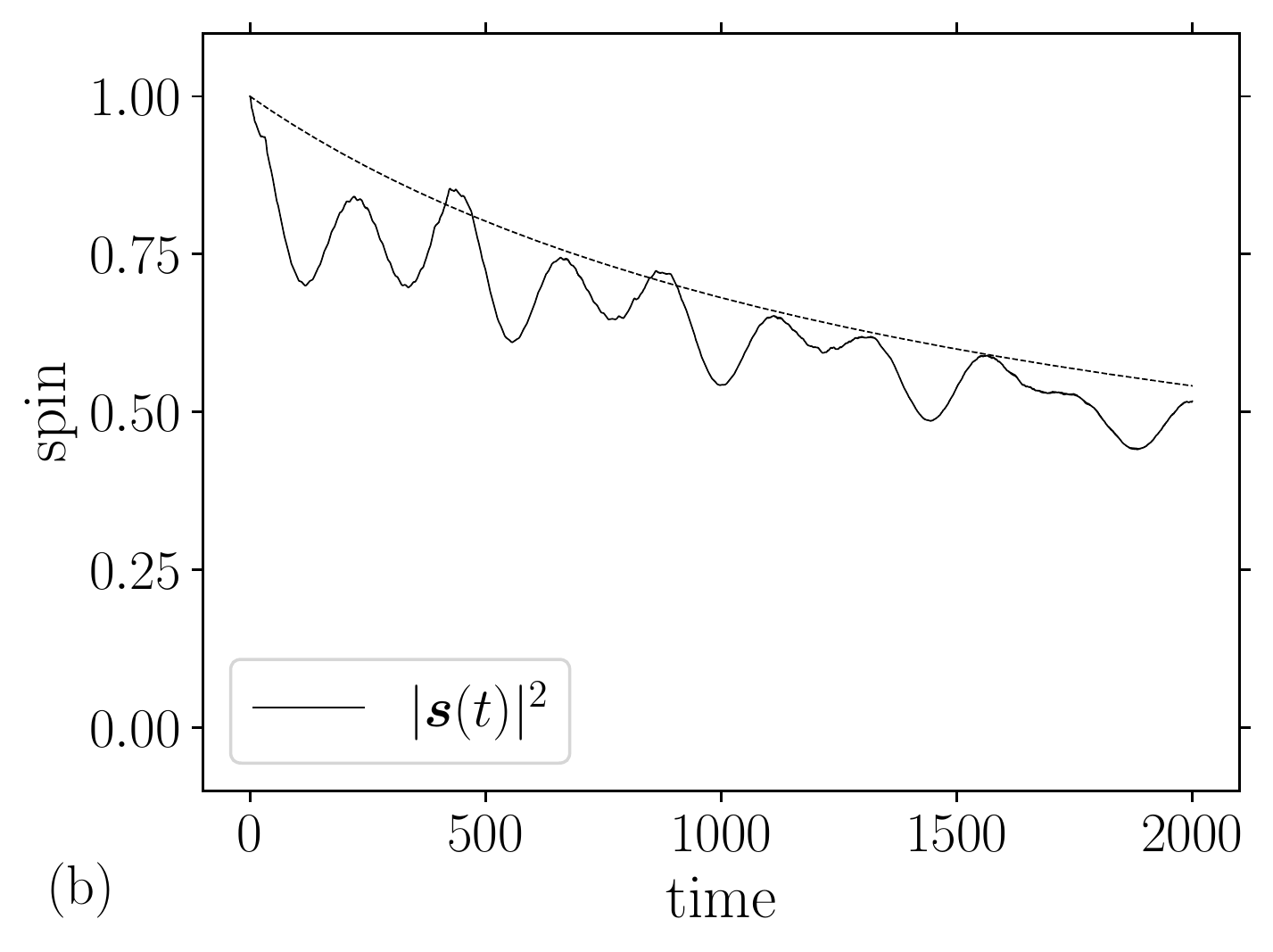}%
  \includegraphics[width=0.5\textwidth]{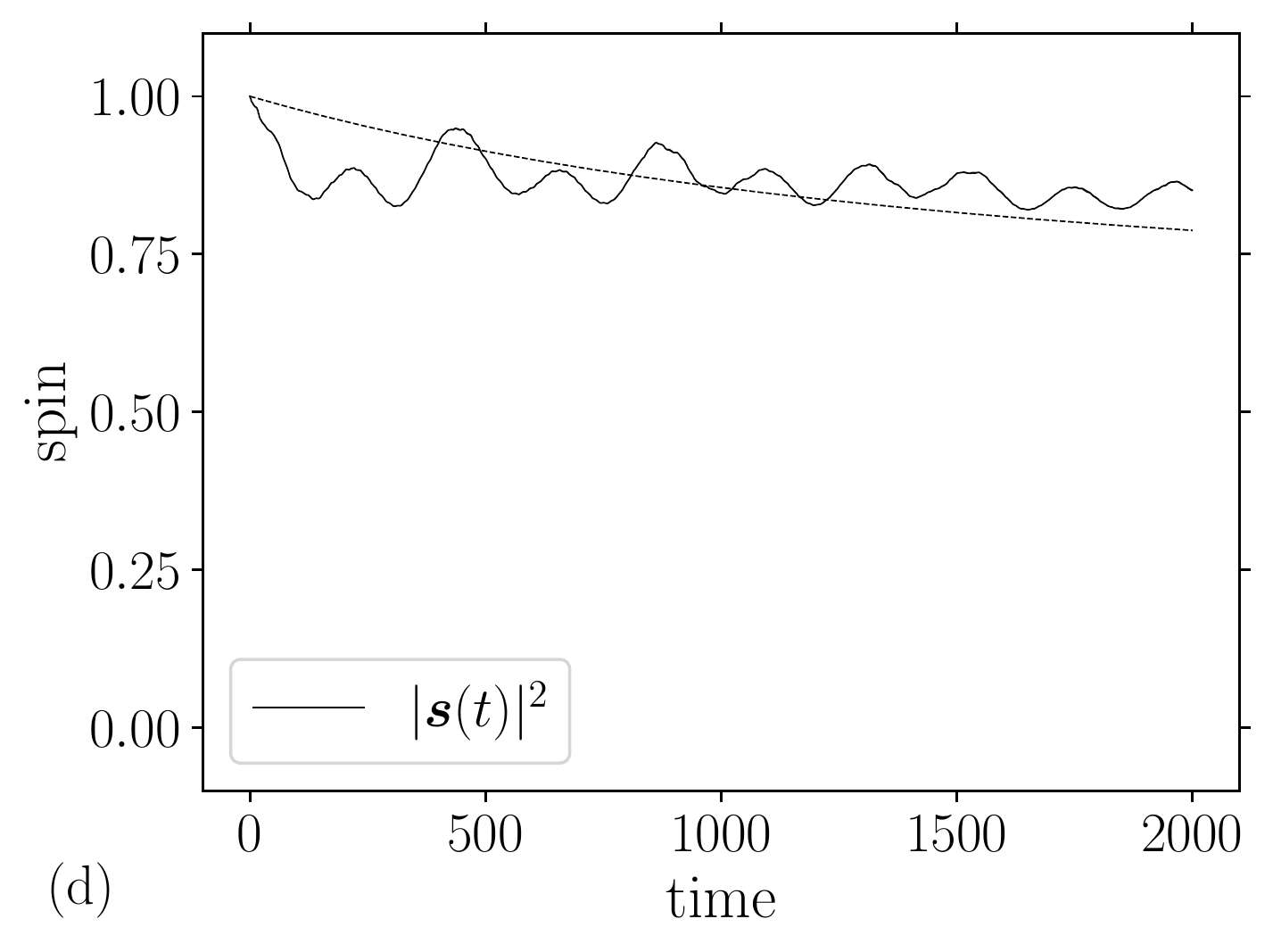}
  \caption{Landau-Lifshitz. Spin oscillations damping in the case of `ix' (a,b) and `iz' (c,d) compared to \protect\eqref{e:LLdiss} (dashed lines). Parameters \(1,0.4\), \(N=13\). The energy of the initial condition corresponds to a wavenumber of \(p = 6\pi/13\).
  \label{f:ixzf}}
\end{figure}

To go further, we investigate the particle-spin coupling in the homogeneous case, using the next order approximation in the small \(J\) limit. We use the standard time dependent perturbation theory to find the \(J\) dependent correction to the wave function of the free particle of momentum \(p\). Taking \(H_J\) as the perturbation of the Dirac Hamiltonian \(H_0\):
\[
  H = H_0 - \frac{J}{4} \bm s(t) \cdot \bm \tau 
\]
where the mean spin density is only time dependent, we may compute the evolution of the state over one step \(\Delta t = 1\), assuming that during this time the spin density is almost constant. We write the correction to the wave function \eqref{e:psi} of a particle of energy \(E_p>0\), as a superposition in the nonperturbed basis \(\psi_E\),
\[\psi(x,t) = \sum_E\E^{-\I E t} \phi_E(t) \psi_E(x)\]
which satisfies,
\begin{equation}
\label{e:1psi}
\phi_E(t+\Delta t) = \phi_E(t) - \frac{J}{4\I} \sum_{E'} \int_t^{t+\Delta t} \D t'\, \E^{\I (E - E') t'} \bm s(t') \cdot \bm \tau \phi_{E'}(t')\,.
\end{equation}
To solve this equation we observe that during one time step the energy exchange \(E - E' \approx \Delta E\) between the mean spin field and the walker must be small, in accordance with the slow variation of \(\bm s(t)\); hence, we put
\[
  \phi_{E'}(t') \approx \E^{- \I \Delta E t'} \phi_{E'}(t) \delta_{EE'} \sim \E^{- \I \Delta E t} \chi_{E'} \delta_{EE'}\,,
\] 
within the integral (the last approximation takes into account the correction to the first order in \(J\), up to a factor of norm one depending on \(t\), which do not contributes to the final result):
\begin{align}
  \phi_E(\Delta t) & = \phi_E(t+\Delta t) - \phi_E(t) = - \frac{J}{4\I} \int_t^{t+\Delta t} \D t'\, \E^{\I (E - E' - \Delta E)t'} \bm s(t') \cdot \tau \chi_{E'} \nonumber \\
                   & \approx - \frac{J}{4\I \sqrt{|V|}} \E^{\I (E-E' - \Delta E) t} \frac{\E^{\I (E-E' - \Delta E) \Delta t} - 1}{\I (E - E' - \Delta E)} \bm s(t) \cdot \bm \tau \chi_{E'}
\label{e:psi1}
\end{align}
where we the energy variation satisfies \(\Delta E \sim J/2|V| \ll 1/\Delta t\) (the time scale of the typical spin oscillations). The expected value of the particle's spin, to first order, is
\begin{align}
\braket{\bm \tau}^{(1)} &= \int \D E f(E) \phi_E^\dagger(\Delta t) \bm \tau \phi_E(\Delta t) \nonumber \\ 
                        &= \left. \frac{J^2 \Delta t}{8|V| v_g} |\bm s(t)|^2 \frac{\bm d}{|\bm d|} \right|_{E-\Delta E}
\label{e:tau1}
\end{align}
where we used the formula \(f(E) = 1/\pi v_g(E)\) for the energy distribution. We suppose that \(\braket{\bm \tau}^{(1)}\) is the dissipation source in the Landau-Lifshitz equation. To the lowest order in \(J\), the form of the dissipation term \(\bm D\) should be \cite{Baryakhtar-2006cq,Verga-2014fk}: 
\begin{equation}
\label{e:dissip}
\bm D = \braket{\dot{\bm s} \times \bm\tau }^{(1)} = \frac{J}{2}\braket{\bm \tau}^{(1)} \times (\braket{\bm \tau} \times \bm s)\,.
\end{equation}
This term has the form of the relativistic correction introduced by Landau and Lifshitz, but with the moment \(\bm s\) replaced here by \(\braket{\bm \tau}\), which introduces a damping of the magnitude of \(\bm s\) (it has a component parallel to \(\bm s\)). The microscopic origin of the dissipation is related to the leaking of torque from the fixed spins to the particle's ``magnetic momentum'', which is here a fast variable and acting as a stochastic torque, effectively giving a term as in \eqref{e:dissip} \cite{Garanin-1997ly}. The Landau-Lifshitz equation becomes,
\begin{equation}
\label{e:LLdiss}
\dot{\bm s} = - \frac{J}{2} \bm s \times \braket{\bm \tau} + \bm D = - \frac{J}{2|V|} \bm s \times \hat{\bm d} + \frac{J^3 \Delta t}{16|V|v_g} |\bm s|^2 \hat{\bm d} \times \left( \hat{\bm d} \times \bm s \right)\,,
\end{equation}
where the precession term may include the correction term \eqref{e:tau1} \(\braket{\bm \tau} = \hat{\bm d}/|V|(1 + O(J^2))\) that we neglect in the numerical comparisons. We show in Fig.~\ref{f:ixzf} a comparison of the approximation \eqref{e:LLdiss} with the exact evolution of the homogeneous cases `ix' and `iz'. The approximation gives satisfactory qualitative results, in particular it contains a mechanism to limit the relaxation to finite values of the spin magnitude due to the presence of the nonlinearity in \(|\bm s|^2\) and the progressive alignement of \(\bm s\) and \(\hat{\bm d}\).

\subsection{Small gradients}

We may extend the analysis of the homogeneous system to the case of small gradients; we naturally have a microscopic scale \(\Delta x = 1\) and a macroscopic one, given by the size of the system \(|V|\), hence small gradients mean spatial variation \(\ell\) of the spin distribution satisfying \(1 \ll \ell \ll |V|\). If the homogeneous case is characterized by a momentum \(p\), the presence of \(\ell\) scale fluctuations, can be taken into account by introducing a modulation \(\Delta p\) of the wave function: \(p \rightarrow p + \Delta p\). We can thus associate the momentum modulation with the spatial gradient, \(\Delta p \rightarrow -\I \partial_x\). The splitting of the one step operator is straightforwardly generalized to the position dependent case, giving the new mean field \(\bm s = \bm s(x,t)\):
\begin{equation}
  \label{e:H}
  H = H_0 + H_J \,, \quad H_J = - \frac{J}{4} \bm s(x,t) \cdot \tau \,,
\end{equation}
and
\(H_0\) the corrected to first order in \(\Delta p\) particle Hamiltonian:
\begin{equation}
\label{e:H01}
H_0 = E(p,\theta) \bm d_0 \cdot \tau + \Delta p \bm d_1 \cdot \tau \,,
\end{equation}
where \(H_0\) is the homogeneous Hamiltonian \eqref{e:H0k} in terms of \(p\),
\[
  \cos E = \cos p \sin\theta \,,
\]
is the energy dispersion relation,
\[
  \bm d_0 = \frac{1}{\sin E(p,\theta)}
  \begin{pmatrix}-\sin p \cos \theta \\
    \cos p \cos \theta \\
    -\sin p \sin \theta
  \end{pmatrix} \,,
\]
the mass term coefficient, and
\[
  \bm d_1 = -\frac{E(p,\theta)}{\sin E(p,\theta)}\begin{pmatrix}\cos p \cos \theta \\
    \sin p \cos \theta\\
    \cos p \sin \theta
  \end{pmatrix} \,,
\]
the coefficient of the momentum term. We remark that \(H_0\) reduces to the continuum limit of the Dirac walk when \(p = 0\) and the energy is supposed to be small \(E\Delta t \rightarrow 0\) \cite{Lee-2015uq}, which is different to the present approximation around a uniform distribution defined by a plane wave of arbitrary momentum \(p\) and energy \(E\). Up to a global rotation transformation, \eqref{e:H01} is a Dirac Hamiltonian with mass term \(\bm d_0\) and kinetic term \(\bm d_1\). Within this `hydrodynamic' like approximation, we actually neglect the distinction between the node and link operators, which will give higher order corrections to the one step evolution operator. One might expect that the validity of the hydrodynamic approximation will be better in the parameter range where the coupling constant is small and particle propagation dominates over localization. The model \eqref{e:H} is reminiscent of the so-called \emph{sd} Hamiltonian which describes the magnetic interaction between independent spins mediated by itinerant electrons \cite{Nolting-2009}. The difference here is that the particle is governed by a massive Dirac equation whose mass and velocity parameters depend on the angle \(\theta\), instead of the usual kinetic energy, quadratic in momentum.

\begin{figure}[tb]
  \centering
  \includegraphics[width=0.7\textwidth]{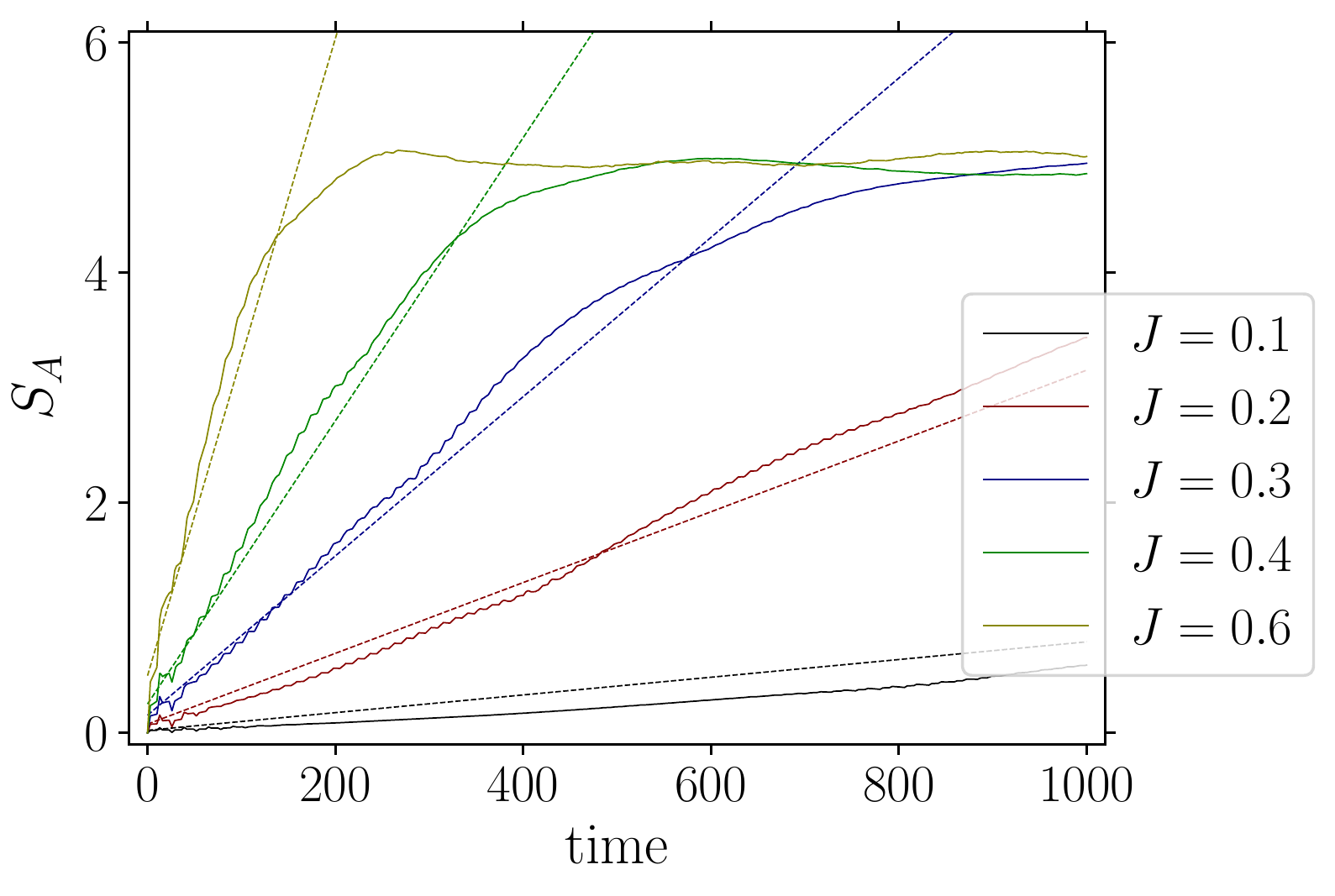}
  \caption{Behavior of the spin set entropy $S_A$ for $|A| = 6$, as a function of the coupling constant: $J=0.1,0.2,0.3,0.4,0.6$ and $\theta = \pi/2$, `x'. The straight lines fit the slope with formula \protect\eqref{e:SAf}.
  \label{f:SAff}}
\end{figure}

In contrast to the homogeneous case, when gradients in the spin distribution mediated by the running particle are present, the system's dynamical properties (oscillations, relaxation and entanglement growth) change qualitatively. For instance, a linear increase of the spin entanglement arises for times shorter than the relaxation time, in the propagation dominated case. An asymptotic relaxation towards a zero spin state is also observed, while persistent oscillations prevailed in the homogeneous case (for similar parameters). An illustration is given in Fig.~\ref{f:SAff}, where we picture the spin entanglement of a set of \(|A|=6\) spins in a system of \(|V|=13\) sites. We fit the linear part of the entanglement growth by the interpolation formula,
\begin{equation}
\label{e:SAf}
S_A(t) \approx \nu_J t\,, \quad \nu_J = \frac{J^2}{|V|} \,,
\end{equation}
which holds for time \(t\) larger than the initial transitory, whose duration is of the order of the system's size \(t > |V|\), and time \(t\) smaller than the characteristic saturation time \(t \lesssim 1/\nu_J\). (We may think that the slope in \eqref{e:SAf} depends on \(|A|\), but for the moment we stress only its quadratic dependence on \(J\).) We are interested in deriving a Landau-Lifshitz equation for the spin dynamics taking the influence of the gradients into account, in particular to explain the origin of the \(J^2/|V|\) effective parameter \eqref{e:SAf}. To this goal we proceed, as we did in the homogeneous case, with a kind of linear response calculation based on a multiscale expansion.

We write the approximate solution of the Dirac equation of Hamiltonian \eqref{e:H}, for a time of the order of one time step \((t, t+ \Delta t)\), as a spatially modulated plane wave,
\begin{equation}
\label{e:xpsi}
\psi(x,t) = \E^{-\I E t + \I p x} \phi(X,t)\,,
\end{equation}
where \(\phi\) varies significantly over a spatial scale \(X \sim 1/\Delta p \sim \ell\). The form \eqref{e:xpsi} separates fast and slow variables as in the Krilov-Bogoliubov-Mitropolsky method used in dynamical systems \cite{Mitropolsky-1997,Kakutani-1974}. The evolution equation of the modulation is then given by,
\begin{equation}
\label{e:xdphi}
\I \frac{\D \phi}{\D t}+ \I \partial_X \left( \bm d_1 \cdot \bm \tau \phi \right) = - \frac{J}{4} \bm s(X,T) \cdot \bm \tau \phi \,,
\end{equation}
where we explicitly put the slow variables dependence of the spin density, with the slow time scale \(T \sim 1/\Delta E\). We obtain the solution of \eqref{e:xdphi} in terms of the Fourier transform in the slow variables, to first order in \(J\):
\begin{equation}
\label{e:xfourier}
\phi(X,t) = - \frac{J}{4\sqrt{|V|}} \int \frac{\D E}{2\pi} \frac{\D \Delta p}{2\pi} \frac{\E^{-\I E t + \I \Delta p X}}{E - \Delta p \bm d_1 \cdot \bm \tau} \bm s_\Delta \cdot \bm \tau \chi \,,
\end{equation}
where \(\bm s_\Delta\) is the Fourier transform of the spin field. We are only interested in the effect of the gradient; using the inverse of the matrix,
\[
  \frac{1}{E - \Delta p \bm d_1 \cdot \bm \tau} = \frac{E + \Delta p \bm d_1 \cdot \bm \tau}{\sqrt{E^2 - \Delta p^2 |\bm d_1|^2}}\,,
\]
to first order in \(\Delta p\), we obtain,
\begin{equation}
\label{e:xint}
\phi(X,t) = \frac{J}{4\sqrt{|V|}} \int \frac{\D E}{2\pi \I} \frac{\E^{-\I E t}}{E^2} \int \frac{\D \Delta p}{2\pi} \, \I \Delta p \E^{\I \Delta p X} (\bm d_1 \cdot \bm \tau) (\bm s_\Delta \cdot \bm \tau) \chi\,.
\end{equation}
Over the time \(\Delta t\) the spin density is almost constant (\(\Delta t \ll 1/\Delta E\)), we can then evaluate the energy integral to get the final result,
\begin{equation}
\label{e:xphi}
\phi(X, \Delta t) = \frac{J \Delta t}{8 \sqrt{|V|}} \left[ (\bm d_1 \cdot \partial_x \bm s) + \bm d_0 \cdot (\bm d_1 \times \partial_x \bm s) \right] \chi \,,
\end{equation}
(the momentum integral gives the gradient) where we used \(\chi^\dagger \bm \tau \chi = \bm d_0\). We apply now \eqref{e:xphi} to calculate the response of the particle's spin to the spin density gradient:
\begin{equation}
\label{e:xtau1}
\braket{\bm \tau}^{(1)} = \braket{\psi_\PL(x,0)|\bm \tau|\psi(x,\Delta t)} = \frac{J \Delta t}{8 |V|} \left[ (\bm d_1 \cdot \partial_x \bm s) + \bm d_0 \cdot (\bm d_1 \times \partial_x \bm s) \right] \bm d_0\,,
\end{equation}
where we substituted \eqref{e:psi} and \eqref{e:xpsi}, \eqref{e:xphi} to \(\psi_\PL\) and \(\psi\), respectively (note that in the homogeneous case this first order contribution to the torque vanished, giving rise to a higher order correction). Within the same order we compute the correction to the interacting energy,
\begin{equation}
\label{e:xEint}
E_\text{int} = - \frac{J^2}{32|V|} \int \D x \, (\bm s \cdot \bm d_0) \left[ (\bm d_1 \cdot \partial_x \bm s) + \bm d_0 \cdot (\bm d_1 \times \partial_x \bm s) \right] \,.
\end{equation}
Finally, after computing the functional derivative of \eqref{e:xEint},
\begin{equation}
\label{e:dEint}
\dot{\bm s} = 2 \bm s \times \left(- \frac{\delta E_\text{int}}{\delta \bm s} \right) \,,
\end{equation}
we can write the modified Landau-Lifshitz equation:
\begin{equation}
\label{e:xLL}
\dot{\bm s} = -\frac{J}{2|V|} \bm s \times \bm d_0 + \frac{J^2\Delta t}{8|V|} \left[ (\bm d_0 \times \bm d_1 + \bm d_0 \times ( \bm d_0 \times \bm d_1) \right] \times (\bm s \times \bm \partial_x \bm s) + \bm D\,,
\end{equation}
which contains a correction to the precession frequency,
\begin{equation}
\label{e:xLLfreq}
\frac{J}{2|V|} \rightarrow \frac{J}{2|V|} + \frac{J^2\Delta t}{4|V|} (\bm d_1 \cdot \partial_x \bm s) + \frac{J}{2} \bm d_0 \cdot (\bm d_1 \times \partial_x \bm s)\,,
\end{equation}
and a correction to the effective applied field,
\begin{equation}
\label{e:xLLfield}
\bm d_0 \rightarrow \bm d_0 - \frac{J^2 \Delta t}{4|V|} \left[ \bm d_1 + (\bm d_0 \times \bm d_1) \right] (\bm d_0 \cdot \partial_x \bm s) \,,
\end{equation}
and where we added a dissipation term \(\bm D\), which should also be of order \(O(J^2)\)  (c.f. \eqref{e:dissip} with the particle spin given by \eqref{e:xtau1}). Therefore, within this approximation, the effect of the gradient translates into a nonlinear shift of the precession frequency, which contributes to an anharmonic evolution of the spin oscillations, together with a first order in the spin gradient force, whose common physical origin is the scattering of the walker wave function off the mean field spin inhomogeneities. We note a fundamental difference with respect to the usual exchange interaction, which is proportional to the second derivative of the magnetization; the added force is reminiscent to the spin-orbit terms found in some materials \cite{Tserkovnyak-2012ys,Chen-2014db}.

\begin{figure}[tb]
  \centering
  \includegraphics[width=0.517\textwidth]{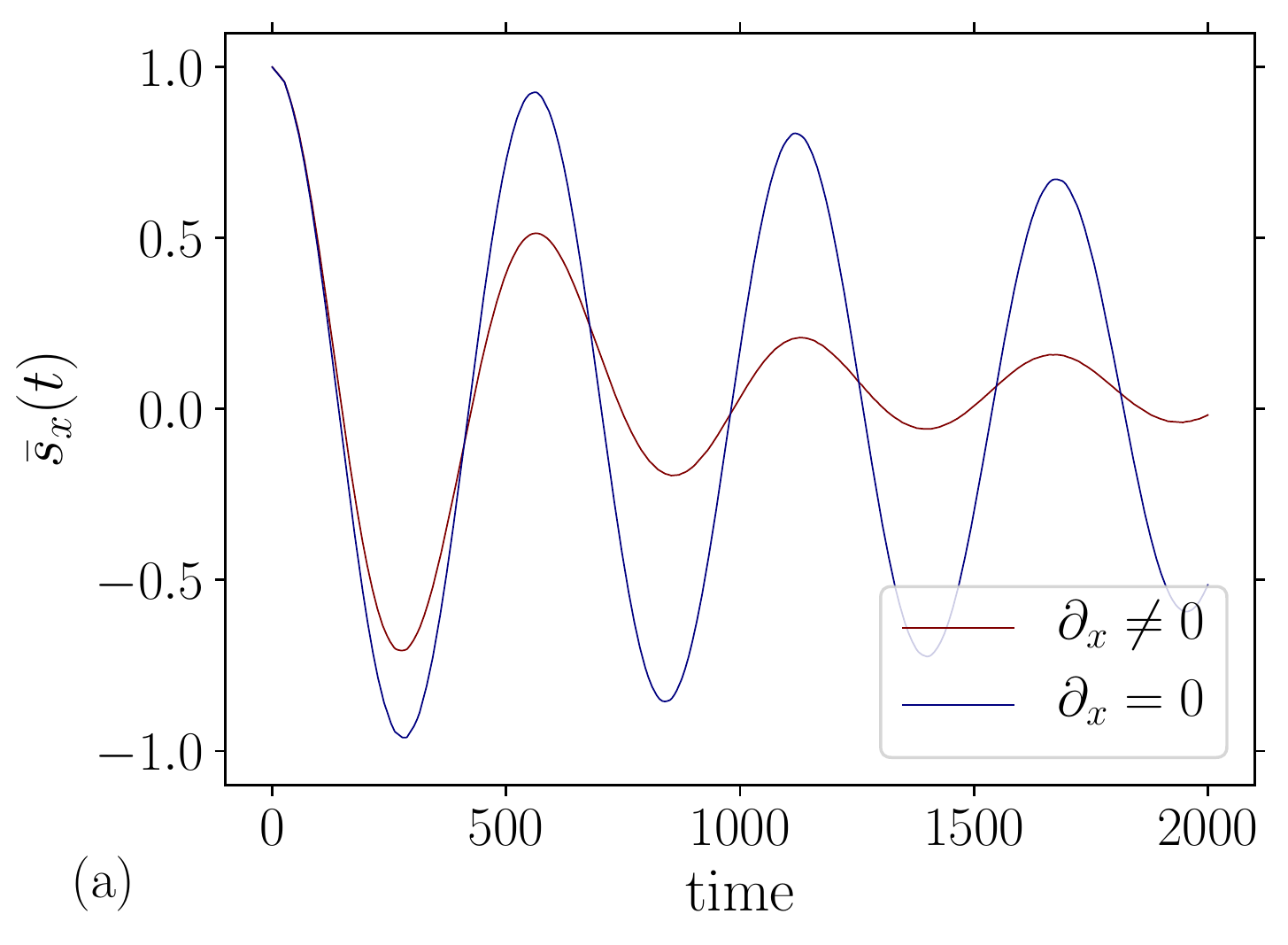}%
  \includegraphics[width=0.483\textwidth]{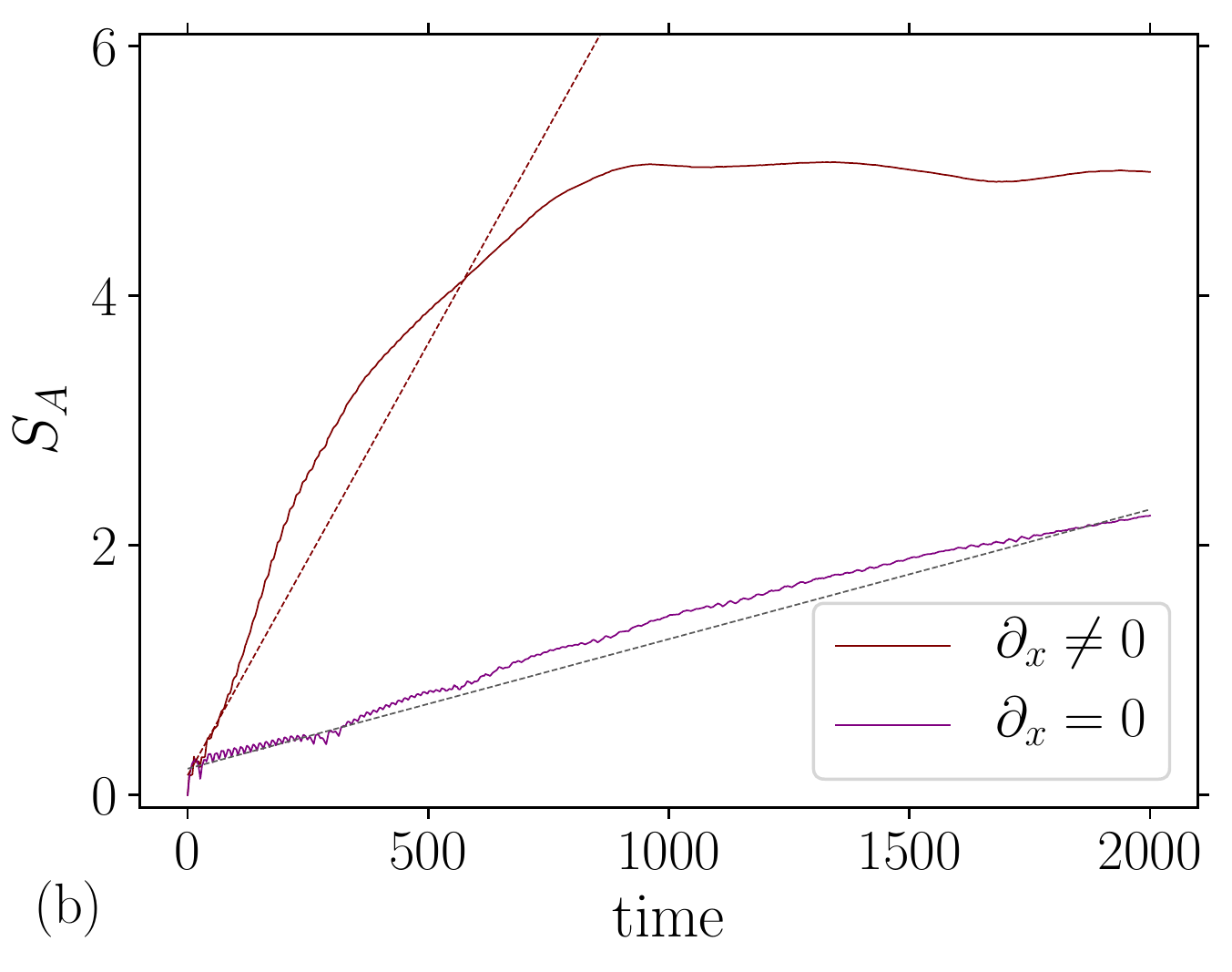}
  \caption{Comparison of the homogeneous $\partial_x = 0$ and inhomogeneous $\partial_x  \ne 0$ evolution of the spin $\bar{s}_x$ (a) and spin entanglement $S_A$ (b). Parameters $(1.5,0.3)$, `x', $|A|=6$, $N=13$. The dashed lines are linear fits with slope $\nu = J^2/N \sim J^2$ ($\partial_x \ne  0$, \protect\eqref{e:SAf}), and $(J/2)\nu\sim J^3$ ($\partial_x = 0$).
\label{f:dx}}
\end{figure}

In a system of only a dozen of sites it is difficult to compare quantitatively \eqref{e:xLL} with the exact evolution of the interacting walk (gradients extend over a few nodes); however, it is possible to test the parameter dependency in \(J^2\) (already observed in Fig.~\ref{f:SAff}). We compare the uniform and nonuniform time evolution of the spin density and spin entanglement entropy in Fig.~\ref{f:dx}. It is remarquable that the simple multiplication \((J/2) \nu\), with \(\nu = J^2/|V|\) (the natural factor coming from the Landau-Lifshitz equation), fits so well the homogeneous case, comforting the relevance of the second order correction. In the same figure we show \(\bar{s}_x\) for `x', which relaxes faster than the `ix' case, in agreement with the behavior of \(S_A\). In addition, the anisotropy of the oscillation  (for instance, the \(z\) component stay at zero in Fig.~\ref{f:o54p}) is a consequence of the anisotropy of the unit vector \(\hat{\bm d} = \bm d_0\) (\(x\) and \(y\) components vanish for \(\theta = \pi/2\)).

Equation \eqref{e:xLL} can describe the oscillation-relaxarion regime in an effective external field which to lowest order \(\bm \tau \sim \bm d_0\) is constant, and whose corrections are perturbative in the coupling \(J\). The strong \(J\) regime would need a different approximation approach, for instance a kind of adiabatic approximation in which the particle's torque follows the spin's one, valid in principle for the strong coupling regime. We leave this investigation for future work. 

\section{Conclusions}
\label{S:concl}

In this paper we investigated a generalization of the quantum walk to a many-body interacting system, in order to build an entangled state from the dynamics of a walker coupled to a network of spins. This approach may be compared with the usual adiabatic model of quantum computing \cite{Kadowaki-1998,Farhi-2000,Aharonov-2008,Albash-2018}. Instead of adiabatically evolving a quantum system from an initial Hamiltonian whose ground state is a product of basis states, to a useful entangled ground state of another Hamiltonian, we obtain the resource state from the unitary evolution of the system towards a stationary state (stationary in the sense of its observables properties: magnetization, entanglement, probability distribution). In its generality the model is implemented on an arbitrary (simple) graph, in which the walker wanders between neighboring nodes interacting with fixed spins located in the links. The fixed spins are then coupled through the particle's degrees of freedom, position and color. The color degree of freedom represents internally the connectivity of the graph. This model, loosely inspired by the condensed matter magnetic metals in which the interaction is of the RKKY type governed by a \emph{sd}-Hamiltonian, is characterized by an exchange interaction between the color of the two nodes of a link, and the local spin. This interaction differs from the more usual Ising-type interaction used to build cluster states, because it depends on the angle between the particle and local spin magnetic moments. In addition, the coin operator that modifies the particle's color, introduces an anisotropic redistribution of the state amplitudes between the incoming edges of a vertex. This anisotropy has important consequences for the dynamics of the system. Another motivation for the present model comes from the experiments in which an entangled state of local spins (nitrogen vacancies in diamond or magnetic moments of trapped cold atoms, for instance) is obtained by an interaction mediated by a nonlocal moment \cite{Bradley-2019,Pagliero-2020}, or by the interaction with the environment in a spin chain \cite{Maier-2019}, which can be useful for the construction of quantum networks \cite{Cirac-1997,Christandl-2004}.

The present work focused on the simplest geometry, the one dimensional lattice with closed or periodic boundary conditions. Nevertheless, we observed a fairly rich dynamical phenomenology even for this simple case in which we consider the Dirac walker coupled to non interacting spins through its color. We could exhibit a variety of behaviors, reminiscent to a magnetic chain in which, taking into account that the motion of the walker is correlated to its internal state, a magnetic interaction arises mediated by a kind of spin-orbit coupling. We determined using direct numerical computation by successive application of the one step evolution operator, the parameter phase space regions of localization and propagation, and explored the weak and strong coupling regimes, the regular and irregular dynamics as well as the influence of the interaction on the walk topological properties. We found that the magnetic dynamics, for homogeneous and weakly inhomogeneous states, is well described by a quasiclassical mean field Landau-Lifshitz equation with dissipation, nonlinerar frequency shift and torques proportional to the magnetization gradient (c.f. Eq.~\eqref{e:xLL}). Both, dissipation and nonlinear torques, arise from the scattering off on the fixed spins of the walker wave function; the hydrodynamic like approximation applies when the temporal and spatial scales characteristics of the walker and the edge spins can be well separated. Let us mention that as a by-product of the low energy approximation we generalized the usual method to derive the continuum limit of the free quantum walk, Eq.~\eqref{e:H01}.

We found that the interacting many-body quantum walk leads for a large range of parameters, to highly nonlocal entangled states, with an entropy proportional to the size of the entangled region. In particular, we demonstrated that it gives rise to localized edge states at the interface between topologically different phases, in contrast to the free walk. The entanglement dynamics and its structure are governed by the interplay of the local degrees of freedom, the fixed spins, and the global one, the walker whose distribution is delocalized over the entire system. We may conclude that this system has properties that are relevant to the quantum simulation of condensed matter systems as well as to the build up of entangled states of interest as a quantum computational resource.

\section*{Acknowledgements}
We benefited from discussions with Álvaro Núñez, Giuseppe Di Molfetta, Gabriel Elías and Pablo Arrighi. Numerical calculations used \texttt{python} with the libraries \texttt{numpy} and \texttt{scipy}. Figures were produced with \texttt{matplotlib} \cite{Hunter-2007}. 

%

\begin{appendix}

\section{The one step operator $U$}
\label{S:U}

In this Appendix we will exhibit the structure of the one step operator \(U\) by applying it to the canonical basis of the Hilbert space \(\ket{xcs} \in \mathcal{H}_G\), in the case where \(G\) is a linear lattice. Let \(\ket{\psi}\) be a general state,
\begin{equation}
\label{e:Upsi}
\ket{\psi} = \sum_{xcs} \psi_{xcs} \ket{xcs}, \quad \psi_{xcs} \in \mathbb{C}\,.
\end{equation}
We compute \(U\ket{\psi} = V(J)MC \ket{\psi}\), using the definitions \eqref{e:M}-\eqref{e:VxyJ},
\begin{equation}
\label{e:UUpsi}
U\ket{\psi} = V(J) \sum_{xs} \big( \cos\theta \, \psi_{x+1\,0s} - \sin\theta \, \psi_{x1s} \big) \ket{x0s} + \big( \sin\theta \, \psi_{x0s} + \cos\theta \, \psi_{x1s}  \big) \ket{x-1\,0s}
\end{equation}
where we used the rotation matrix \eqref{e:C} and the definition of the motion operator \eqref{e:M} (for the one-dimensional case). The action of \(V(J)\) is conveniently expressed in the edge basis \eqref{e:se}, we introduce then the notation \(s_x(0)\) for the set of spins \(s=1,\ldots,2^{|E|}-1\) with the label \(s_x = 0\) (corresponding to the edge \(e = (x,x+1)\)), idem \(s_x = 1\):
\begin{align}
\label{e:UVJpsi}
U \ket{\psi} = &  \E^{-\I J/4} \sum_{x,s} \left\{ 
  \E^{\I J/2} \big( \sin\theta \, \psi_{x+1\,0s_x(0)} + \cos\theta \, \psi_{x+1\,1s_x(0)} \big) \ket{x\,0s_x(0)} + \right. \nonumber \\
        & \left[ 
      \I \sin(J/2) \big( \cos\theta \, \psi_{x\,0s_x(0)} - \sin\theta \, \psi_{x\,1s_x(0)} \big) + \right. \nonumber \\
        & \left. 
  \cos(J/2) \big( \sin\theta \, \psi_{x+1\,0s_x(1)} + \cos\theta \, \psi_{x+1\,1s_x(1)} \big)  \right] \ket{x0s_x(1)}  + \nonumber \\
        & \left[ 
   \cos(J/2) \big( \cos\theta \, \psi_{x\,0s_x(0)} - \sin\theta \, \psi_{x\,1s_x(0)} \big)+ \right. \nonumber \\
        & \left. 
     \I \sin(J/2) \big( \sin\theta \, \psi_{x+1\,0s_x(1)} + \cos\theta \, \psi_{x+1\,1s_x(1)} \big)  \right] \ket{x+1\,1s_x(0)} + \nonumber \\
        & \left. \E^{\I J/2} \big( \cos\theta \, \psi_{x\,0s_x(1)} - \sin\theta \, \psi_{x\,1s_x(1)} \big) \ket{x+1\,1s_x(1)}  
\right\} \,.
\end{align}
We observe that each term of the sum contains the basis vectors of the edge \eqref{e:Vop} with coefficients depending on the eight amplitudes corresponding to the two nodes and edge spin, therefore, coupling effectively the edge \(e_x\) with its two neighbors \(e_{x \pm 0}\). One may condensate this structure into a formula which gives the amplitudes of edge \(e_x\) in terms of the amplitudes of its neighbors:
\begin{equation}
\label{e:Uedge}
\ket{\psi_{e_x, s_x}(t+1)} = A \ket{\psi_{e_{x-1}, s_x}(t)} + B \ket{\psi_{e_x, s_x}(t)} + C \ket{\psi_{e_{x+1}, s_x}(t)} \,,
\end{equation}
where,
\begin{equation*}
  \ket{\psi_{e_x,s_x}} = \begin{pmatrix}
      \psi_{x\,00_x} \\
      \psi_{x\,01_x} \\
      \psi_{x+1\,10_x} \\
      \psi_{x+1\,11_x} 
      \end{pmatrix}
\end{equation*}
and the matrices \((A,B,C)\) are given by,
\begin{equation}
\label{e:UA}
  A = \E^{-\I J/4} \begin{pmatrix} 
0 & 0 & 0 & 0 \\
0 & 0 & -\I \sin(J/2) \sin(\theta) & 0\\
0 & 0 & -\cos(J/2) \sin(\theta) & 0\\
0 & 0 & 0 & -\E^{\I J/2} \sin(\theta)
\end{pmatrix},
\end{equation}
\begin{equation}
\label{e:UC}
  C = \E^{-\I J/2} \begin{pmatrix} 
\E^{\I J/2} \sin(\theta) & 0 & 0 & 0 \\
0 & \cos(J/2) \sin(\theta) & 0 & 0\\
0 & \I \sin(J/2) \sin(\theta) &0 & 0\\
0 & 0 & 0 & 0
\end{pmatrix},
\end{equation}
and
\begin{equation}
\label{e:UB}
B = \E^{-\I J/4} \begin{pmatrix} 
0 & 0 & \E^{\I J/4} \cos(\theta) & 0 \\
\I \sin(J/2) \sin(\theta) & 0 & 0 & \cos(J/2) \cos(\theta) \\
\cos(J/2) \cos(\theta) & 0 & 0 & \I \sin(J/2) \sin(\theta) \\
0 & e^{i\frac{J}{2}}\cos(\theta) & 0 & 0
\end{pmatrix}\,,
\end{equation}
note that the spin index is local to the \(e_x\) edge. The fact that the spin in \(e_x\) appears in the terms of the \(A\) and \(C\) matrices, associated respectively to the \(e_{x-1}\) and \(e_{x-1}\) neighbors, shows the typical mixing of amplitudes of a unitary operator. Two adjacent spins are indirectly coupled by the color of their common node, which splits between both and forbids writing \(U\) in a simple block form.

\section{$|V| = 2$, full coupling}
\label{S:A}

To enlighten the structure of the operator \(U\) defining the step of the interacting walk \eqref{e:U1}, and to study the mechanisms underlying the particle-spin entanglement it is convenient to reduce the system to its simplest nontrivial form. In this appendix we investigate the \(|V| = |E| = 2\) case, which can be worked out analytically for a few steps.

The Hilbert space dimension is 16. The explicit expression of \(U\) in the \(\ket{xcs}\) basis is useful for the computation of the observables, but do not highlight the interaction structure which is related to the edges. In the edge basis, consisting in sets of four labels of the form,
\[e_x(s) = \{x0s_x,(x+1)1s_x\}\,, \quad x=0,1, \; s = 0,1\]
\(U\) can be build up form block of \(4\times 4\) matrices; here \(0_0 = \{00,01\}\), \(0_1 = \{00,10\}\), etc., for example:
\[
  e_0(1)= \begin{pmatrix} 0010\\ 0011\\  1110 \\ 1111 \end{pmatrix}\,,
\]
(compare with \eqref{e:Vop}). To construct the operator \(U\) we use the method of Appendix~\ref{S:U}. We write the recurrence \eqref{e:Uedge} in the mixed base \(e_x(s)\) and \(e_x(\bar{s})\), where \(\bar{s}_x = s_{x+1}\) is a edge vector with the spin of the next edge:
\begin{equation}
\label{e:AABC}
\psi_{t+1} \!\begin{pmatrix} 
  e_0(0) \\ e_0(1) \\ e_1(0) \\ e_1(1)
\end{pmatrix}
=
(A+C)\, \psi_t\! \begin{pmatrix} 
  e_1(\bar{0}) \\ e_1(\bar{1}) \\ e_0(\bar{0}) \\ e_0(\bar{1})
\end{pmatrix}
+
B\, \psi_t\! \begin{pmatrix} 
  e_0(0) \\ e_0(1) \\ e_1(0) \\ e_1(1)
\end{pmatrix}
\end{equation}
where we used the periodicity of the chain. Now the problem is to express \(e_x(\bar{s}) = (e_x, s_{x+1})\) in the edge basis; 
\begin{align}
\label{e:Utransf}
e_1(\bar{0}) & = M_{00} e_1(0) + M_{01} e_1(1) \nonumber \\
e_1(\bar{1}) & = M_{10} e_1(0) + M_{11} e_1(1) \nonumber \\
e_0(\bar{0}) & = M_{00} e_0(0) + M_{01} e_0(1) \nonumber \\
e_0(\bar{1}) & = M_{10} e_0(0) + M_{11} e_0(1)
\end{align}
where the transformation matrices \(M_{sx}\) are given by,
\[
  M_{00} = \begin{pmatrix}
    1 & 0 & 0 & 0 \\ 0 & 0 & 0 & 0 \\ 0 & 0 & 1 & 0 \\ 0 & 0 & 0 & 0
  \end{pmatrix}, \quad 
  M_{01} = \begin{pmatrix}
    0 & 0 & 0 & 0 \\ 1 & 0 & 0 & 0 \\ 0 & 0 & 0 & 0 \\ 0 & 0 & 1 & 0
  \end{pmatrix}, \quad 
\]
and,
\[
  M_{10} = \begin{pmatrix}
    0 & 1 & 0 & 0 \\ 0 & 0 & 0 & 0 \\ 0 & 0 & 0 & 1 \\ 0 & 0 & 0 & 0
  \end{pmatrix}, \quad 
  M_{11} = \begin{pmatrix}
    0 & 0 & 0 & 0  \\ 0 & 1 & 0 & 0 \\ 0 & 0 & 0 & 0 \\ 0 & 0 & 0 & 1
  \end{pmatrix}\,.
\]
Substituting this transformation into \eqref{e:AABC} we obtain an expression in the edge basis, with \(U\):
\begin{equation}
\label{e:A_U1}
U = \begin{pmatrix}
  B & 0 & A_{00} & A_{01} \\
  0 & B & A_{10} & A_{11} \\
  A_{00} & A_{01} & B & 0 \\
  A_{10} & A_{11} & 0 & B
\end{pmatrix}
\end{equation}
where
\[
  A_{sx} = (A + C) M_{sx}, \quad 
\]
\((A,B,C\) are defined in \eqref{e:UA}-\eqref{e:UB}. We have, on the diagonal, the (non unitary) matrix \(B\) proportional to \(\cos \theta\), and off diagonal matrices \(A,C\), proportional to \(\sin \theta\): this block three diagonal form become much more cumbersome for an arbitrary number of nodes (in the sense of Appendix~\ref{S:U}); it reflects the locality of the original operators, and implies that at each step two new edges enter into play.

\begin{figure}[tpb]
  \centering
  \includegraphics[width=0.5\textwidth]{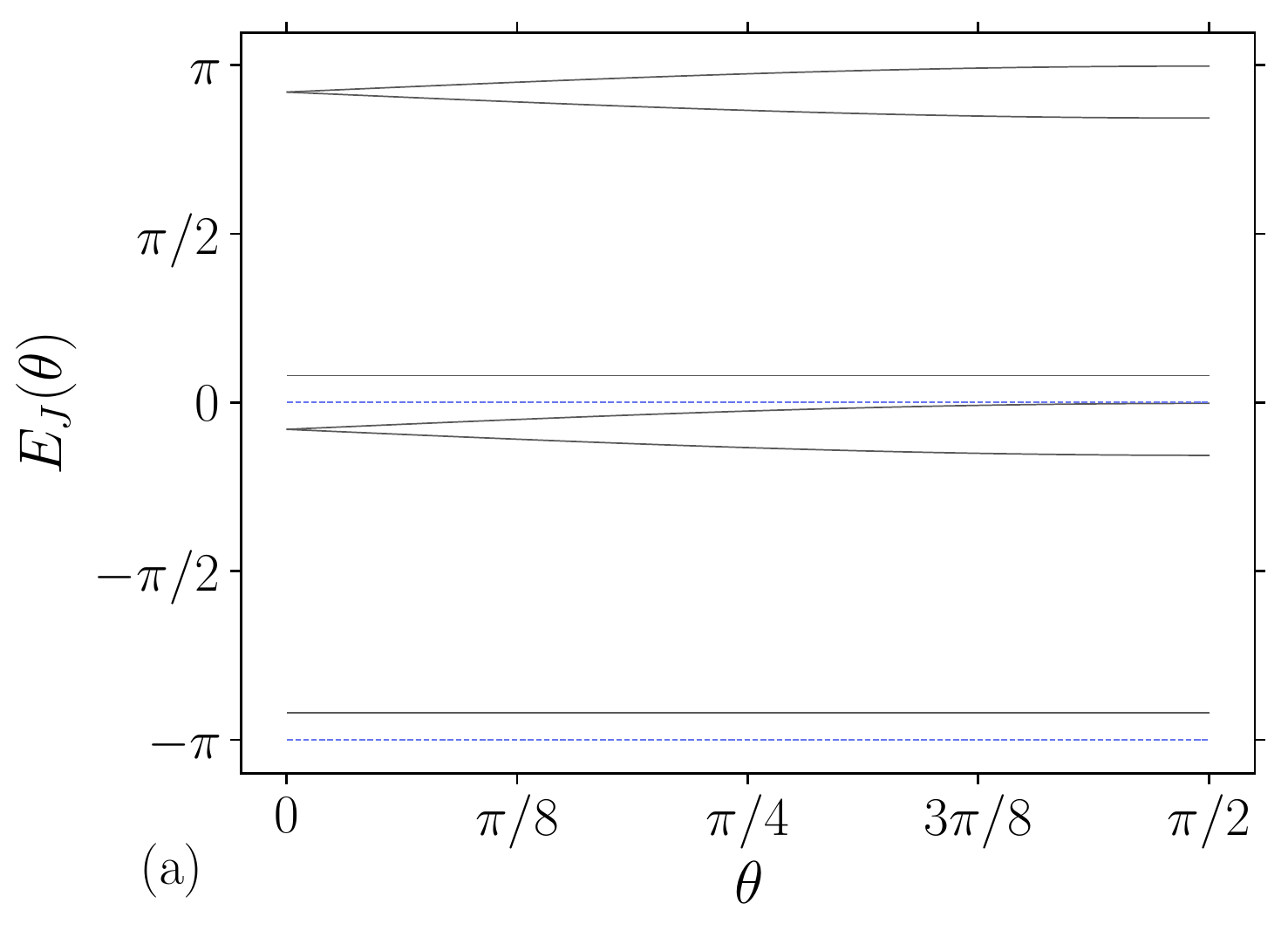}%
  \includegraphics[width=0.5\textwidth]{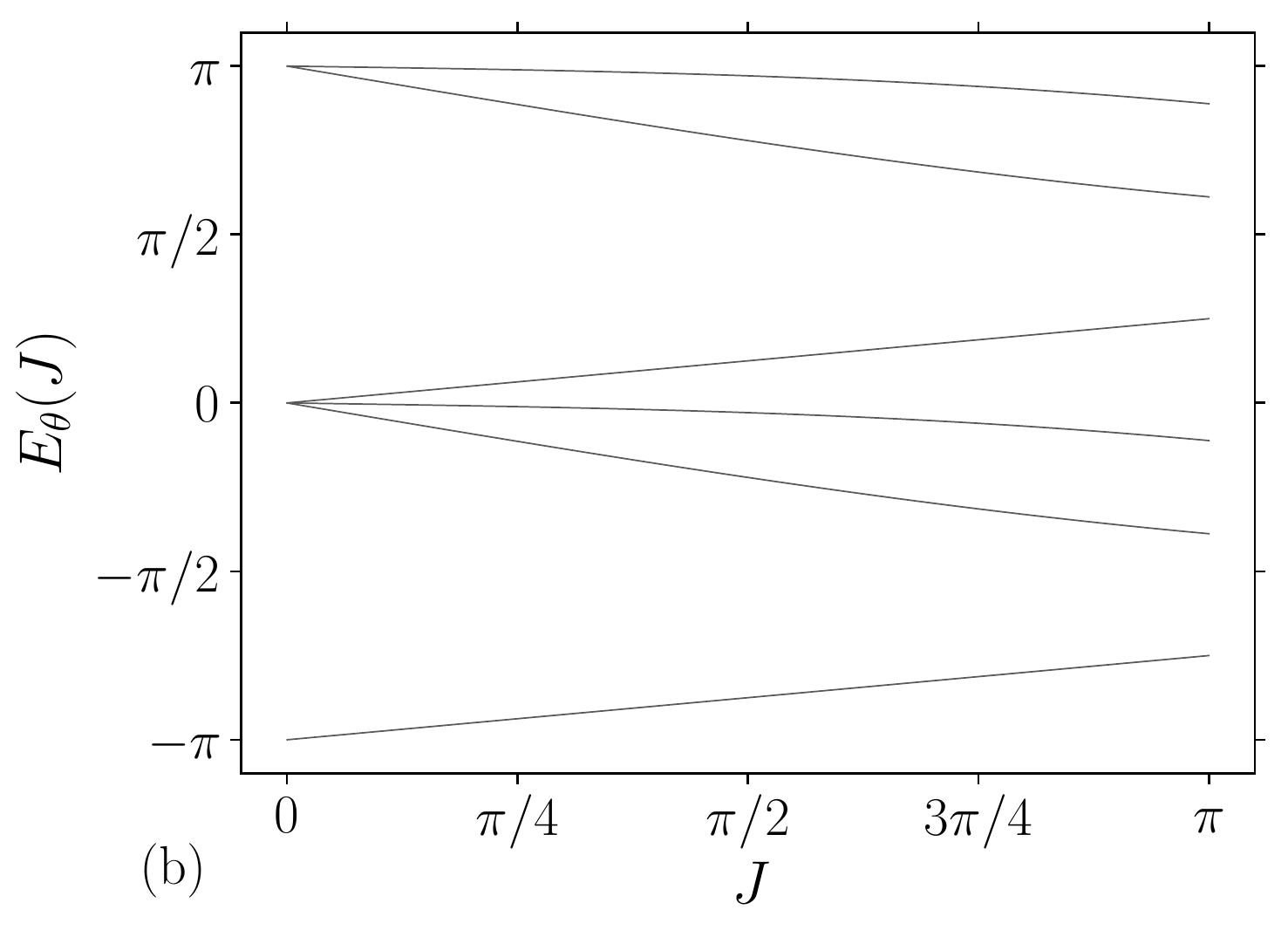}
  \caption{Spectrum \protect\eqref{e:eigenU} of the $|V| = 2$ one step operator. (a) $\arg\lambda$ as a function of $\theta$ for $J = 1$ (the dashed lines correspond to $J=0$); (b) $\arg\lambda$ as a function of $J$ for $\theta = 1$. The curves are symmetric with respect to the vertical axes $\theta=0$ and $J = 0$, respectively.
  \label{f:l6}}
\end{figure}

The eigenvalues of \(U\) are readily computed:
\begin{align}
  \lambda_\pm &= \pm \E^{\I J/4} \\
  \lambda_n &= \pm \frac{ \E^{-\I J/4} }{ \sqrt{2} } \sqrt{2 - \sin^2(J/2) \sin^2\theta \pm \I |\sin(J/2) \sin\theta|\sqrt{ 4 - \sin^2(J/2) \sin^2\theta }}
\label{e:eigenU}
\end{align}
where each \(\lambda_\pm\) is four fold degenerated, and each \(\lambda_n\) (with \(n = --,-+,+-,++\)) is twice degenerated. Note that \(\theta = 0\) gives a trivial system (off diagonal terms vanish), but \(\theta = \pi/2\) contains information about the interaction. In Fig.~\ref{f:l6} we plot the argument of the eigenvalues (quasienergies \(E = \arg \lambda\)) of \(U\), as a function of the parameters; the interaction partially lifts the degeneracy creating gaps that grow with both \(\theta\) and \(J\). In this simple limit levels do not cross at finite values of the interaction, therefore, the long time behavior of the \(|V|=2\) system is essentially quasiperiodic, with a basic period of \(T = 8\pi/J\) (in accordance with \eqref{e:oscT}).

We compute now the entanglement entropy and a spin correlation function, in order to investigate the way in which the particle-spin interaction propagates. We take the particle initially on the node 0 and the spins in the state \(\ket{+}\) (`x' initial condition):
\[\ket{\psi(0)} = (1,1,1,1,0,0,0,0,0,0,0,0,0,0,0,0)^\textsc{T}\,.\]
After one time step the state becomes
\[\ket{\psi(1)} = U \ket{\psi(0)}\,,\]
and the corresponding density matrix \(\rho = \ket{\psi(1)}\bra{\psi(1)}\), from which we compute the partial trace over the particle degrees of freedom:
\[\rho_s(1) = \Tr_{xc} \rho\,, \quad x = 0,1, \; c = 0, 1\]
to get the spin state density matrix. For example, putting \(\theta = \pi/2\), we find the eigenvalues of \(\rho_s\) to be:
\begin{equation*}
  r_0 = 0 \,, \quad
  r_\pm = \frac{1}{2} \pm \frac{1}{4} \sqrt{4 - \sin^2J} \left[ 1 - 2 \sin(2J - \arg(8J))   \right]\,.
\end{equation*}
the first eigenvalue is twice degenerated. The corresponding entropy is
\begin{equation}
\label{e:ssrn}
S_s(J) = - \sum_n r_n \log r_n\,,
\end{equation}
which vanishes for \(J=0, \pi\), and has a maximum for \(J=\pi/2\): the number of maxima increases with \(t\), and the shape of \(S_s\) may become highly oscillatory for large \(t\). However, we are only interested in the properties of the case \(|V|=2\) that may be of useful to understand larger systems. What we see is that after a step particle and spins become entangled, but the entanglement is a nontrivial function of \((\theta,\phi)\).

\begin{figure}[tb]
  \centering
  \includegraphics[width=0.5\textwidth]{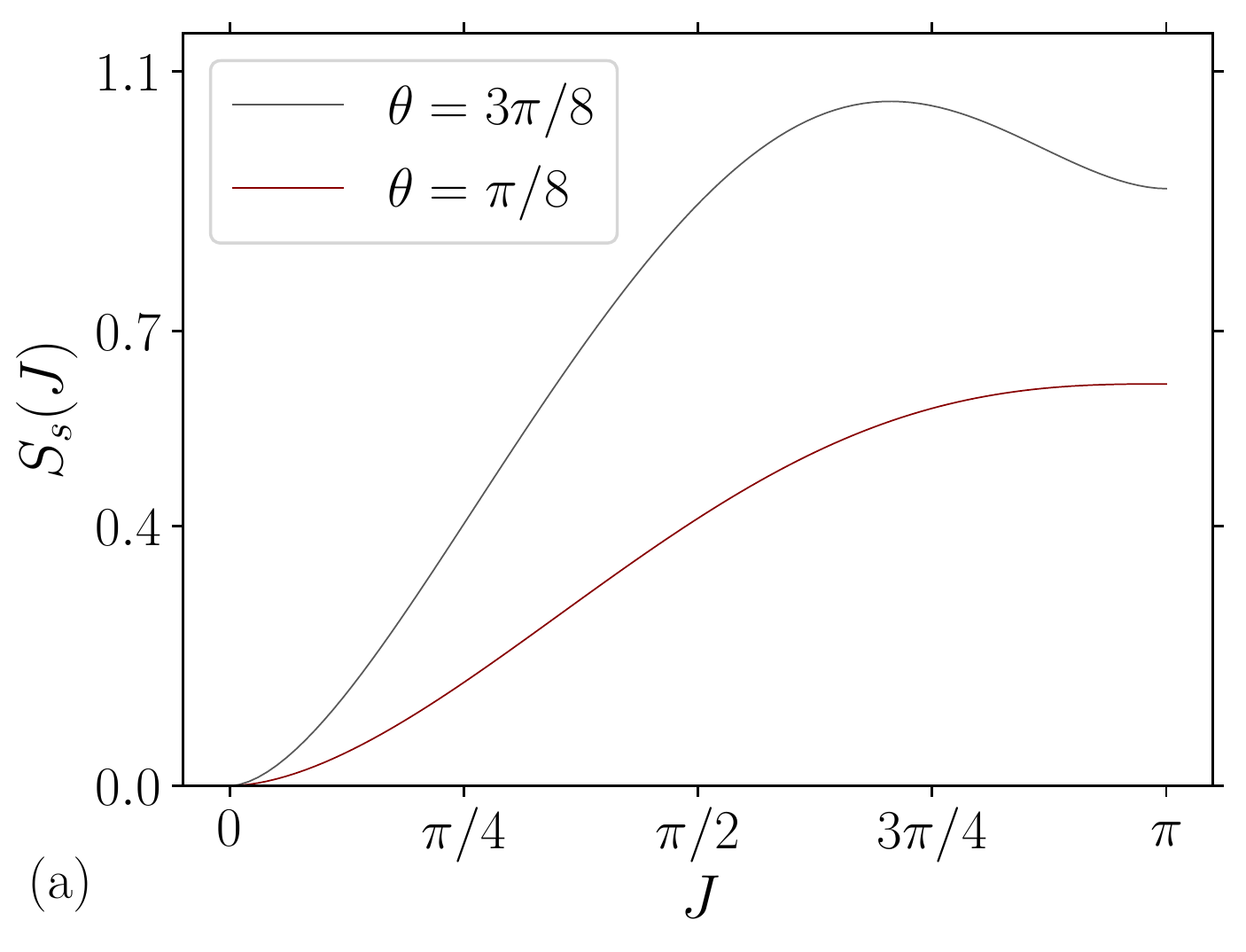}%
  \includegraphics[width=0.5\textwidth]{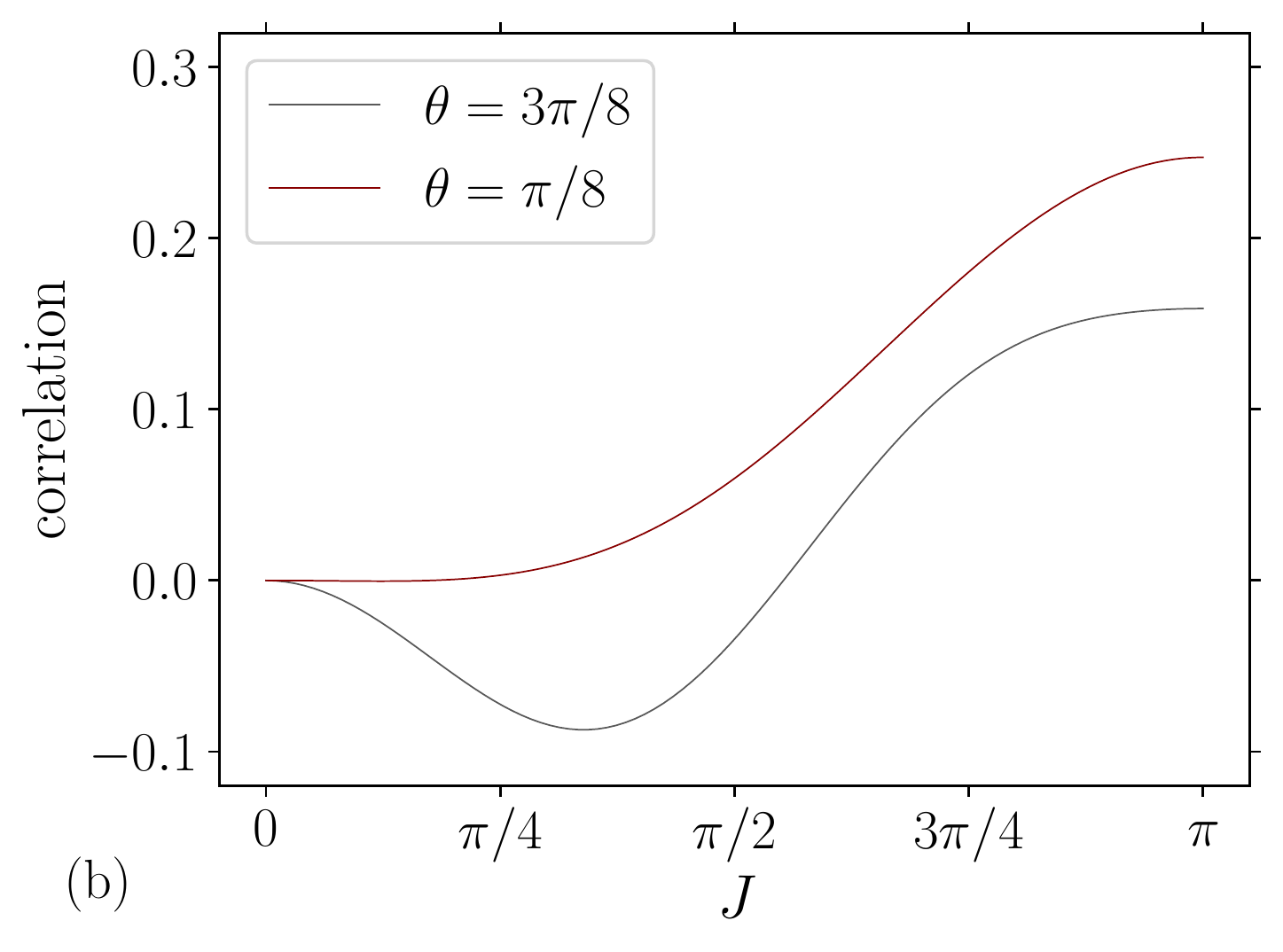}
  \caption{Spin correlation. (a) Particle-spin entropy; (b) spin correlation, as a function of the coupling constant for two angles $\theta=\pi/8, 3\pi/8$. Parameters $t = 2$, `x', $|V| = 2$.
  \label{f:corr}}
\end{figure}

To assert the spin-spin effective interaction we compute the correlation function (in the form of an interaction energy between adjacent spins):
\begin{equation}
\label{e:corr}
C_t(\theta, J) = \braket{\bm \sigma_1 \cdot \bm \sigma_2} - \braket{\bm \sigma_1} \cdot \braket{\bm \sigma_2} \,,
\end{equation}
where the expected value refers to \(\braket{\ldots} = \Tr[\rho_s(t) \ldots]\). For the first time step it gives,
\begin{equation}
\label{e:corr1}
C_1(\theta, J) = \frac{\sin^2(2\theta)}{16} \sin^2 J \,,
\end{equation}
which has a maximum \(C_1 = 1/16\) for \((\theta=\pi/4, J = \pi/2)\); significant correlation is obtained with \(t = 2\), once the particle amplitudes were distributed over the two sites (initially the amplitudes on node 1 are 0). In Fig.~\ref{f:corr} we show \eqref{e:ssrn} and \eqref{e:corr} for \(t=2\), computed numerically. The entropy is not a monotone function of \(J\), as explained before; idem for the correlation function, which also is strongly dependent on the color angle. We note that, for small \(J\), the correlation behaves as \(C\sim J^2\), which is consistent with the mediated character of the spin-spin interaction.

\end{appendix}




\begin{thebibliography}{100}
\providecommand{\url}[1]{\texttt{#1}}
\providecommand{\urlprefix}{URL }
\expandafter\ifx\csname urlstyle\endcsname\relax
  \providecommand{\doi}[1]{doi:\discretionary{}{}{}#1}\else
  \providecommand{\doi}{doi:\discretionary{}{}{}\begingroup
  \urlstyle{rm}\Url}\fi
\providecommand{\eprint}[2][]{\url{#2}}

\bibitem{Szilard-1929kb}
L.~Szilard,
\newblock \emph{Über die {{Entropieverminderung}} in einem thermodynamischen
  {{System}} bei {{Eingriffen}} intelligenter {{Wesen}}},
\newblock Z. Für Phys. \textbf{53}(11), 840 (1929),
\newblock \doi{10.1007/BF01341281}.

\bibitem{Landauer-1961uq}
R.~Landauer,
\newblock \emph{Irreversibility and heat generation in the computing process},
\newblock IBM J. Res. Dev. \textbf{5}(3), 183 (1961),
\newblock \doi{10.1147/rd.53.0183}.

\bibitem{Schumacher-1995uq}
B.~Schumacher,
\newblock \emph{Quantum coding},
\newblock Phys Rev A \textbf{51}, 2738 (1995),
\newblock \doi{10.1103/PhysRevA.51.2738}.

\bibitem{Shannon-1948fj}
C.~E. Shannon,
\newblock \emph{A mathematical theory of communication},
\newblock Bell Syst. Tech. J. \textbf{27}(3), 379 (1948),
\newblock \doi{10.1002/j.1538-7305.1948.tb01338.x}.

\bibitem{Bennett-1998jk}
C.~H. Bennett and P.~W. Shor,
\newblock \emph{Quantum information theory},
\newblock IEEE Trans. Inf. Theory \textbf{44}(6), 2724 (1998),
\newblock \doi{10.1109/18.720553}.

\bibitem{Preskill-1998cl}
J.~Preskill,
\newblock \emph{Lecture notes for physics 229: {{Quantum}} information and
  computation},
\newblock Calif. Inst. Technol.  (1998).

\bibitem{Watrous-2018}
J.~Watrous,
\newblock \emph{The Theory of Quantum Information},
\newblock {Cambridge University Press},
\newblock \doi{10.1017/9781316848142} (2018).

\bibitem{Vedral-2006}
V.~Vedral,
\newblock \emph{Introduction to Quantum Information Science},
\newblock {Oxford University Press},
\newblock \doi{10.1093/acprof:oso/9780199215706.001.0001} (2006).

\bibitem{Schumacher-2010zl}
B.~Schumacher and M.~Westmoreland,
\newblock \emph{Quantum {{Processes}}, {{Systems}}, and {{Information}}},
\newblock {Cambridge University Press}, {Cambridge},
\newblock \doi{10.1017/CBO9780511814006} (2010).

\bibitem{Amico-2008zj}
L.~Amico, R.~Fazio, A.~Osterloh and V.~Vedral,
\newblock \emph{Entanglement in many-body systems},
\newblock Rev. Mod. Phys. \textbf{80}, 517 (2008),
\newblock \doi{10.1103/RevModPhys.80.517}.

\bibitem{Laflorencie-2016fk}
N.~Laflorencie,
\newblock \emph{Quantum entanglement in condensed matter systems},
\newblock Phys. Rep. \textbf{646}, 1 (2016),
\newblock \doi{http://dx.doi.org/10.1016/j.physrep.2016.06.008}.

\bibitem{Abanin-2019}
D.~A. Abanin, E.~Altman, I.~Bloch and M.~Serbyn,
\newblock \emph{Colloquium: {{Many}}-body localization, thermalization, and
  entanglement},
\newblock Rev. Mod. Phys. \textbf{91}(2), 021001 (2019),
\newblock \doi{10.1103/RevModPhys.91.021001}.

\bibitem{Zeng-2019}
B.~Zeng, X.~Chen, D.-L. Zhou and X.-G. Wen,
\newblock \emph{Quantum {{Information Meets Quantum Matter}}},
\newblock {Springer}, {New York, NY} (2019), \eprint{1508.02595}.

\bibitem{Kempe-2003fk}
J.~Kempe,
\newblock \emph{Quantum random walks: {{An}} introductory overview},
\newblock Contemp. Phys. \textbf{44}(4), 307 (2003),
\newblock \doi{10.1080/00107151031000110776}.

\bibitem{Kitagawa-2012fk}
T.~Kitagawa,
\newblock \emph{Topological phenomena in quantum walks: Elementary introduction
  to the physics of topological phases},
\newblock Quantum Inf. Process. \textbf{11}(5), 1107 (2012),
\newblock \doi{10.1007/s11128-012-0425-4}.

\bibitem{Portugal-2013}
R.~Portugal,
\newblock \emph{Quantum Walks and Search Algorithms},
\newblock Quantum {{Science}} and {{Technology}}. {Springer, Cham},
\newblock \doi{10.1007/978-1-4614-6336-8} (2013).

\bibitem{Aharonov-1993fk}
Y.~Aharonov, L.~Davidovich and N.~Zagury,
\newblock \emph{Quantum random walks},
\newblock Phys Rev A \textbf{48}, 1687 (1993),
\newblock \doi{10.1103/PhysRevA.48.1687}.

\bibitem{Meyer-1996sf}
D.~A. Meyer,
\newblock \emph{From quantum cellular automata to quantum lattice gases},
\newblock J Stat Phys \textbf{85}(5), 551 (1996),
\newblock \doi{10.1007/BF02199356}.

\bibitem{Feynman-1986kx}
R.~P. Feynman,
\newblock \emph{Quantum mechanical computers},
\newblock Fond. Phys. \textbf{16}(6), 507 (1986),
\newblock \doi{10.1007/BF01886518}.

\bibitem{Deutsch-1989}
D.~E. Deutsch,
\newblock \emph{Quantum computational networks},
\newblock Proc. R. Soc. Lond. A \textbf{425}, 73 (1989),
\newblock \doi{10.1098/rspa.1989.0099}.

\bibitem{Lovett-2010rm}
N.~B. Lovett, S.~Cooper, M.~Everitt, M.~Trevers and V.~Kendon,
\newblock \emph{Universal quantum computation using the discrete-time quantum
  walk},
\newblock Phys Rev A \textbf{81}, 042330 (2010),
\newblock \doi{10.1103/PhysRevA.81.042330}.

\bibitem{Kendon-2020}
V.~Kendon,
\newblock \emph{How to {{Compute Using Quantum Walks}}},
\newblock Electron. Proc. Theor. Comput. Sci. \textbf{315}, 1 (2020),
\newblock \doi{10.4204/EPTCS.315.1},
\newblock \eprint{2004.01329}.

\bibitem{Kitagawa-2010jk}
T.~Kitagawa, M.~S. Rudner, E.~Berg and E.~Demler,
\newblock \emph{Exploring topological phases with quantum walks},
\newblock Phys Rev A \textbf{82}, 033429 (2010),
\newblock \doi{10.1103/PhysRevA.82.033429}.

\bibitem{Broome-2010}
M.~A. Broome, A.~Fedrizzi, B.~P. Lanyon, I.~Kassal, A.~{Aspuru-Guzik} and A.~G.
  White,
\newblock \emph{Discrete {{Single}}-{{Photon Quantum Walks}} with {{Tunable
  Decoherence}}},
\newblock Phys. Rev. Lett. \textbf{104}(15), 153602 (2010),
\newblock \doi{10.1103/PhysRevLett.104.153602}.

\bibitem{Kitagawa-2012xy}
T.~Kitagawa, M.~A. Broome, A.~Fedrizzi, M.~S. Rudner, E.~Berg, I.~Kassal,
  A.~{Aspuru-Guzik}, E.~Demler and A.~G. White,
\newblock \emph{Observation of topologically protected bound states in photonic
  quantum walks},
\newblock Nat. Commun \textbf{3}, 882 (2012),
\newblock \doi{10.1038/ncomms1872}.

\bibitem{Flurin-2017}
E.~Flurin, V.~V. Ramasesh, S.~{Hacohen-Gourgy}, L.~S. Martin, N.~Y. Yao and
  I.~Siddiqi,
\newblock \emph{Observing {{Topological Invariants Using Quantum Walks}} in
  {{Superconducting Circuits}}},
\newblock Phys. Rev. X \textbf{7}(3), 031023 (2017),
\newblock \doi{10.1103/PhysRevX.7.031023}.

\bibitem{Cardano-2017}
F.~Cardano, A.~D’Errico, A.~Dauphin, M.~Maffei, B.~Piccirillo, C.~{de Lisio},
  G.~De~Filippis, V.~Cataudella, E.~Santamato, L.~Marrucci, M.~Lewenstein and
  P.~Massignan,
\newblock \emph{Detection of {{Zak}} phases and topological invariants in a
  chiral quantum walk of twisted photons},
\newblock Nat. Commun. \textbf{8}, 15516 (2017),
\newblock \doi{10.1038/ncomms15516}.

\bibitem{Xie-2020}
D.~Xie, T.-S. Deng, T.~Xiao, W.~Gou, T.~Chen, W.~Yi and B.~Yan,
\newblock \emph{Topological {{Quantum Walks}} in {{Momentum Space}} with a
  {{Bose}}-{{Einstein Condensate}}},
\newblock Phys. Rev. Lett. \textbf{124}(5), 050502 (2020),
\newblock \doi{10.1103/PhysRevLett.124.050502}.

\bibitem{Omar-2006vn}
Y.~Omar, N.~{Paunkovi ć}, L.~Sheridan and S.~Bose,
\newblock \emph{Quantum walk on a line with two entangled particles},
\newblock Phys Rev A \textbf{74}, 042304 (2006),
\newblock \doi{10.1103/PhysRevA.74.042304}.

\bibitem{Ahlbrecht-2012ec}
A.~Ahlbrecht, A.~Alberti, D.~Meschede, V.~B. Scholz, A.~H. Werner and R.~F.
  Werner,
\newblock \emph{Molecular binding in interacting quantum walks},
\newblock New J Phys \textbf{14}(7), 073050 (2012),
\newblock \doi{10.1088/1367-2630/14/7/073050}.

\bibitem{Bisio-2018}
A.~Bisio, G.~M. D’Ariano, N.~Mosco, P.~Perinotti and A.~Tosini,
\newblock \emph{Solutions of a {{Two}}-{{Particle Interacting Quantum Walk}}},
\newblock Entropy \textbf{20}(6), 435 (2018),
\newblock \doi{10.3390/e20060435}.

\bibitem{Verga-2018}
A.~D. Verga and R.~G. Elias,
\newblock \emph{Entanglement and interaction in a topological quantum walk},
\newblock SciPost Phys. \textbf{5}(2), 019 (2018),
\newblock \doi{10.21468/SciPostPhys.5.2.019}.

\bibitem{Sansoni-2012kx}
L.~Sansoni, F.~Sciarrino, G.~Vallone, P.~Mataloni, A.~Crespi, R.~Ramponi and
  R.~Osellame,
\newblock \emph{Two-{{Particle Bosonic}}-{{Fermionic Quantum Walk}} via
  {{Integrated Photonics}}},
\newblock Phys Rev Lett \textbf{108}, 010502 (2012),
\newblock \doi{10.1103/PhysRevLett.108.010502}.

\bibitem{Schreiber-2012fk}
A.~Schreiber, A.~Gábris, P.~P. Rohde, K.~Laiho, M.~Štefaňák, V.~Potoček,
  C.~Hamilton, I.~Jex and C.~Silberhorn,
\newblock \emph{A {{2D Quantum Walk Simulation}} of {{Two}}-{{Particle
  Dynamics}}},
\newblock Science \textbf{336}(6077), 55 (2012),
\newblock \doi{10.1126/science.1218448}.

\bibitem{Crespi-2013xe}
A.~Crespi, R.~Osellame, R.~Ramponi, V.~Giovannetti, R.~Fazio, L.~Sansoni,
  F.~De~Nicola, F.~Sciarrino and P.~Mataloni,
\newblock \emph{Anderson localization of entangled photons in an integrated
  quantum walk},
\newblock Nat. Photon \textbf{7}(4), 322 (2013),
\newblock \doi{10.1038/nphoton.2013.26}.

\bibitem{Aharonov-2001ty}
D.~Aharonov, A.~Ambainis, J.~Kempe and U.~Vazirani,
\newblock \emph{Quantum {{Walks}} on {{Graphs}}},
\newblock In \emph{Proceedings of the {{Thirty}}-Third {{Annual ACM Symposium}}
  on {{Theory}} of {{Computing}}}, {{STOC}} '01, pp. 50--59. {ACM}, {New York,
  NY, USA},
\newblock \doi{10.1145/380752.380758} (2001).

\bibitem{Tregenna-2003rp}
B.~Tregenna, W.~Flanagan, R.~Maile and V.~Kendon,
\newblock \emph{Controlling discrete quantum walks: Coins and initial states},
\newblock New J Phys \textbf{5}(1), 83 (2003),
\newblock \doi{10.1088/1367-2630/5/1/383}.

\bibitem{Szegedy-2004ul}
M.~Szegedy,
\newblock \emph{Quantum speed-up of {{Markov}} chain based algorithms},
\newblock In \emph{Foundations of {{Computer Science}}, 2004. {{Proceedings}}.
  45th {{Annual IEEE Symposium}}.}, pp. 32--41,
\newblock \doi{10.1109/FOCS.2004.53} (2004).

\bibitem{Grover-1997}
L.~K. Grover,
\newblock \emph{Quantum {{Mechanics Helps}} in {{Searching}} for a {{Needle}}
  in a {{Haystack}}},
\newblock Phys. Rev. Lett. \textbf{79}(2), 325 (1997),
\newblock \doi{10.1103/PhysRevLett.79.325}.

\bibitem{Shenvi-2003fk}
N.~Shenvi, J.~Kempe and K.~B. Whaley,
\newblock \emph{Quantum random-walk search algorithm},
\newblock Phys Rev A \textbf{67}, 052307 (2003),
\newblock \doi{10.1103/PhysRevA.67.052307}.

\bibitem{Ambainis-2004qq}
A.~Ambainis, J.~Kempe and A.~Rivosh,
\newblock \emph{Coins {{Make Quantum Walks Faster}}},
\newblock Proc 16th ACM-SIAM SODA 2005 pp. 1099--1108 (2004),
\newblock \eprint{quant-ph/0402107}.

\bibitem{Stefanak-2016}
M.~Štefaňák and S.~Skoupý,
\newblock \emph{Perfect state transfer by means of discrete-time quantum walk
  search algorithms on highly symmetric graphs},
\newblock Phys. Rev. A \textbf{94}(2), 022301 (2016),
\newblock \doi{10.1103/PhysRevA.94.022301}.

\bibitem{Berry-2011qq}
S.~D. Berry and J.~B. Wang,
\newblock \emph{Two-particle quantum walks: {{Entanglement}} and graph
  isomorphism testing},
\newblock Phys Rev A \textbf{83}, 042317 (2011),
\newblock \doi{10.1103/PhysRevA.83.042317}.

\bibitem{Wang-2015}
H.~Wang, J.~Wu, X.~Yang and X.~Yi,
\newblock \emph{A graph isomorphism algorithm using signatures computed via
  quantum walk search model},
\newblock J. Phys. A: Math. Theor. \textbf{48}(11), 115302 (2015),
\newblock \doi{10.1088/1751-8113/48/11/115302}.

\bibitem{Mills-2019}
P.~W. Mills, R.~P. Rundle, J.~H. Samson, S.~J. Devitt, T.~Tilma, V.~M. Dwyer
  and M.~J. Everitt,
\newblock \emph{Quantum invariants and the graph isomorphism problem},
\newblock Phys. Rev. A \textbf{100}(5), 052317 (2019),
\newblock \doi{10.1103/PhysRevA.100.052317}.

\bibitem{Farrelly-2017}
T.~Farrelly,
\newblock \emph{Insights from {{Quantum Information}} into {{Fundamental
  Physics}}},
\newblock ArXiv170808897 Quant-Ph  (2017),
\newblock \eprint{1708.08897}.

\bibitem{Bisio-2018nr}
A.~Bisio, G.~M. D'Ariano, P.~Perinotti and A.~Tosini,
\newblock \emph{Thirring quantum cellular automaton},
\newblock Phys Rev A \textbf{97}, 032132 (2018),
\newblock \doi{10.1103/PhysRevA.97.032132}.

\bibitem{Arrighi-2019a}
P.~Arrighi,
\newblock \emph{An overview of quantum cellular automata},
\newblock Nat Comput \textbf{18}(4), 885 (2019),
\newblock \doi{10.1007/s11047-019-09762-6}.

\bibitem{Nielsen-2006fv}
M.~A. Nielsen,
\newblock \emph{Cluster-state quantum computation},
\newblock Rep. Math. Phys. \textbf{57}(1), 147 (2006),
\newblock \doi{http://dx.doi.org/10.1016/S0034-4877(06)80014-5}.

\bibitem{Raussendorf-2001uq}
R.~Raussendorf and H.~J. Briegel,
\newblock \emph{A {{One}}-{{Way Quantum Computer}}},
\newblock Phys Rev Lett \textbf{86}, 5188 (2001),
\newblock \doi{10.1103/PhysRevLett.86.5188}.

\bibitem{Raussendorf-2003rm}
R.~Raussendorf, D.~E. Browne and H.~J. Briegel,
\newblock \emph{Measurement-based quantum computation on cluster states},
\newblock Phys Rev A \textbf{68}, 022312 (2003),
\newblock \doi{10.1103/PhysRevA.68.022312}.

\bibitem{Nielsen-2003}
M.~A. Nielsen,
\newblock \emph{Quantum computation by measurement and quantum memory},
\newblock Physics Letters A \textbf{308}(2), 96 (2003),
\newblock \doi{10.1016/S0375-9601(02)01803-0}.

\bibitem{Barenco-1995}
A.~Barenco, C.~H. Bennett, R.~Cleve, D.~P. DiVincenzo, N.~Margolus, P.~Shor,
  T.~Sleator, J.~A. Smolin and H.~Weinfurter,
\newblock \emph{Elementary gates for quantum computation},
\newblock Phys. Rev. A \textbf{52}(5), 3457 (1995),
\newblock \doi{10.1103/PhysRevA.52.3457}.

\bibitem{Lloyd-1996}
S.~Lloyd,
\newblock \emph{Universal {{Quantum Simulators}}},
\newblock Science \textbf{273}(5278), 1073 (1996),
\newblock \doi{10.1126/science.273.5278.1073}.

\bibitem{DiVincenzo-2000cr}
D.~P. DiVincenzo, D.~Bacon, J.~Kempe, G.~Burkard and K.~B. Whaley,
\newblock \emph{Universal quantum computation with the exchange interaction},
\newblock Nature \textbf{408}, 339 (2000),
\newblock \doi{10.1038/35042541}.

\bibitem{Briegel-2001fk}
H.~J. Briegel and R.~Raussendorf,
\newblock \emph{Persistent {{Entanglement}} in {{Arrays}} of {{Interacting
  Particles}}},
\newblock Phys Rev Lett \textbf{86}, 910 (2001),
\newblock \doi{10.1103/PhysRevLett.86.910}.

\bibitem{Hein-2006eu}
M.~Hein, W.~Dür, J.~Eisert, R.~Raussendorf, M.~Nest and H.-J. Briegel,
\newblock \emph{Entanglement in graph states and its applications},
\newblock ArXiv Prepr.  (2006),
\newblock \eprint{quant-ph/0602096}.

\bibitem{Ambainis-2003if}
A.~Ambainis,
\newblock \emph{Quantum walks and their algorithmic applications},
\newblock Int J Quantum Inf. \textbf{01}(04), 507 (2003),
\newblock \doi{10.1142/S0219749903000383}.

\bibitem{Bose-2003}
S.~Bose,
\newblock \emph{Quantum {{Communication}} through an {{Unmodulated Spin
  Chain}}},
\newblock Phys. Rev. Lett. \textbf{91}(20), 207901 (2003),
\newblock \doi{10.1103/PhysRevLett.91.207901}.

\bibitem{Christandl-2004}
M.~Christandl, N.~Datta, A.~Ekert and A.~J. Landahl,
\newblock \emph{Perfect {{State Transfer}} in {{Quantum Spin Networks}}},
\newblock Phys. Rev. Lett. \textbf{92}(18), 187902 (2004),
\newblock \doi{10.1103/PhysRevLett.92.187902}.

\bibitem{Bradley-2019}
C.~E. Bradley, J.~Randall, M.~H. Abobeih, R.~C. Berrevoets, M.~J. Degen, M.~A.
  Bakker, M.~Markham, D.~J. Twitchen and T.~H. Taminiau,
\newblock \emph{A {{Ten}}-{{Qubit Solid}}-{{State Spin Register}} with
  {{Quantum Memory}} up to {{One Minute}}},
\newblock Phys. Rev. X \textbf{9}(3), 031045 (2019),
\newblock \doi{10.1103/PhysRevX.9.031045}.

\bibitem{Pfaff-2014}
W.~Pfaff, B.~J. Hensen, H.~Bernien, S.~B. van Dam, M.~S. Blok, T.~H. Taminiau,
  M.~J. Tiggelman, R.~N. Schouten, M.~Markham, D.~J. Twitchen and R.~Hanson,
\newblock \emph{Unconditional quantum teleportation between distant solid-state
  quantum bits},
\newblock Science \textbf{345}(6196), 532 (2014),
\newblock \doi{10.1126/science.1253512}.

\bibitem{Waldherr-2014}
G.~Waldherr, Y.~Wang, S.~Zaiser, M.~Jamali, T.~{Schulte-Herbrüggen}, H.~Abe,
  T.~Ohshima, J.~Isoya, J.~F. Du, P.~Neumann and J.~Wrachtrup,
\newblock \emph{Quantum error correction in a solid-state hybrid spin
  register},
\newblock Nature \textbf{506}(7487), 204 (2014),
\newblock \doi{10.1038/nature12919}.

\bibitem{Verga-2019}
A.~D. Verga,
\newblock \emph{Interacting quantum walk on a graph},
\newblock Phys. Rev. E \textbf{99}(1), 012127 (2019),
\newblock \doi{10.1103/PhysRevE.99.012127}.

\bibitem{Verga-2019b}
A.~D. Verga and R.~G. Elías,
\newblock \emph{Thermal state entanglement entropy on a quantum graph},
\newblock Phys. Rev. E \textbf{100}(6), 062137 (2019),
\newblock \doi{10.1103/PhysRevE.100.062137}.

\bibitem{Ruderman-1954}
M.~A. Ruderman and C.~Kittel,
\newblock \emph{Indirect {{Exchange Coupling}} of {{Nuclear Magnetic Moments}}
  by {{Conduction Electrons}}},
\newblock Phys. Rev. \textbf{96}(1), 99 (1954),
\newblock \doi{10.1103/PhysRev.96.99}.

\bibitem{Kasuya-1956}
T.~Kasuya,
\newblock \emph{A {{Theory}} of {{Metallic Ferro}}- and {{Antiferromagnetism}}
  on {{Zener}}'s {{Model}}},
\newblock Prog Theor Phys \textbf{16}(1), 45 (1956),
\newblock \doi{10.1143/PTP.16.45}.

\bibitem{Nolting-2009}
W.~Nolting and A.~Ramakanth,
\newblock \emph{Quantum Theory of Magnetism},
\newblock {Springer Science \& Business Media} (2009).

\bibitem{Lauchli-2008}
A.~M. Läuchli and C.~Kollath,
\newblock \emph{Spreading of correlations and entanglement after a quench in
  the one-dimensional {{Bose}}–{{Hubbard}} model},
\newblock J. Stat. Mech. \textbf{2008}(05), P05018 (2008),
\newblock \doi{10.1088/1742-5468/2008/05/P05018}.

\bibitem{Zhang-2015}
L.~Zhang, H.~Kim and D.~A. Huse,
\newblock \emph{Thermalization of entanglement},
\newblock Phys. Rev. E \textbf{91}(6), 062128 (2015),
\newblock \doi{10.1103/PhysRevE.91.062128}.

\bibitem{Nahum-2017qf}
A.~Nahum, J.~Ruhman, S.~Vijay and J.~Haah,
\newblock \emph{Quantum {{Entanglement Growth}} under {{Random Unitary
  Dynamics}}},
\newblock Phys Rev X \textbf{7}, 031016 (2017),
\newblock \doi{10.1103/PhysRevX.7.031016}.

\bibitem{Hackl-2018}
L.~Hackl, E.~Bianchi, R.~Modak and M.~Rigol,
\newblock \emph{Entanglement production in bosonic systems: {{Linear}} and
  logarithmic growth},
\newblock Phys. Rev. A \textbf{97}(3), 032321 (2018),
\newblock \doi{10.1103/PhysRevA.97.032321}.

\bibitem{DErrico-2020}
A.~D’Errico, F.~Cardano, M.~Maffei, A.~Dauphin, R.~Barboza, C.~Esposito,
  B.~Piccirillo, M.~Lewenstein, P.~Massignan and L.~Marrucci,
\newblock \emph{Two-dimensional topological quantum walks in the momentum space
  of structured light},
\newblock Optica, OPTICA \textbf{7}(2), 108 (2020),
\newblock \doi{10.1364/OPTICA.365028}.

\bibitem{Di-Molfetta-2016kn}
G.~Di~Molfetta and A.~Pérez,
\newblock \emph{Quantum walks as simulators of neutrino oscillations in a
  vacuum and matter},
\newblock New J Phys \textbf{18}(10), 103038 (2016),
\newblock \doi{10.1088/1367-2630/18/10/103038}.

\bibitem{Bisio-2017}
A.~Bisio, G.~M. D’Ariano and P.~Perinotti,
\newblock \emph{Quantum {{Walks}}, {{Weyl Equation}} and the {{Lorentz
  Group}}},
\newblock Found Phys \textbf{47}(8), 1065 (2017),
\newblock \doi{10.1007/s10701-017-0086-3}.

\bibitem{Arrighi-2018}
P.~Arrighi, G.~Di~Molfetta, I.~{Márquez-Martín} and A.~Pérez,
\newblock \emph{Dirac equation as a quantum walk over the honeycomb and
  triangular lattices},
\newblock Phys. Rev. A \textbf{97}(6), 062111 (2018),
\newblock \doi{10.1103/PhysRevA.97.062111}.

\bibitem{Landau-1935fk}
L.~D. Landau and E.~M. Lifshitz,
\newblock \emph{On the theory of the dispersion of magnetic permeability in
  ferromagnetic bodies},
\newblock Phys Zeitsch Sow \textbf{8}, 153 (1935).

\bibitem{Slonczewski-1996lq}
J.~C. Slonczewski,
\newblock \emph{Current-driven excitation of magnetic multilayers},
\newblock J. Magn. Magn. Mater. \textbf{159}(1-2), L1 (1996),
\newblock \doi{10.1016/0304-8853(96)00062-5}.

\bibitem{Tatara-2019}
G.~Tatara,
\newblock \emph{Effective gauge field theory of spintronics},
\newblock Physica E: Low-dimensional Systems and Nanostructures \textbf{106},
  208 (2019),
\newblock \doi{10.1016/j.physe.2018.05.011}.

\bibitem{Strauch-2007}
F.~W. Strauch,
\newblock \emph{Relativistic effects and rigorous limits for discrete- and
  continuous-time quantum walks},
\newblock Journal of Mathematical Physics \textbf{48}(8), 082102 (2007),
\newblock \doi{10.1063/1.2759837}.

\bibitem{Di-Molfetta-2012fv}
G.~Di~Molfetta and F.~Debbasch,
\newblock \emph{Discrete-time quantum walks: {{Continuous}} limit and
  symmetries},
\newblock J Math Phys \textbf{53}(12), 123302 (2012),
\newblock \doi{http://dx.doi.org/10.1063/1.4764876}.

\bibitem{Jozsa-1999a}
R.~Jozsa,
\newblock \emph{Quantum {{Effects}} in {{Algorithms}}},
\newblock In C.~P. Williams, ed., \emph{Quantum {{Computing}} and {{Quantum
  Communications}}}, Lecture {{Notes}} in {{Computer Science}}, pp. 103--112.
  {Springer}, {Berlin, Heidelberg},
\newblock ISBN 978-3-540-49208-5,
\newblock \doi{10.1007/3-540-49208-9_7} (1999).

\bibitem{Bennett-1992}
C.~H. Bennett and S.~J. Wiesner,
\newblock \emph{Communication via one- and two-particle operators on
  {{Einstein}}-{{Podolsky}}-{{Rosen}} states},
\newblock Phys. Rev. Lett. \textbf{69}(20), 2881 (1992),
\newblock \doi{10.1103/PhysRevLett.69.2881}.

\bibitem{Shor-1994qr}
P.~W. Shor,
\newblock \emph{Algorithms for quantum computation: Discrete logarithms and
  factoring},
\newblock In \emph{Foundations of {{Computer Science}}, 1994 {{Proceedings}}.,
  35th {{Annual Symposium}} On}, pp. 124--134,
\newblock \doi{10.1109/SFCS.1994.365700} (1994).

\bibitem{Harrow-2009}
A.~W. Harrow, A.~Hassidim and S.~Lloyd,
\newblock \emph{Quantum {{Algorithm}} for {{Linear Systems}} of {{Equations}}},
\newblock Phys. Rev. Lett. \textbf{103}(15), 150502 (2009),
\newblock \doi{10.1103/PhysRevLett.103.150502}.

\bibitem{Kendon-2005fk}
V.~Kendon and B.~C. Sanders,
\newblock \emph{Complementarity and quantum walks},
\newblock Phys Rev A \textbf{71}, 022307 (2005),
\newblock \doi{10.1103/PhysRevA.71.022307}.

\bibitem{Kendon-2011}
V.~M. Kendon and C.~Tamon,
\newblock \emph{Perfect state transfer in quantum walks on graphs},
\newblock J. Comput. Theor. Nanosci. \textbf{8}(3), 422 (2011),
\newblock \doi{10.1166/jctn.2011.1706}.

\bibitem{Berger-1984uq}
L.~Berger,
\newblock \emph{Exchange interaction between ferromagnetic domain wall and
  electric current in very thin metallic films},
\newblock J. Appl. Phys. \textbf{55}(6), 1954 (1984),
\newblock \doi{10.1063/1.333530}.

\bibitem{Horodecki-1994}
R.~Horodecki and P.~Horodecki,
\newblock \emph{Quantum redundancies and local realism},
\newblock Physics Letters A \textbf{194}(3), 147 (1994),
\newblock \doi{10.1016/0375-9601(94)91275-0}.

\bibitem{Bennett-1996fr}
C.~H. Bennett, H.~J. Bernstein, S.~Popescu and B.~Schumacher,
\newblock \emph{Concentrating partial entanglement by local operations},
\newblock Phys Rev A \textbf{53}, 2046 (1996),
\newblock \doi{10.1103/PhysRevA.53.2046}.

\bibitem{Wootters-1998}
W.~K. Wootters,
\newblock \emph{Entanglement of {{Formation}} of an {{Arbitrary State}} of
  {{Two Qubits}}},
\newblock Phys. Rev. Lett. \textbf{80}(10), 2245 (1998),
\newblock \doi{10.1103/PhysRevLett.80.2245}.

\bibitem{Horodecki-2009yg}
R.~Horodecki, P.~Horodecki, M.~Horodecki and K.~Horodecki,
\newblock \emph{Quantum entanglement},
\newblock Rev Mod Phys \textbf{81}, 865 (2009),
\newblock \doi{10.1103/RevModPhys.81.865}.

\bibitem{Asboth-2012qy}
J.~K. Asbóth,
\newblock \emph{Symmetries, topological phases, and bound states in the
  one-dimensional quantum walk},
\newblock Phys Rev B \textbf{86}, 195414 (2012),
\newblock \doi{10.1103/PhysRevB.86.195414}.

\bibitem{Wang-2004fk}
X.~Wang, S.~Ghose, B.~C. Sanders and B.~Hu,
\newblock \emph{Entanglement as a signature of quantum chaos},
\newblock Phys Rev E \textbf{70}, 016217 (2004),
\newblock \doi{10.1103/PhysRevE.70.016217}.

\bibitem{Shen-2011kx}
S.-Q. Shen, W.-Y. Shan and H.-Z. Lu,
\newblock \emph{Topological {{Insulator}} and the {{Dirac Equation}}},
\newblock SPIN \textbf{01}(01), 33 (2011),
\newblock \doi{10.1142/S2010324711000057}.

\bibitem{Page-1993nr}
D.~N. Page,
\newblock \emph{Average entropy of a subsystem},
\newblock Phys Rev Lett \textbf{71}, 1291 (1993),
\newblock \doi{10.1103/PhysRevLett.71.1291}.

\bibitem{Ralph-2008ly}
D.~C. Ralph and M.~D. Stiles,
\newblock \emph{Spin transfer torques},
\newblock J Magn Magn Mater \textbf{320}(7), 1190 (2008),
\newblock \doi{10.1016/j.jmmm.2007.12.019}.

\bibitem{Sayad-2016a}
M.~Sayad, R.~Rausch and M.~Potthoff,
\newblock \emph{Inertia effects in the real-time dynamics of a quantum spin
  coupled to a {{Fermi}} sea},
\newblock EPL \textbf{116}(1), 17001 (2016),
\newblock \doi{10.1209/0295-5075/116/17001}.

\bibitem{Baryakhtar-2006cq}
V.~G. Baryakhtar and A.~G. Danielevich,
\newblock \emph{Spin-wave damping at spin-orientation phase transitions},
\newblock Low Temp. Phys. \textbf{32}(8), 768 (2006),
\newblock \doi{10.1063/1.2219498}.

\bibitem{Verga-2014fk}
A.~D. Verga,
\newblock \emph{Skyrmion to ferromagnetic state transition: {{A}} description
  of the topological change as a finite-time singularity in the skyrmion
  dynamics},
\newblock Phys Rev B \textbf{90}, 174428 (2014),
\newblock \doi{10.1103/PhysRevB.90.174428}.

\bibitem{Garanin-1997ly}
D.~A. Garanin,
\newblock \emph{Fokker-{{Planck}} and {{Landau}}-{{Lifshitz}}-{{Bloch}}
  equations for classical ferromagnets},
\newblock Phys Rev B \textbf{55}(5) (1997),
\newblock \doi{10.1103/PhysRevB.55.3050}.

\bibitem{Lee-2015uq}
C.-W. Lee, P.~Kurzynski and H.~Nha,
\newblock \emph{Quantum walk as a simulator of nonlinear dynamics: {{Nonlinear
  Dirac}} equation and solitons},
\newblock Phys Rev A \textbf{92}, 052336 (2015),
\newblock \doi{10.1103/PhysRevA.92.052336}.

\bibitem{Mitropolsky-1997}
Y.~A. Mitropolsky and N.~Van~Dao,
\newblock \emph{Applied Asymptotic Methods in Nonlinear Oscillations},
\newblock {Kluwer Academic (Springer, Dordrecht)},
\newblock \doi{10.1007/978-94-011-5752-0} (1997).

\bibitem{Kakutani-1974}
T.~Kakutani and N.~Sugimoto,
\newblock \emph{Krylov‐{{Bogoliubov}}‐{{Mitropolsky}} method for nonlinear
  wave modulation},
\newblock The Physics of Fluids \textbf{17}(8), 1617 (1974),
\newblock \doi{10.1063/1.1694942}.

\bibitem{Tserkovnyak-2012ys}
Y.~Tserkovnyak and D.~Loss,
\newblock \emph{Thin-{{Film Magnetization Dynamics}} on the {{Surface}} of a
  {{Topological Insulator}}},
\newblock Phys Rev Lett \textbf{108}, 187201 (2012),
\newblock \doi{10.1103/PhysRevLett.108.187201}.

\bibitem{Chen-2014db}
J.~Chen, M.~B. Abdul~Jalil and S.~G. Tan,
\newblock \emph{Current-{{Induced Spin Torque}} on {{Magnetization Textures
  Coupled}} to the {{Topological Surface States}} of {{Three}}-{{Dimensional
  Topological Insulators}}},
\newblock J. Phys. Soc. Jpn. \textbf{83}(6), 064710 (2014),
\newblock \doi{10.7566/JPSJ.83.064710}.

\bibitem{Kadowaki-1998}
T.~Kadowaki and H.~Nishimori,
\newblock \emph{Quantum annealing in the transverse {{Ising}} model},
\newblock Phys. Rev. E \textbf{58}(5), 5355 (1998),
\newblock \doi{10.1103/PhysRevE.58.5355}.

\bibitem{Farhi-2000}
E.~Farhi, J.~Goldstone, S.~Gutmann and M.~Sipser,
\newblock \emph{Quantum {{Computation}} by {{Adiabatic Evolution}}},
\newblock ArXiv Quant-Ph0001106  (2000),
\newblock \eprint{quant-ph/0001106}.

\bibitem{Aharonov-2008}
D.~Aharonov, W.~{van Dam}, J.~Kempe, Z.~Landau, S.~Lloyd and O.~Regev,
\newblock \emph{Adiabatic {{Quantum Computation Is Equivalent}} to {{Standard
  Quantum Computation}}},
\newblock SIAM Rev. \textbf{50}(4), 755 (2008),
\newblock \doi{10.1137/080734479}.

\bibitem{Albash-2018}
T.~Albash and D.~A. Lidar,
\newblock \emph{Adiabatic quantum computation},
\newblock Rev. Mod. Phys. \textbf{90}(1), 015002 (2018),
\newblock \doi{10.1103/RevModPhys.90.015002}.

\bibitem{Pagliero-2020}
D.~Pagliero, P.~R. Zangara, J.~Henshaw, A.~Ajoy, R.~H. Acosta, J.~A. Reimer,
  A.~Pines and C.~A. Meriles,
\newblock \emph{Optically pumped spin polarization as a probe of many-body
  thermalization},
\newblock Sci. Adv. \textbf{6}(18), eaaz6986 (2020),
\newblock \doi{10.1126/sciadv.aaz6986}.

\bibitem{Maier-2019}
C.~Maier, T.~Brydges, P.~Jurcevic, N.~Trautmann, C.~Hempel, B.~P. Lanyon,
  P.~Hauke, R.~Blatt and C.~F. Roos,
\newblock \emph{Environment-{{Assisted Quantum Transport}} in a 10-qubit
  {{Network}}},
\newblock Phys. Rev. Lett. \textbf{122}(5), 050501 (2019),
\newblock \doi{10.1103/PhysRevLett.122.050501}.

\bibitem{Cirac-1997}
J.~I. Cirac, P.~Zoller, H.~J. Kimble and H.~Mabuchi,
\newblock \emph{Quantum {{State Transfer}} and {{Entanglement Distribution}}
  among {{Distant Nodes}} in a {{Quantum Network}}},
\newblock Phys. Rev. Lett. \textbf{78}(16), 3221 (1997),
\newblock \doi{10.1103/PhysRevLett.78.3221}.

\bibitem{Hunter-2007}
J.~D. Hunter,
\newblock \emph{Matplotlib: {{A 2D Graphics Environment}}},
\newblock Comput. Sci. Eng. \textbf{9}(3), 90 (2007),
\newblock \doi{10.1109/MCSE.2007.55}.

\end{thebibliography}

\nolinenumbers

\end{document}